\documentclass[12pt]{article}

\usepackage{float}
\usepackage[fleqn]{amsmath}
\usepackage{amsfonts}
\usepackage{amssymb}
\usepackage{epsfig}
\usepackage{geometry}
\usepackage{ctable}

\DeclareMathAlphabet{\mathpzc}{OT1}{pzc}{m}{it}

\newcommand{\limpt}{\mathrm{limpt}\,}

\newtheorem{theo}{Theorem}[section]
\newtheorem{prop}[theo]{Proposition}
\newtheorem{lemm}[theo]{Lemma}
\newtheorem{coro}[theo]{Corollary}
\newtheorem{defi}[theo]{Definition}
\newtheorem{conj}[theo]{Conjecture}

\newtheorem{rema}[theo]{Remark}

\renewcommand{\iint}{{\int\!\!\!\!\int}}

\newcommand{\sign}{\mathrm{sign}}

\newcommand{\vareps}{\varepsilon}

\newcommand\gFUNC{{\mathfrak{g}}}
\newcommand\hFUNC{{\mathfrak{h}}}

\newcommand\eett{{\mathfrak{e}}}

\newcommand{\sV}{\mathbf{s}}           

\newcommand{\uli}[1]{\underline #1 }

\newcommand{\dd}{\mathrm{d}}

\newcommand{\Cset}{\mathbb{C}}

\newcommand{\Nset}{\mathbb{N}}

\newcommand{\Rset}{\mathbb{R}}
\newcommand{\Sset}{\mathbb{S}}
\newcommand{\Zset}{\mathbb{Z}}

\newcommand{\cS}{\mathcal{S}}

\newcommand{\dst}{\displaystyle}
\newcommand{\sst}{\scriptstyle}

\newcommand{\tst}{\textstyle}

\newcommand{\sfrNT}{{\sst{\frac{\beta}{N}}}}

\begin{document}
\title{Heuristic Relative Entropy Principles with Complex \\
Measures: Large-Degree Asymptotics of a Family  \\ 
$\!\!\!\!$of Multi-Variate Normal Random Polynomials$\!\!\!$\vspace{-17pt}}
\author{\sc{Michael Karl-Heinz Kiessling\vspace{-8pt}}\\ 
{\tiny{Department of Mathematics, Rutgers University, Piscataway, NJ 08854}}\vspace{-10pt}}
\date{\small Dedidated to the memory of my mother,\\ Elisabeth Kiessling, n\'ee Appeltrath.\vspace{-22pt}} 
\maketitle

\begin{abstract}\vspace{-2pt}
\noindent
 Let $z\in \Cset$, let $\sigma^2>0$ be a variance, and for $N\in\Nset$ define the integrals\vspace{-5pt}
\begin{equation}\notag
\hspace{-17pt}
E_N^{}(z;\sigma) := 
\left\{\hspace{-7.5pt}
 \begin{array}{ll}
&\displaystyle\!\!\!\!\!
{\textstyle\frac{1}{\sigma}}
\!\!\!\int_{\Rset}\! (x^2+z^2) \frac{e^{-\frac{1}{2\sigma^2} x^2}}{\sqrt{2\pi}}dx\ ..................................................\ 
\mbox{if}\, N=1,\cr
&\displaystyle\!\!\!\!\!
{\textstyle\frac{1}{\sigma}} \!\!\!\int_{\Rset^N}\!
\prod\!\!\prod_{\hspace{-16pt}1\leq k<l\leq N}\!\! e^{-\frac{1}{2N}(1-\sigma^{-2}) (x_k-x_l)^2}
\!\!\!\!\!\prod_{1\leq n\leq N}\!\!\!\!(x_n^2+z^2) \frac{e^{-\frac{1}{2\sigma^2} x_n^2}}{\sqrt{2\pi}}dx_n \ \mbox{if}\, N>1.
\vspace{-8pt}
\end{array}\right.\!\!\!
\end{equation}
 These are expected values of the polynomials $P_N^{}(z)=\prod_{1\leq n\leq N}(X_n^2+z^2)$ whose
$2N$ zeros $\{\pm i X_k\}^{}_{k=1,...,N}$ are generated by $N$ identically distributed multi-variate mean-zero normal random variables 
$\{X_k\}^{N}_{k=1}$ with co-variance 
${\rm{Cov}}_N^{}(X_k,X_l)=(1+\frac{\sigma^2-1}{N})\delta_{k,l}+\frac{\sigma^2-1}{N}(1-\delta_{k,l})$.
 The $E_N^{}(z;\sigma)$ are polynomials in $z^2$, explicitly computable for arbitrary $N$,
yet a list of the first three $E_N^{}(z;\sigma)$ shows that the expressions become unwieldy already for moderate $N$ ---
unless $\sigma = 1$, in which case $E_N^{}(z;1) = (1+z^2)^N$ for all $z\in\Cset$ and $N\in\Nset$.
 (Incidentally, commonly available computer algebra evaluates the integrals $E_N^{}(z;\sigma)$ only for $N$ up to a dozen,
due to memory constraints).
 Asymptotic evaluations are needed for the large-$N$ regime. 
 For general complex $z$ these have traditionally been limited to analytic 
expansion techniques; several rigorous results are proved for complex $z$ near $0$.
 Yet if $z\in\Rset$ one can also compute this ``infinite-degree'' limit with the
help of the familiar relative entropy principle for probability measures; a rigorous proof of this fact is supplied.
 Computer algebra-generated evidence is presented in support of a conjecture that a generalization of the
relative entropy principle to \emph{signed or complex measures} governs the $N\to\infty$ asymptotics of the regime $iz\in\Rset$. 
 Potential generalizations, in particular to 
point vortex ensembles and the prescribed Gauss curvature problem, and to random matrix ensembles, are emphasized.
\smallskip\hrule\medskip

\noindent
Original: August 30, 2016; Revised: June 11, 2017; Final: August 01, 2017
\end{abstract}
\vfill

\hrule\smallskip
\copyright{2017} The author. Reproduction for non-commercial purposes is permitted.\vspace{-10pt}
\newpage

\section{Introduction}
\vspace{-10pt}

 The primary purpose of this article is to introduce the novel notion of 
an \emph{entropy of a signed or complex measure, relative to a signed a-priori measure}.
 We will present empirical evidence for its potential usefulness as a statistical tool in the asymptotic evaluation of 
sign-changing (or complex) expected value functionals which occur in science and mathematics.

 In this introduction we first recall the original motivation for contemplating such an 
extension of the familiar statistical mechanics entropy principles to signed measures, namely a question
which Alice Chang asked the author about proving existence of solutions to the sign-changing prescribed Gauss 
curvature equation of differential geometry using a continuum limit $N\to\infty$ of Onsager's \cite{Lars} 
equilibrium statistical distributions of $N\in\Nset$ point vortices in two-dimensional Euler fluids. 
 We next formulate some questions about the zeros of complex random polynomials which likewise suggest to look
for a signed, or complex relative entropy principle to help answering (some of) them.
 The study of these complex random polynomials grew out of attempts to make progress on 
Alice Chang's question by studying simpler, one-dimensional, models with quadratic rather than logarithmic
pair interactions --- which then turned out to be of independent interest.
 In the main part of this article we focus entirely on such a Gaussian ensemble of complex random polynomials.
 This facilitates the explanation of the key ideas, which will be backed up with graphical evidence produced by computer-algebra.
 From there it will be only a small step to conceive of other potential applications, see our section: ``Summary and Outlook.''
\vspace{-15pt}
\subsection{Prescribing Gauss curvature using point vortices}
 \vspace{-5pt}

 Suppose $N$ point vortices with positions $\sV_k\in\Rset^2$, $k=1,...,N$, are distributed by a canonical ensemble 
probability measure 
\begin{equation}
	\dd\mu^{(N)}_{\beta}
= \label{holoMU}
\frac{\qquad e^{- {\sfrNT}H^{(N)}}\dd^N\mu_0}{\int_{\Rset^{2N}}\! e^{-  {\sfrNT}H^{(N)}} \dd^N\mu_0},
\end{equation}
where $\dd\mu_0$ is a two-dimensional a-priori measure (e.g. $\dd\mu_0 = f(\sV)\dd^2 s$, where $f(\sV)$ is
a positive Schwartz function and $\dd^2s$ Lebesgue measure), and
\begin{equation}
	H^{(N)}(\sV_1,...,\sV_N) 
\ =\ \label{hamiltonian}
	\tst\frac{c^2}{2\pi}\; {\tst{\sum\!\!\!\sum\limits_{\hskip-14pt 1\leq j<k\leq N}^{}}}   \ln \frac{1}{|\sV_j -\sV_k|} 
\end{equation}
is Kirchhoff's Hamilton function;\footnote{Kirchhoff's Hamilton function generates point vortex motion in $\Rset^2$
without boundaries or externally produced stream functions. The a-priori measure $\mu_0$ adds an external 
stream function $\sum_k\ln f(\sV_k)$ to $-\sfrNT H^{(N)}$ to prevent the vortices from escaping to spatial $\infty$.}
 here, $c\in\Rset$ is the circulation of each vortex, which we now set equal to unity (defining pertinent physical units).
 Lastly, $N/\beta$ is the \emph{Onsager temperature} of a ``vortex heat bath,'' satisfying $\beta > - 8\pi$ (more precisely:
$c^2\beta>-8\pi$, if $c$ is not unity).
 The Onsager temperature of the heat bath in- or decreases (according as $\beta >0$ or $\beta <0$, with $\beta$ fixed) 
proportional to the number $N$ of vortices in the system in order to counter the superlinear growth of the vortex energy
given by the $N(N-1)/2$ pair interaction terms. 
 This so-called mean-field scaling gives rise to an interesting continuum limit $N\to\infty$
(see \cite{CLMPa}, \cite{KiesslingCPAM}, \cite{ChaKieDMJ}) in which 
the canonical measure concentrates on normalized point vortex densities $\rho(\sV)$ relative to 
$\dd\uli{\mu}_0 := \dd\mu_0/\int \dd\mu_0$ (i.e. $\int\!\! \rho(\sV)\dd\uli\mu_0=1$) which satisfy
\begin{equation}
	\rho(\sV) 
=\label{fixPOINTEQrho}
	\frac{ \dst 
	e^{- \frac{\beta}{2\pi}{\tst\int}\!\rho(\tilde{\sV}) \ln \frac{1}{|\sV -\tilde{\sV}|}\dd\uli\mu_0(\tilde{\sV})}}
   {\dst\int_{\Rset^2}e^{-\frac{\beta}{2\pi}{\tst\int}\!\rho(\tilde{\sV}) \ln \frac{1}{|\hat{\sV} -\tilde{\sV}|}\dd\uli\mu_0(\tilde{\sV})}
			\dd\uli\mu_0(\hat\sV)};
\end{equation}
as explained below, only maximum relative entropy solutions of this fixed point equation are obtained in this limit.
 Now defining 
\begin{equation}
	u(\sV) 
:= \label{uDEF}
 - \frac{\beta}{\pi}\int_{\Rset^2}\!\rho(\tilde{\sV}) \ln \frac{1}{|\sV -\tilde{\sV}|}\dd\uli\mu_0(\tilde{\sV})
\end{equation} 
and recalling that 
\begin{equation}
	\Delta_{\sV}  \ln {|\sV -\sV'|}
= \label{GREENdef}
  {2\pi}\delta_{\sV'}(\sV) 
\end{equation}
where $\delta_{\sV'}$ is the Dirac measure supported at $\sV'$, from (\ref{fixPOINTEQrho})  we see that 
$u(\sV)$ satisfies
\vspace{-5pt}
\begin{equation}
	-\Delta_{\sV} u (\sV) 
= \label{fixPOINTu}
	-2\beta \frac{ \dst f(\sV) e^{2u(\sV)}  }
   {\dst\int_{\Rset^2} f(\hat{\sV})e^{2 u(\hat{\sV})}\dd^2\hat {s}}.
\end{equation}

\vspace{-5pt}
\noindent
 Finally, since (\ref{fixPOINTu}) is invariant under the map $u\mapsto u+C$ with an arbitrary constant $C$, 
without loss of generality we can ask that solutions $u$ satisfy
\vspace{-5pt}
\begin{equation}
   \dst\int_{\Rset^2} f({\sV})e^{2 u({\sV})}\dd^2{s} 
= \label{GaussCONSTR}
1.
\end{equation}

\vspace{-5pt}
\noindent
 Thus we find that (\ref{fixPOINTu}) is equivalent to the \emph{prescribed Gauss curvature equation} 
\vspace{-5pt}
\begin{equation}
	-\Delta_{\sV} u (\sV) 
= \label{prescrGAUSScurvEQN}
	 K(\sV) e^{2u(\sV)} 
\end{equation}

\vspace{-5pt}
\noindent
with Gauss curvature function $K(\sV):= -2\beta f(\sV)$, constrained by (\ref{GaussCONSTR}).

 Since under the stated assumptions (in particular: $\beta>-8\pi$; the assumption that
$f(\sV)$ be a positive Schwartz function can be considerably weakened --- see \cite{ChaKieDMJ})
the limit $N\to\infty$ for (\ref{holoMU}) does exist (in a suitable sense), it follows that
(\ref{fixPOINTEQrho}) has a solution, which implies that the prescribed Gauss curvature equation
(\ref{prescrGAUSScurvEQN}) has a solution (satisfying (\ref{GaussCONSTR})).
 Thus Onsager's statistical mechanics of point vortices has an unintended but welcome side effect: it contributes
some partial yet explicit answers to Nirenberg's question ``Which functions $K(\sV)$ are Gauss curvatures?'',
the complete answer to which requires the characterization of all functions $K(\sV)$ for which the prescribed 
Gauss curvature equation $-\Delta_{\sV} u (\sV) = K(\sV) e^{2u(\sV)}$ has a solution.\footnote{Louis Nirenberg
 originally posed the problem for the prescribed Gauss curvature equation on the sphere $\Sset^2$, which 
 is much harder to answer due to some topological obstructions (see, e.g., \cite{KazdanWarner,ChangYangA,Han}).
 One such obstruction translates into the interesting requirement that the (rescaled) reciprocal Onsager temperature 
$\beta=-8\pi$, but this is exactly the borderline value where the canonical ensemble becomes a singular measure 
and therefore \emph{fails} to supply partial answers to Nirenberg's question. 
 However, as explained in \cite{KiesslingPHYSICA,KiesslingWangJSP}, the microcanonical point vortex ensemble
\cite{Lars} will produce solutions to the prescribed Gauss curvature equation on $\Sset^2$ whenever some exist,
although only maximum entropy solutions can be produced.}

 It is clear from this brief synopsis that the canonical
point vortex ensembles only yield Gauss curvature functions $K(\sV)=-\beta f(\sV)$ which do not change sign. 
 However, differential geometers are also interested in sign-changing Gauss curvature functions, and 
whether the statistical mechanics technique could somehow be generalized to also cover sign-changing $K(\sV)$
is precisely the question raised by Alice Chang.\footnote{Personal communication from A.C. to M.K.; ca. 2000.}

 A few symbolic manipulations suggest what needs to be done.
 Replace the a-priori measure $\dd\mu_0=f(\sV)\dd^2s$ with an a-priori signed measure $\dd\varsigma_0 = g(\sV)\dd^2s$
which differs from $\dd\mu_0$ merely in the sense that $g(\sV)$ is a sign-changing (say: Schwartz) function, while $f(\sV)$ 
is positive. 
 We will use the notation $\dd\varsigma_\beta^{(N)}$ for (\ref{holoMU}) with $\dd\mu_0$ replaced by $\dd\varsigma_0$. 
 The limit $N\to\infty$ should then lead to (\ref{fixPOINTEQrho}) with $\dd\uli\mu_0$ replaced by $\dd\uli\varsigma_0$, 
where $\dd\uli\varsigma_0 := \dd\varsigma_0/\int\dd\varsigma_0$. 
 The same steps that lead from (\ref{fixPOINTEQrho}) to (\ref{prescrGAUSScurvEQN}), (\ref{GaussCONSTR}) again lead
to (\ref{prescrGAUSScurvEQN}) but now with $K(\sV)=-\beta g(\sV)$, constrained by (\ref{GaussCONSTR}) with $f$ replaced
by $g$. 
 So all that needs to be done, or so it would seem, is to prove that the limit $N\to\infty$ for $\dd\varsigma_\beta^{(N)}$
given by  (\ref{holoMU}) with $\dd\mu_0$ replaced by $\dd\varsigma_0$ yields existence of a solution to (\ref{fixPOINTEQrho}) 
with $\dd\uli\mu_0$ replaced by $\dd\uli\varsigma_0$. 
 Can this be rigorously shown?

 We recall that the usual proof that (\ref{holoMU}) concentrates on (certain) solutions of (\ref{fixPOINTEQrho}) in 
the limit $N\to\infty$ relies heavily on a maximum entropy principle, which states that 
$\mu_\beta^{(N)}$ is the (unique) maximizer of the relative Gibbs entropy functional\footnote{Probabilists prefer the opposite 
sign convention for the relative entropy; cf. \cite{EllisBOOK}.} 
\vspace{-10pt}
\begin{equation}
 \cS(\mu^{(N)}|	\mu^{(N)}_\beta) 
: =
- \int_{\Rset^{2N}} \ln \tfrac{\dd\mu^{(N)}}{\dd\mu^{(N)}_\beta} \dd\mu^{(N)}
\label{GibbsENTROPYrelative}
\end{equation}
among all permutation symmetric $N$-point probability measures which are absolutely continuous w.r.t. 
$\dd\mu^{(N)}_\beta$ and for which r.h.s.(\ref{GibbsENTROPYrelative}) is finite.
 Note that $\cS(\mu^{(N)}_\beta|\mu^{(N)}_\beta) =0$.
 In a nutshell, the proof proceeds as follows: first one shows that the strict lower bound 
$N^{-1}\sup_{\varrho} \cS(\varrho^{\otimes N}|\mu^{(N)}_\beta) (< 0)$
on $N^{-1}\max_{\mu^{(N)}}  \cS(\mu^{(N)}|\mu^{(N)}_\beta) (=0)$ converges to $0$ in the limit $N\to\infty$; 
here, $\varrho$ denotes probability measures on $\Rset^2$. 
 Note that the symbolic Euler--Lagrange equation for $\cS(\varrho^{\otimes N}|\mu^{(N)}_\beta)$ is essentially 
(\ref{fixPOINTEQrho}), up to terms of $O(N^{-1})$.
 A tightness result for the sequences of the marginals of the maximizers $\dd\mu^{_{(N)}}_\beta$ together with sub-additivity, 
concavity, and weak lower semi-continuity estimates of the relative entropies of the marginals establishes that a maximizer 
$\varrho$ of the restricted variational principle exists in the limit.  
 For the details see the already cited literature.
 Also see \cite{MesserSpohn} for the origin of this strategy.

 By analogy one would expect that the \emph{signed relative entropy} functional
\vspace{-5pt}
\begin{equation}
 \cS(\varsigma^{(N)}|\varsigma^{(N)}_\beta) 
: =
- \int_{\Rset^{2N}} \ln \tfrac{\dd\varsigma^{(N)}}{\dd\varsigma^{(N)}_\beta} \dd\varsigma^{(N)}
\label{signedENTROPYrelative}
\end{equation}

\vspace{-5pt}
\noindent
will play a decisive role in the desired proof that 
the limit $N\to\infty$ for $\dd\varsigma_\beta^{(N)}$ yields the existence of a solution to (\ref{fixPOINTEQrho}) 
with $\dd\uli\mu_0$ replaced by $\dd\uli\varsigma_0$. 
 Here $\varsigma^{(N)}$ denotes a signed $N$-point measure which is absolutely continuous w.r.t. $\varsigma_\beta^{(N)}$ 
and such that the Radon--Nikodym derivative ${\dd\varsigma^{(N)}}/{\dd\varsigma^{(N)}_\beta} \geq 0$. 
 It is straightforward to show that the Euler--Lagrange equation for critical points of 
(\ref{signedENTROPYrelative}) is uniquely solved by $\varsigma_\beta^{(N)}$.
 Furthermore, the formal Euler--Lagrange equation for critical points of 
 $\cS(\rho^{\otimes N}\dd^N\uli\varsigma_0|\varsigma^{(N)}_\beta)$, where $\rho\geq 0$ and $\int \rho\dd\uli\varsigma_0 =1$, yields 
in the symbolic limit $N\to\infty$ precisely (\ref{fixPOINTEQrho}) with $\dd\uli\mu_0$ replaced by $\dd\uli\varsigma_0$. 
 All this is very encouraging. 

 However, to rigorously establish a signed relative entropy principle has been an elusive goal.
 The proof of the large $N$ limit of the traditional relative entropy principle for probability measures
makes heavy use of the non-positivity and the concavity of the relative entropy functional with a-priori 
probability measure, and also of the subadditivity of this traditional entropy functional --- none of these 
technical ingredients are available when the a-priori ``measure'' is not a (positive) measure!

 Technically we are therefore facing a measure-theoretical problem without convexity and sub-additivity to aid our control, 
and no proof of the desired result has been forthcoming. 
 Worse, the finite-$N$ expressions have so far resisted all attempts to evaluate them exactly.
 If such expressions were available one could hope to take their limit $N\to\infty$ explicitly and compare with solutions of the 
putative limiting prescribed sign-changing Gauss curvature equation,\footnote{Such a strategy may 
       not be futile, for similar integrals occur in the theory of random matrices and have been evaluated exactly for all $N$;
       see \cite{MehtaBOOK}, \cite{ForresterBOOK}, and see section \ref{Outlook}.
        However, the $N$-scaling of the Onsager temperature in our setting has been an obstacle so far.}
but so far this ``pedestrian'' way of checking the surmised limit result has been out of reach.

 At this point it was prudent to simplify the problem so that explicit finite-$N$ calculations became feasible which would
allow the checking of the key ideas.
 Thus we reduced the problem from two- to one-dimensional random positions, and we replaced the logarithmic with quadratic
pair interactions. 
 Moreover, instead of an arbitrary sign-changing function $g$ we chose $g$ to be quadratic, too. 
 This multivariate Gaussian problem was finally amenable to some explicit computations which, happily, supported our ideas.
 A major part of the present article is devoted to reporting our results about this multivariate Gaussian model.

 But first, we explain that our problem is not only of (tangential) interest to 
experts in differential geometry / geometric partial differential equations (as a curious technique for finding answers to 
Nirenberg's problem), or possibly to statisticians working with multivariate normal random variables. 
 Rather it paves the ground for answering a whole class of statistical questions.
 
\vspace{-15pt}
 
\subsection{Statistical significance of ``signed ensemble measures''}
\vspace{-5pt}

 While the signed a-priori measure $\dd\varsigma_0$ could be given a physical meaning as a signed 
vortex (or charge) density, a signed measure on $N$-point configuration space has no such interpretation.
 And since ``negative probabilities'' literally make no sense, 
there is no obvious statistical / probabilistical interpretation of what na\"{\i}vely one could call 
a ``signed ensemble measure $\dd\varsigma_\beta^{(N)}$,'' so readers in a statistical mechanics frame of mind may
rightfully ask what's in it for them.

 Fortunately there is a way out of this dilemma: the ratio of the normalizing integrals for $\varsigma_\beta^{(N)}$ and
$\mu_\beta^{(N)}$ can be interpreted as the expected value of a sign-changing product random variable w.r.t. the 
probability measure $\mu_\beta^{(N)}$, viz.
\vspace{-10pt}
\begin{equation}
\frac{\int_{\Rset^{2N}}\! e^{-  {\sfrNT}H^{(N)}} \dd^N\varsigma_0}{\int_{\Rset^{2N}}\! e^{-  {\sfrNT}H^{(N)}} \dd^N\mu_0} 
 = \label{ExpOFgOVERfPROD}
\int_{\Rset^{2N}}\! \prod_{k=1}^N \frac{g(\sV_k)}{f(\sV_k)} \dd\mu_\beta^{(N)} ,
\end{equation}

\vspace{-5pt}
\noindent
where $f>0$ is a Schwartz function and $\dd\mu_0 = f(\sV)\dd^2s$ as before.
 The ``statistical meaning'' of ``signed ensemble measures'' is provided in terms of the averages 
(\ref{ExpOFgOVERfPROD}); this points to an open field of probabilistic and statistical applications.

 In particular, our multi-variate Gaussian problem, phrased in terms of an expected value (\ref{ExpOFgOVERfPROD}),
reveals that it is not only of interest as a ``mock Gaussian curvature'' problem but also as part of an inquiry into 
random polynomials; see next.

\vspace{-17pt}
 
\subsection{Random polynomials and the Riemann hypothesis}
\vspace{-7pt}

 Consider the family of random polynomials $P_N^{}(z):=\prod_{1\leq k\leq N} (X_k^2+z^2)$, $N\in\Nset$, with 
$z\in\Cset$, whose $2N$ zeros $\{\pm i X_k\}^{N}_{k=1}$ are generated by $N$ identically distributed, generally correlated,
centered real random variables $\{X_k\}^{N}_{k=1}$, the law of which should be permutation symmetric
with non-zero $2N$-th moments. 
 Irrespectively of the law, the zeros of each such random polynomial all lie on the 
imaginary axis, reflection-symmetrical w.r.t. the real axis.
 Now take the expected value of $P_N^{}(z)$ w.r.t. the pertinent law of the $\{X_k\}^{}_{k=1,...,N}$, 
denoted $\langle{P_N^{}}\rangle_N^{}(z):={\rm Exp}_N^{}\!\left[\prod_{1\leq k\leq N} (X_k^2+z^2)\right]$. 
 The subscript ${}_N$ at the expected values is meant to avoid confusing
${\rm{Exp}}_N\left[f(X_1,...,X_n)\right]$ with ${\rm{Exp}}_n\left[f(X_1,...,X_n)\right]$ if $n<N$; note that
the distribution of the $\{X_k\}_{k=1}^{N}$ generally depends on $N$.
 Expanding, and using permutation invariance, one finds that \vspace{-15pt}
\begin{equation}\label{EXPECTofPOLYexpand}
\hspace{-10pt}
\langle{P_N^{}}\rangle_N^{}(z)
=
z^{2N} + N{\rm{Var}}_N\!\left[X_1\right] z^{2(N-1)} +
{\textstyle\sum\limits_{j=0}^{N-2}} 
\genfrac{(}{)}{0pt}{0}{N}{j} {\rm{Exp}}_N\!\left[\,{\textstyle\prod\limits_{m=1}^{N-j}} X_m^2\right]z^{2j} 
\end{equation}

\vspace{-7pt}
\noindent
is also a polynomial of degree $2N$ containing only even powers of $z$.
 Moreover, it is clear that 
$\langle{P_N^{}}\rangle_N^{}(z) ={\rm Exp}_N^{}\!\left[\prod_{1\leq k\leq N} (X_k^2+z^2)\right]$ has no real zeros. 
 On the other hand, it is not clear whether 
$\langle{P_N^{}}\rangle_N^{}(z)$
has any purely imaginary zeros --- unless $N$ is odd, in which case it is easy to see that $\langle{P_N^{}}\rangle_N^{}(z)$ must 
have at least two such zeros. 
 Since each ${P_N^{}}(z)$ has only imaginary zeros, the first interesting question is: 

\smallskip\noindent
{\bf{Q1}}: 
\emph{Can one identify all the permutation-symmetric laws for the random variables
$\{X_k\}^{}_{k=1,...,N}$ for which $\langle{P_N^{}}\rangle_N^{}(z)$ has only purely imaginary zeros?}
\smallskip

 Since for i.i.d. random variables we have $\langle{P_N^{}}\rangle_N^{}(z) = ({\rm{Var}}_N\!\left[X_1\right]+z^2)^N$, which clearly 
has $2N$ purely imaginary zeros $\pm i\surd{\rm{Var}}_N\!\left[X_1\right]$, counted in multiplicity, the existence of 
favorable laws is not in question. 
 However, this example makes it plain that to have an \emph{interesting} (i.e. non-degenerate) set of $2N$ purely imaginary zeros 
one needs to consider dependent random variables. 
 Since for $z=iy$ with $y\in\Rset$, r.h.s.(\ref{EXPECTofPOLYexpand}) $\in\Rset$ has alternating-sign coefficients, it is certainly 
conceivable that permutation-symmetric non-i.i.d. laws exist for which $\langle{P_N^{}}\rangle_N^{}(z)$ has only purely imaginary zeros.

 Indeed, there are infinitely many such correlated random variable laws. 
 Namely, for any real sequence $\{x_n\}_{n\in\Nset}$ the permutation-invariant empirical $N$-point measures
$\frac{1}{N!}\sum_{\varpi\in S_N} \prod_{n=1}^N \delta_{\varpi(n)}(x_n)$ define joint distributions for $N$ random variables
$\{X_n\}_{n=1}^N$ which for each $N$ yield $\langle{P_N^{}}\rangle_N^{}(z) = \prod_{1\leq k\leq N} (x_k^2+z^2)$,
manifestly having only purely imaginary zeros. 
 Unfortunately, the only ``randomness'' in these random variables is in their
labelling, and the labelling of point particles has no physical significance. 
 Factoring out the symmetry group $S_N$ yields a distribution without dispersion, and even though technically such an unlabeled 
nondispersive $N$-point configuration qualifies as a ``random configuration'' (in the same sense in which the number 1 is formally 
a random variable), this is not very interesting.
 Thus, to make question ${\bf{Q1}}$ \emph{really interesting} we have to ask for correlated, permutation-symmetrically distributed
random variables which are dispersed after factoring out the symmetry group $S_N$. 

 The large $N$ limit raises a second interesting question:

\noindent
{\bf{Q2}}:
\emph{Amongst the laws for the random variables $\{X_k\}_{k=1}^{N}$ 
for which $\langle{P_N^{}}\rangle_N^{}(z)$ has only purely imaginary zeros, are there any for which 
$\{\langle{P_N^{}}\rangle_N^{}(z)\}_{N\in\Nset}$ (suitably rescaled) or 
$\{\langle{P_N^{}}\rangle_N^{1/N}(z)\}_{N\in\Nset}$ have limit points when $iz\in\Rset$? If so, can they be determined?}
\newpage

 Again, the question is only interesting for dependent random variables. 
 Namely, suppose the random variables $\{X_k\}_{k=1}^{N}$ are
i.i.d., so $\langle{P_N^{}}\rangle_N^{}(z) = ({\rm{Var}}_N\!\left[X_1\right]+z^2)^N$; and so,
whenever ${\rm{Var}}_N\!\left[X_1\right]\to \sigma^2_\infty$ if $N\to\infty$, then
$\{\langle{P_N^{}}\rangle_N^{1/N}(z)\}_{N\in\Nset}$ has exactly two limit points when $iz\in\Rset$:
$\sigma_\infty^2+z^2$ for the odd-$N$, and $|\sigma_\infty^2+z^2|$ for the even-$N$ subsequence --- and
in this case also ${\rm{Var}}_N\!\left[X_1\right]^{-N}\langle{P_N^{}}\rangle_N^{}(z/\surd{N}) \to e^{z^2/\sigma_\infty^2}$.
 Furthermore, amongst the dependent random variables those distributed by a permutation-symmetric empirical law as explained
above are not interesting either, for in this case the answer is the foregone conclusion that the set of zeros of 
$\{\langle{P_N^{}}\rangle_N^{}(z)\}_{N\in\Nset}$ 
becomes $\{x_n\}_{n\in\Nset}$ in the limit $N\to\infty$, trivially. 

 If the answer to {\bf{Q2}} is ``yes'' also for some {dispersively correlated} laws, then for a subset of
these laws limit points of the sequence $\{\langle{P_N^{}}\rangle_N^{}(z/\surd{N})/\langle{P_N^{}}\rangle_N^{}(0)\}_{N\in\Nset}$,
or perhaps limit points of the sequence $\{\langle{P_N^{}}\rangle_N^{1/N}(z)\}_{N\in\Nset}$, 
may have countably many imaginary zeros, all with multiplicity 1, and no other zeros.
 For instance, a law which $\forall\;j\in\Nset\cup\{0\}$ yields
${\rm Exp}_N^{}\!\bigl[\prod_{k=1}^{N-j} X^2_k\bigr]/{\rm Exp}_N^{}\!\left[\prod_{k=1}^{N} X^2_k\right]\to j!/(2j)!$
when $N\to\infty$ forces
$\langle{P_N^{}}\rangle_N^{}(0)^{-1}\langle{P_N^{}}\rangle_N^{}(z/\surd{N}) \to \cosh(z)$ ($= \cos(y)$ for $z=iy$),
even if none of the polynomials $\langle{P_N^{}}\rangle_N(z)$ has any zeros on the imaginary axis.
 Whether this particular scenario is realizable I don't know, but I would be surprised if not.

 In any event, it does not take much imagination now to ask the inevitable:

\smallskip\noindent
{\bf{Q3}}: 
\emph{If there are any {dispersive, correlated} laws for which a limit point of the sequence 
$\{\langle{P_N^{}}\rangle_N^{}(z)\}_{N\in\Nset}$, after suitably rescaling $z$ and $\langle P_N^{}\rangle$ with $N$, 
has countably many imaginary zeros all with multiplicity 1, and no other zeros, is amongst these 
a law which reproduces the non-trivial zeros, translated by $-1/2$, of Riemann's $\zeta$ function \cite{Edwards,Henryk}?
 (The same question may be asked about the zeros of each member of the family of $\zeta$ functions in the generalized Riemann hypothesis
\cite{Conrey}.)}
\smallskip

 If the answer to ${\bf{Q3}}$ is ``Yes,'' then Riemann's hypothesis is true.

 Lest we give the reader the false impression that we were trying to suggest that a generalization to signed measures of the 
familiar notion of relative entropy for probability measures would answer all these questions, we emphasize that this generalized
notion of relative entropy would only help to establish and determine any limit points of the sequences
$\{\langle{P_N^{}}\rangle_N^{1/N}(z)\}_{N\in\Nset}$ and $\{\langle{P_N^{}}\rangle_N^{}(z)\}_{N\in\Nset}$ (rescaled)
 --- irrespectively of whether $\langle{P_N^{}}\rangle_N^{}(z)$ has~only~imaginary~zeros~or~not.
 This is of quite some interest on its own; of course, this is also important for 
answering questions {\bf{Q2}} and {\bf{Q3}}.

 In this vein, before answering question {\bf{Q2}} one first should answer question:

\smallskip\noindent
{\bf{Q2}${}^\prime$}: 
\emph{For which permutation-symmetrically distributed random variables $\{X_k\}^{N}_{k=1}$ does
$\{\langle{P_N^{}}\rangle_N^{}(z/\surd{N})/\langle{P_N^{}}\rangle_N^{}(0)\}_{N\in\Nset}$ or $\{\langle{P_N^{}}\rangle_N^{1/N}(z)\}_{N\in\Nset}$ 
have limit points? And which functions of $z$ are limit points?}

\hspace{-8pt} We now study certain multivariate normal random variables in the context of {\bf{Q2}${}^\prime$}.
\subsection{Statement of the main results}
 We carry out our inquiry into {\bf{Q2}${}^\prime$} 
with a one-parameter family of identically distributed multi-variate mean-zero normal random variables $\{X_k\}^{N}_{k=1}$ 
with co-variance matrix given by ${\rm{Cov}}_N(X_k,X_l) = ({1} +\frac{\sigma^2-1}{N})\delta_{k,l} +\frac{\sigma^2-1}{N}(1-\delta_{k,l})$.
 When $\sigma^2=1$ the multi-variate normal random variables are i.i.d. standard normal, and thus amongst the ``trivial''
answers to {\bf{Q1}} and {\bf{Q2}} identified above.
 This will be a helpful ``anchor'' for exploring the interesting parameter regime $\sigma^2\neq 1$.

 To reveal the significance of the parameter $\sigma^2$, we note that by an $SO(N)$ transformation one can rotate $\{X_k\}_{k=1}^N$ 
into a system $\{Y_k\}_{k=1}^N$ of $N$ independent multi-variate mean-zero normal random variables, of which $\{Y_k\}_{k=2}^N$ are 
i.i.d. standard normal random variables, and $Y_1 :={\textstyle{\frac{1}{\sqrt{N}}}}\sum_{1\leq k \leq N} X_k$ has variance $\sigma^2>0$; 
see section \ref{exactRESULTSfiniteNa}.
 Thus, when $\sigma^2 >1$, respectively $\sigma^2 <1$, their constant probability density level surfaces are prolate, respectively oblate 
hyper-ellipsoids $\frac{1}{\sigma^2}y_1^2+\sum_{k=2}^N y_k^2 = C>0$, whose symmetry axis points along the diagonal in the first ``$2^N$-ant'' 
of the Cartesian $\{x_1,x_2,...,x_N\}$ coordinate system; this is illustrated in Fig.~\ref{normalISOsurfaces}~for~$N=3$:

\begin{figure}[H]\vspace{-11pt}
\includegraphics[scale=0.4]{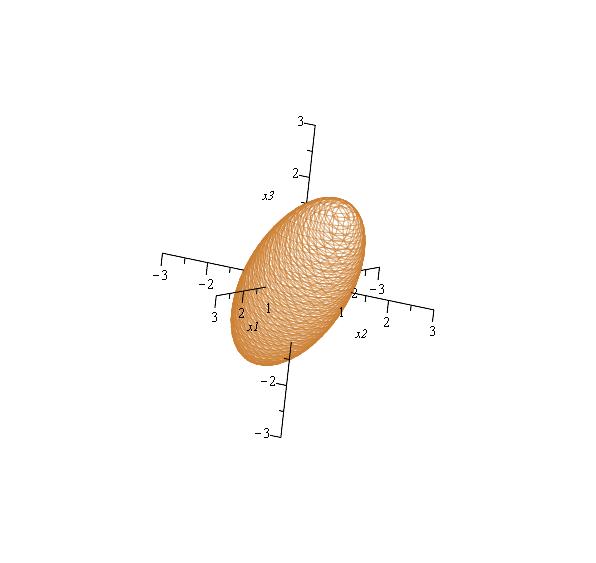}
\hspace{-60pt}
\includegraphics[scale=0.4]{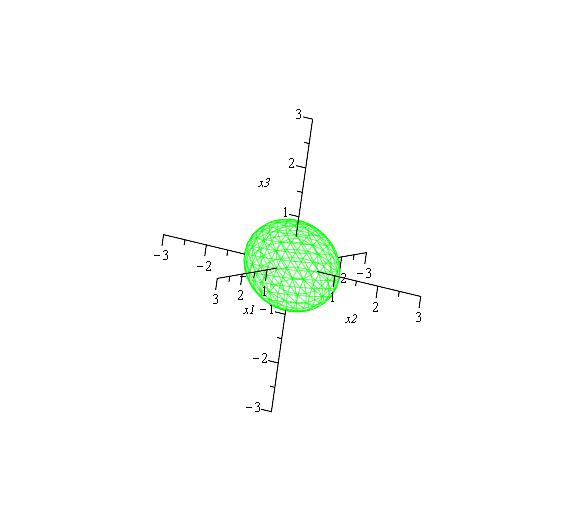}\vspace{-25pt}
\caption{\small For $N=3$, one example each of the level surfaces pdf $=1$ in $(x_1^{},x_2^{},x_3^{})$ space
for the prolate regime $\sigma^2>1$ (left panel) and the oblate regime $0<\sigma^2<1$ (right panel).}
\label{normalISOsurfaces}\vspace{-10pt}
\end{figure} 
 Incidentally, the fact that the system $\{Y_k\}_{k=1}^N$ consists of independent multi-variate mean-zero normal random variables
does not imply that the expected values of the polynomials $P_N(z)=\prod_{1\leq n\leq N}(X_n^2+z^2)$ can be factored into a product
of $N$ independent integrals by this change of random variables, for the $X_n^{}$ are still dependent; they are linear functions 
of typically all the $Y_k^{}$. 
\newpage

 For the stipulated $N$-variate normal random variables $X_k$, the expected polynomials 
$\langle{P_N^{}}\rangle_N^{}(z)\ \Big(\!\!= {\rm Exp}_N^{}\left[\prod_{k=1}^{N} (X_k^2+z^2)\right]\Big) =:E_N^{}(z;\sigma)$, $N\in\Nset$ 
are given by 
\vspace{-5pt}
\begin{equation}\label{EXPECToneX}
E_1^{}(z;\sigma) := 
\displaystyle
\int_{\Rset} (x^2+z^2) \frac{e^{-\frac{1}{2\sigma^2} x^2}}{\sqrt{2\pi}\sigma}dx ,
\end{equation}

\vspace{-5pt}
\noindent
which manifestly evaluates to $E_1^{}(z;\sigma)= z^2 + \sigma^2$, and for $N>1$ (see section \ref{exactRESULTSfiniteNa}) by
\vspace{-5pt}
\begin{equation}\label{EXPECTmoreTHANoneX}
E_N^{}(z;\sigma)\! := 
\displaystyle
\!\!\int_{\Rset^N} \!
\Bigl[\;\prod\!\!\prod_{\hspace{-16pt}1\leq k<l\leq N}\sigma^2 e^{-(\sigma^2-1)\frac{1}{2\sigma^2}(x_k-x_l)^2}\Bigr]{\!}^{^\frac1N}
\!\!\!\!\prod_{1\leq n\leq N}(x_n^2+z^2) \frac{e^{-\frac{1}{2\sigma^2}{x_n^2}}}{\sqrt{2\pi}\sigma}dx_n, 
\end{equation}

\vspace{-5pt}
\noindent
which are explicitly computable, too.
 Namely, by Isserlis' theorem, the general $r$-th centered moment of a multi-variate normal 
distribution is an explicitly computable polynomial of the elements of the covariance matrix, with rational coefficients.
 In our special case the expected values ${\rm{Exp}}_N\!\Bigl[\prod_{m=1}^{N-j}\, X_m^2\Bigr]$ for $j=0,...,N-1$ 
are explicitly computable polynomials of degree $N-j$ in $\sigma^2$.
 The $j=N$ term is trivial (i.e. $=1$).
 The simplest non-trivial one is the $j=N-1$ term, 
 ${\rm{Exp}}_N\!\left[X_1^2\right] ={\rm{Var}}_N\!\left[X_1\right]$, evaluating to 
${\rm{Var}}_N\!\left[X_1\right] = {1} +({\sigma^2-1})/{N}$ (see also Coro.~\ref{coro:var}) --- both terms are
already displayed separately in (\ref{EXPECTofPOLYexpand}).
 The higher moments can be computed by taking derivatives of the explicitly known moment-generating function, 
but this is inefficient.
 A more efficient method was communicated by one of the referees; see section \ref{exactRESULTSfiniteNb}.

 When $N$ is large enough the evaluation of (\ref{EXPECTmoreTHANoneX}) (more generally (\ref{EXPECTofPOLYexpand}))
is more efficiently done by asymptotic expansion in powers of $1/N$.
 Question {\bf{Q2}${}^\prime$} basically asks about the leading order terms.
 The only regime for which we can rigorously establish some asymptotic results 
\emph{for all} $z\in \Cset$ is the limit $N\to\infty$ of $E_N^{}(z/\surd{N};\sigma)/E_N^{}(0;\sigma)$.
\begin{theo}\label{thmCOMPLEXzSCALED}
 Let $z\in\Cset$ be fixed.
 Then, whenever $\sigma^2 \leq 3/2$, one has 
\begin{equation}\label{EsubNasNtoINFINITYscaledA}
\frac{E_N^{}(z/\surd{N};\sigma)}{E_N^{}(0;\sigma)}
 \stackrel{\scriptstyle{N\to\infty}}{\longrightarrow } 
\exp(z^2);
\end{equation}
and when $\sigma^2>3/2$, one has 
\begin{equation}\label{EsubNasNtoINFINITYscaledB}
\frac{E_N^{}(z/\surd{N};\sigma)}{E_N^{}(0;\sigma)}
\stackrel{\scriptstyle{N\to\infty}}{\longrightarrow } 
\exp\left( \frac{z^2}{2(\sigma^2-1)}\right),
\end{equation}
in the sense that the sequence of the partial sums of the Maclaurin expansion of the left- converges to the pertinent
sequence of partial sums of the right-hand sides.
\end{theo}

 Theorem~\ref{thmCOMPLEXzSCALED} can in principle be proved solely with classical analysis techniques. 
 However, we find it interesting to
 note that the Taylor expansion coefficients of 
$E_N^{}(z/\surd{N};\sigma)/E_N^{}(0;\sigma)$ are the special case $z=0$ of the limit of $E_N^{1/N}(z;\sigma)$, 
established  in section \ref{ASYMPzREAL} by using the (physicists') maximum relative entropy~principle~for
probability measures. 
 Our proof of Theorem~\ref{thmCOMPLEXzSCALED} in section \ref{ASYMPzCOMPLEXrescaled} invokes this principle.
 \newpage

 The convergence of $E_N^{}(z/\surd{N};\sigma)/E_N^{}(0;\sigma)$ will also be illustrated graphically.

 Due to the scaling $z\to z/\surd{N}$ in the polynomials $E_N^{}(z;\sigma)/E_N^{}(0;\sigma)$, Theorem~\ref{thmCOMPLEXzSCALED} 
captures the large-$N$ behavior of $E_N^{}(z;\sigma)/E_N^{}(0;\sigma)$ only in the ``infinitesimal vicinity'' of $z=0$. 
 To explore the large-$N$ behavior of $E_N^{}(z;\sigma)$ in a finite vicinity of $z=0$, one needs to study the sequence 
$N\mapsto E_N^{1/N}(z;\sigma)$ for $N\in\Nset$; note that without the power ``${}^{1/N}$'' at $E_N^{}(z;\sigma)$ its magnitude
will typically go to $\infty$ or to $0$ when $N\to\infty$.
 Alas, controlling the Taylor expansion about $z=0$ of $E_N^{1/N}(z;\sigma)$ is considerably more involved, 
and we have not yet been able to establish control over the large $N$ behavior when $z\in\Cset$ and $|z|$
is ``sufficiently small.'' 
 We will be able, though, to rigorously control the real-$z$ regime without invoking Taylor series arguments; we will also
formulate a conjecture about the regime  $iz\in\Rset$.

 In section \ref{ASYMPzREAL} we study the limit  $N\to\infty$ of $E_N^{1/N}(z;\sigma)$ rigorously for arbitrary $z\in\Rset$.
 Although, as pointed out already, the random polynomials and their expected value have no real zeros, section \ref{ASYMPzREAL} 
serves a useful purpose by paving the ground for the introduction of the relative entropy principle for signed measures. 
 Namely, in section \ref{ASYMPzREAL} we will show that the maximum relative entropy principle 
for probability measures governs the large-$N$ limit of $E_N^{}(z;\sigma)^{{1/N}}$ when $z\in\Rset$.
 We prove
\begin{theo}\label{thmREz}
 Let $z\in\Rset$.
 Then, pointwise, whenever $z^2 \geq 2\sigma^2-3$, one has 
\begin{equation}\label{lnEXPECTfONEfTWOasNtoINFINITYexistsA}
E_N^{}(z;\sigma)^{\frac1N} \stackrel{\scriptstyle{N\to\infty}}{\longrightarrow } 
1+z^2;
\end{equation}
but when $z^2 < 2\sigma^2-3$, which is possible iff $\sigma^2>3/2$, one has 
\begin{equation}\label{lnEXPECTfONEfTWOasNtoINFINITYexistsB}
E_N^{}(z;\sigma)^{\frac1N} \stackrel{\scriptstyle{N\to\infty}}{\longrightarrow } 
2(\sigma^2 -1) \exp\left( \frac{1+z^2}{2(\sigma^2-1)} -1\right).
\end{equation}
\end{theo}

 We will also illustrate the convergence of $E_N^{}(z;\sigma)^{1/N}$ graphically.
\begin{rema}
 Two aspects of Theorem \ref{thmREz} deserve highlighting: 

a) 
for $\sigma^2\leq 3/2$ the limit $N\to\infty$ of $E_N^{}(z;\sigma)^{1/N}$ with $z\in\Rset$ is an entire real 

\hspace{8pt} 
 analytic function, while for $\sigma^2> 3/2$ it is merely piecewise real analytic --- 

\hspace{8pt} 
a ``phase transition'' occurs at $z^2=2\sigma^2-3$; 

 b) note that $1+z^2=E_N^{}(z;1)^{1/N}$ --- thus, the limit $N\to\infty$ of $E_N^{}(z;\sigma)^{1/N}$ is 

\hspace{8pt} 
indistinguishable from the analogous limit with i.i.d. standard normal random

\hspace{8pt} 
variables, $\sigma=1$, when $z^2 \geq 2\sigma^2-3$, 
but if $\sigma^2>3/2$ and $z^2 < 2\sigma^2-3$, then 

\hspace{8pt} 
it retains information about the dependence of the random variables.

\end{rema}

 In section \ref{ASYMPzIMAG} we study the large-$N$ asymptotics of (\ref{EXPECTmoreTHANoneX}) with $iz\in\Rset$.
 Our earlier remarks concerning the large-$N$ asymptotics of (\ref{EXPECTmoreTHANoneX}) with $iz\in\Rset$ for
i.i.d. random variables make it plain that it will be necessary to discuss the even-$N$ and odd-$N$ subsequences separately. 
 One reason is that $\{E_N^{}(z;\sigma)\}_{N\in\Nset}^{}$, although well-defined $\forall\;z\in\Cset$, may change
sign alternatingly with $N$ for certain subsets of $iz\in\Rset$; and an alternating sign sequence, unless converging to zero,
would not converge at all. 
 Another reason is that only for odd $N=2K-1$, $K\in\Nset$, is $E_{2K-1}^{}(z;\sigma)^{1/(2K-1)}$ a-priori well-defined when $iz\in\Rset$.
 If a negative sign of $E_{N}^{}(z;\sigma)^{}$ occurs for certain $iz\in\Rset$ when $N=2K$ 
then $\{E_{2K}^{}(z;\sigma)^{1/2K}\}_{K\in\Nset}^{}$  would a-priori not be defined for those $iz\in\Rset$. 
 Using analytic continuation to define $E_{2K}^{}(z;\sigma)^{1/2K}$ for these $n$ and $iz\in\Rset$, 
the resulting complex functions would generally live on different Riemann surfaces, depending on $n$; this would take us in a different
direction.
 The first thing we will prove in section \ref{ASYMPzIMAG} is that $E_{2K}^{}(z;\sigma)\geq 0$ for $K\in\Nset$ and $iz\in\Rset$, which 
establishes that both the even-$N$ and the odd-$N$ subsequences of $\{E_N^{1/N}(z;\sigma)\}_{N\in\Nset}^{}$ are well-defined when 
$iz\in\Rset$ (incidentally, the same is trivially true when $z\in\Rset$.)

 As will be clear from section \ref{ASYMPzREAL}, the \emph{technique of proving} Theorem \ref{thmREz}
does not apply to the domain $z\not\in\Rset$.
 Nevertheless, the proof of Theorem \ref{thmREz} in concert with our discussion of the special case of the i.i.d. zeros
has inspired the surmise that the analytical extension from $z\in\Rset$ to $iz\in\Rset$ of the limit
functions given at r.h.s.(\ref{lnEXPECTfONEfTWOasNtoINFINITYexistsA}) and r.h.s.(\ref{lnEXPECTfONEfTWOasNtoINFINITYexistsB}) 
might capture the large-$N$ asymptotics of $\{E_{2K-1}^{}(z;\sigma)^{\frac{1}{2K-1}}\}_{K\in\Nset}$, and
their absolute value might capture the asymptotics of $\{E_{2K}^{}(z;\sigma)^{\frac{1}{2K}}\}_{K\in\Nset}$, when $iz\in\Rset$.
 We alert the reader that the accurate statement will be much more refined, with piecewise real-analytical limit curves featuring
multiple phase transitions!

 Our surmise about the even-$N$ and odd-$N$ subsequences of 
$\{E_N^{}(z;\sigma)^{\frac1N}\}_{N\in\Nset}$ with $iz\in\Rset$
is investigated in section \ref{ASYMPzIMAG}, where we will present graphical evidence in its support, obtained 
with the help of computer-algebraic evaluations of $E_N(z;\sigma)$ for imaginary $z$ with $N$ up to a dozen. 
 This has turned our surmise into the refined\vspace{-4pt}
\begin{conj}\label{conjIMz}
Let $iz\in\Rset$. 
 Then the even-$N$ and odd-$N$ subsequences of the sequence
$\{E_N^{}(z;\sigma)^{\frac1N}\}_{N\in\Nset}$ do have limits, characterized as follows.
  Define $z^2_*(\sigma^2) := 2\kappa_*^{} (1- \sigma^2) -1$, where
$\kappa_*$ is the unique solution of the fixed point equation $\kappa = e^{-\kappa-1}$; numerically, $\kappa_*^{}=0.27846454276...$. 
 Note that $z_*^2<0$; more precisely, $z_*^2<-1$ for $\sigma^2> 1$ and $-1<z_*^2<0$ for $0\leq \sigma^2< 1$, with 
$z_*^2=-1$ exactly when $\sigma^2= 1$.

 The limits for the even-$N$ and odd-$N$ subsequences are now listed separately.

\noindent{\sc{The sequence}} $\{E_{2K}^{}(z;\sigma)^{\frac{1}{2K}}\}_{K\in\Nset}$:

1. Let $\sigma^2\geq 3/2$.
 Then, if $-z^2 \leq -z_*^2$ one has 
\begin{equation}\label{IMzNevenLIMITa}
E_{2K}^{}(z;\sigma)^{\frac{1}{2K}}
\stackrel{\scriptstyle{n\to\infty}}{\longrightarrow } 
2(\sigma^2 -1) \exp\left( \frac{1+z^2}{2(\sigma^2-1)} -1\right),
\end{equation}

whereas if $-z^2 > -z_*^2$  one has 
\begin{equation}\label{IMzNevenLIMITb}
E_{2K}^{}(z;\sigma)^{\frac{1}{2K}} \stackrel{\scriptstyle{n\to\infty}}{\longrightarrow } 
-(1+z^2);
\end{equation}
\newpage

2.  Let $1<\sigma^2 < 3/2$.
 Then, if $-z^2 \leq 3-2\sigma^2$ one has 
\begin{equation}\label{IMzNevenLIMITc}
E_{2K}^{}(z;\sigma)^{\frac{1}{2K}}
\stackrel{\scriptstyle{n\to\infty}}{\longrightarrow } 
1+z^2,
\end{equation}

whereas if $3-2\sigma^2< -z^2 < -z_*^2$ one has 
\begin{equation}\label{IMzNevenLIMITd}
E_{2K}^{}(z;\sigma)^{\frac{1}{2K}} 
\stackrel{\scriptstyle{n\to\infty}}{\longrightarrow } 
2(\sigma^2 -1) \exp\left( \frac{1+z^2}{2(\sigma^2-1)} -1\right),
\end{equation}

while for $-z^2 \geq -z_*^2$  one has 
\begin{equation}\label{IMzNevenLIMITe}
E_{2K}^{}(z;\sigma)^{\frac{1}{2K}} \stackrel{\scriptstyle{n\to\infty}}{\longrightarrow } 
-(1+z^2);
\end{equation}

3.  Let $\sigma^2 =1$.
    Then one has 
\begin{equation}\label{IMzNevenLIMITf}
E_{2K}^{}(z;\sigma)^{\frac{1}{2K}}
\stackrel{\scriptstyle{n\to\infty}}{\longrightarrow } 
|1+z^2|;
\end{equation}

4.  Let $0\leq \sigma^2 < 1$.
    Then, if $-z^2 \leq -z_*^2$  one has 
\begin{equation}\label{IMzNevenLIMITg}
E_{2K}^{}(z;\sigma)^{\frac{1}{2K}}
\stackrel{\scriptstyle{n\to\infty}}{\longrightarrow } 
1+z^2,
\end{equation}

whereas if $-z_*^2 < -z^2 < 3-2\sigma^2$ one has 
\begin{equation}\label{IMzNevenLIMITh}
E_{2K}^{}(z;\sigma)^{\frac{1}{2K}} 
\stackrel{\scriptstyle{n\to\infty}}{\longrightarrow } 
2(1-\sigma^2) \exp\left( \frac{1+z^2}{2(\sigma^2-1)} -1\right),
\end{equation}

while for $-z^2 \geq 3-2\sigma^2$ one has 
\begin{equation}\label{IMzNevenLIMITi}
E_{2K}^{}(z;\sigma)^{\frac{1}{2K}} \stackrel{\scriptstyle{n\to\infty}}{\longrightarrow } 
-(1+z^2);
\end{equation}

\noindent{\sc{The sequence}} $\{E_{2K-1}^{}(z;\sigma)^{\frac{1}{2K-1}}\}_{K\in\Nset}$:

1. Let $\sigma^2\geq 3/2$.
 Then, if $-z^2 \leq -z_*^2$  one has 
\begin{equation}\label{IMzNoddLIMITa}
E_{2K-1}^{}(z;\sigma)^{\frac{1}{2K-1}}
\stackrel{\scriptstyle{n\to\infty}}{\longrightarrow } 
2(\sigma^2 -1) \exp\left( \frac{1+z^2}{2(\sigma^2-1)} -1\right),
\end{equation}

and if $-z^2 > -z_*^2$  one has 
\begin{equation}\label{IMzNoddLIMITb}
E_{2K-1}^{}(z;\sigma)^{\frac{1}{2K-1}} \stackrel{\scriptstyle{n\to\infty}}{\longrightarrow } 
1+z^2;
\end{equation}

2.  Let $1<\sigma^2 < 3/2$.
 Then, if $-z^2 \leq 3-2\sigma^2$ one has 
\begin{equation}\label{IMzNoddLIMITc}
E_{2K-1}^{}(z;\sigma)^{\frac{1}{2K-1}}
\stackrel{\scriptstyle{n\to\infty}}{\longrightarrow } 
1+z^2,
\end{equation}

and if $3-2\sigma^2< -z^2 < -z_*^2$ one has 
\begin{equation}\label{IMzNoddLIMITd}
E_{2K-1}^{}(z;\sigma)^{\frac{1}{2K-1}} 
\stackrel{\scriptstyle{n\to\infty}}{\longrightarrow } 
2(\sigma^2 -1) \exp\left( \frac{1+z^2}{2(\sigma^2-1)} -1\right),
\end{equation}

while if $-z^2 \geq -z_*^2$  one has 
\begin{equation}\label{IMzNoddLIMITe}
E_{2K-1}^{}(z;\sigma)^{\frac{1}{2K-1}} \stackrel{\scriptstyle{n\to\infty}}{\longrightarrow } 
1+z^2;
\end{equation}

3.  Let $\sigma^2 =1$.
 Then, one has 
\begin{equation}\label{IMzNoddLIMITf}
E_{2K-1}^{}(z;\sigma)^{\frac{1}{2K-1}}
\stackrel{\scriptstyle{n\to\infty}}{\longrightarrow } 
1+z^2;
\end{equation}

4.  Let $0\leq \sigma^2 < 1$.
 Then, if $-z^2 \leq -z_*^2$  one has 
\begin{equation}\label{IMzNoddLIMITg}
E_{2K-1}^{}(z;\sigma)^{\frac{1}{2K-1}}
\stackrel{\scriptstyle{n\to\infty}}{\longrightarrow } 
1+z^2,
\end{equation}

and if $-z_*^2 < -z^2 < 3-2\sigma^2$ one has 
\begin{equation}\label{IMzNoddLIMITh}
E_{2K-1}^{}(z;\sigma)^{\frac{1}{2K-1}} 
\stackrel{\scriptstyle{n\to\infty}}{\longrightarrow } 
2(\sigma^2-1) \exp\left( \frac{1+z^2}{2(\sigma^2-1)} -1\right),
\end{equation}

while if $-z^2 \geq 3-2\sigma^2$ one has 
\begin{equation}\label{IMzNoddLIMITi}
E_{2K-1}^{}(z;\sigma)^{\frac{1}{2K-1}} \stackrel{\scriptstyle{n\to\infty}}{\longrightarrow } 
1+z^2.
\end{equation}
\end{conj}

\begin{rema}
 Note that the even-$N$ and odd-$N$ limit points with $iz\in\Rset$ coincide as long as $-z^2 < \min\{-z_*^2,3-2\sigma^2\}$, and otherwise
they are negatives of one another.
\end{rema}

 In section \ref{RelEntSignMeas} we show that all the limit formulas stated in Conjecture \ref{conjIMz}
can be obtained from a heuristic extension of the relative entropy principle to normalized signed or complex measures 
\emph{relative to a signed a-priori measure}.
 Since the limit point behavior with $iz\in\Rset$ is much more complicated, and therefore more interesting, than the 
limit with $z\in\Rset$ and its na$\ddot{\mbox{\i}}$ve analytical extension, it is self-evident that a relative entropy principle 
for signed or complex measures which is capable of capturing such complicated scenarios is a potentially powerful tool. 

 Lastly, in section \ref{Outlook} we conclude the paper with a to-do list of open problems, and by emphasizing potential 
applications of a relative entropy principle with signed a-priori measures in various fields of science and mathematics, besides
differential geometry and random polynomials, also random matrices and mathematical biology. 

\vfill
\medskip
\noindent
{\bf{Acknowledgement}:} I am grateful to Alice Chang for her question about the connection between statistical mechanics of point vortices
and Nirenberg's problem with signed Gaussian curvature, which started this inquiry. 
 I also thank Roger Nussbaum, Alex Kontorovich, and Shadi Tahvildar-Zadeh for patiently listening to my reports of progress
 which helped me obtaining greater clarity in this writeup.
 Finally I thank both referees for their constructive criticisms.

\newpage
\section{The $N$-variate normal pdf}\label{exactRESULTSfiniteNa}

 Here we supply the proof that (\ref{EXPECTmoreTHANoneX}) is indeed the expected value it is claimed to be (note that
for (\ref{EXPECToneX}) this is obvious).
 This is accomplished by first proving 
\begin{prop}\label{prop:pdf}
If $\sigma>0$, then 
\begin{equation}\label{EXPECTfTWO}
\displaystyle
\int_{\Rset^N} 
\Bigl[\;\prod\!\!\prod_{\hspace{-16pt}1\leq k<l\leq N} {\sigma}^2 e^{-\frac{1}{2}(1-\sigma^{-2}) (x_k-x_l)^2}\Bigr]{\!}^{^{\frac1N}}
\! \prod_{1\leq n\leq N}\frac{e^{-\frac{1}{2\sigma^2} x_n^2}}{\sqrt{2\pi}\sigma}dx_n 
=
1.
\end{equation}
\end{prop}

\smallskip
\noindent
\textit{Proof of Identity (\ref{EXPECTfTWO})}:

 We rewrite 
\begin{equation}\label{multiVARnormalREWRITE}
\begin{array}{rl}
&
\displaystyle
\prod\!\!\prod_{\hspace{-16pt}1\leq k<l\leq N}\! e^{-\frac{1}{2N}(1-\sigma^{-2}) (x_k-x_l)^2}\!\!\!
\prod_{1\leq n\leq N}\!\!e^{-\frac{1}{2\sigma^2} x_n^2}
=\cr
&\hspace{20pt}
\exp\biggl(\!-\frac{1}{2\sigma^2}
\biggl[{{({\sigma^{2}-1})\frac{1}{N}}}\;\sum\!\!\!\sum\limits_{\hspace{-16pt}1\leq k<l\leq N} (x_k-x_l)^2 
  + \sum\limits_{1\leq n \leq N} x_n^2 \biggr]\biggr)
\end{array}
\end{equation}
and compute
\begin{equation}\label{harmonicIDENTITY}
\frac{1}{N}
\sum\!\!\sum_{\hspace{-16pt}1\leq k<l\leq N} (x_k-x_l)^2
=
\frac{1}{2N}\sum\!\sum_{\hspace{-16pt}1\leq k,l\leq N} (x_k-x_l)^2
=
\sum_{1\leq k \leq N} x_k^2 - 
\biggl({\textstyle{\frac{1}{\sqrt{N}}}}\sum_{1\leq k \leq N} x_k\biggr)^2.
\end{equation}
 Thus,
\begin{equation}\label{harmonicIDENTITYconcl}
(\sigma^{2}-1)\frac{1}{N}
\sum\!\!\sum_{\hspace{-16pt}1\leq k<l\leq N} (x_k-x_l)^2 + \sum_{1\leq n \leq N} x_n^2 
=
\sigma^{2}\sum_{1\leq k \leq N} x_k^2 - (\sigma^{2}-1)\biggl({\textstyle{\frac{1}{\sqrt{N}}}}\sum_{1\leq k \leq N} x_k\biggr)^2\!.
\end{equation}
 Next we note that we can rotate the Cartesian coordinates
$\{x_1,x_2,...,x_N\}$ 
into a Cartesian coordinate system $\{y_1,y_2,...,y_N\}$ 
with $y_1:={\textstyle{\frac{1}{\sqrt{N}}}}\sum_{1\leq k \leq N} x_k$ (whose coordinate axis points along the diagonal in the first 
``$2^N$-ant'' of the $\{x_1,x_2,...,x_N\}$ system).
 Since Euclidean distances are preserved under Euclidean transformations, rotations amongst them, 
we have that $\sum_{1\leq k \leq N} x_k^2 = \sum_{1\leq k \leq N} y_k^2$, so that
\begin{equation}
\sigma^{2}\!\!\sum_{1\leq k \leq N} x_k^2
 - (\sigma^{2}-1)\biggl({\textstyle{\frac{1}{\sqrt{N}}}}\sum_{1\leq k \leq N} x_k\biggr)^2
=
\sigma^{2}\!\!\sum_{1\leq k \leq N} y_k^2 - (\sigma^{2}-1)y_1^2 = y_1^2 + \sigma^{2}\!\!\sum_{2\leq k \leq N} y_k^2;
\end{equation}
similarly, Euclidean volumes are invariant, i.e.
$
\prod_{1\leq n\leq N}dx_n = \prod_{1\leq n\leq N}dy_n.
$
 Thus the rotated variables $\{Y_1,...,Y_n\}$ are independent normal random variables, each having mean 0, but they 
are not identically distributed. 
 The component $Y_1$ is a normal random variable with mean zero and variance $\sigma^{2}$, while the $Y_k, k=2,...,N$, 
are i.i.d. standard normal random variables. 
 This proves (\ref{EXPECTfTWO}).\hfill QED

 This proposition establishes that the manifestly positive integrand of (\ref{EXPECTfTWO}) is a pdf.
 Obviously, this pdf is  an $N$-variate normal pdf.

 Next we compute the covariance matrix of $X_k$ and $X_l$. 
\begin{prop}\label{prop:cov}
For $\sigma>0$, we have  ${\rm{Cov}}_N(X_k,X_l) = ({1} +\frac{\sigma^2-1}{N})\delta_{k,l} +\frac{\sigma^2-1}{N}(1-\delta_{k,l})$.
\end{prop}

\smallskip
\noindent
\textit{Proof of Prop. \ref{prop:cov}}:

 We rewrite 
\begin{equation}\label{SIGMAnotation}
\displaystyle
\exp\biggl(\!-\frac{1}{2\sigma^2}
\biggl[{{({\sigma^{2}-1})\frac{1}{N}}}\;\sum\!\!\!\sum\limits_{\hspace{-16pt}1\leq k<l\leq N} (x_k-x_l)^2 
  + \sum\limits_{1\leq n \leq N} x_n^2 \biggr]\biggr)= 
 \exp\biggl(\!-\frac{1}{2} {\bf{x}}^T{\bf\Sigma}_N^{-1}{\bf{x}}\biggr),
\end{equation}
where we introduced the matrix notation ${\bf{x}}^T :=(x_1^{},...,x_N^{})$, with ${\bf\Sigma}_N^{}$ the co-variance matrix.
 We read off of (\ref{SIGMAnotation}) that
\begin{equation}\label{SIGMAinverse}
{\bf\Sigma}_N^{-1} = {\boldsymbol{\rm{D}}_N}(1) - {\textstyle(1-{\sigma^{-2}})\frac1N} {\boldsymbol{\rm{A}}_N}(1) ,
\end{equation}
where ${\boldsymbol{\rm{D}}_N}(1)$ is the (\emph{diagonal}) $N\times N$ identity matrix, and 
${\boldsymbol{\rm{A}}_N}(1)$ is the $N\times N$ matrix having \emph{all} its elements equal to 1.
 For $\sigma>0$ the inverse of ${\bf\Sigma}_N^{-1}$ exists and is readily computed from its Neumann series, yielding
\begin{equation}\label{SIGMA}
{\bf\Sigma}_N^{} = {\boldsymbol{\rm{D}}_N}(1) - (1-\sigma^2){\textstyle\frac1N} {\boldsymbol{\rm{A}}_N}(1) .
\end{equation}
 This completes the proof. \hfill QED

 The variance of $X_k$ is the $k$-th diagonal element of the covariance, i.e. we have
\begin{coro}\label{coro:var}
For $\sigma>0$, we have ${\rm{Var}}_N\!\left[X_1\right] = {1} +({\sigma^2-1})/{N}$.
\end{coro}

 We remark that the variance of $X_k$ can be directly computed with the help of
the $SO(N)$ transformation employed in the proof of proposition \ref{prop:pdf}; see Appendix A.

 We end this section by giving also the moment-generating function 
$M_N^{}({\bf{t}}) := {\rm{Exp}}_N\!\left[{\textstyle\prod_{n=1}^{N}} e^{t_nX_n}\right]$
for our multi-variate normal distribution, which reads
\begin{equation}\label{MOMgenFUNC}
M_N^{}({\bf{t}}) 
= \exp\left({\tfrac12 {\bf t}^T {\bf\Sigma}_N^{}{\bf t}}\right),
\end{equation}
where ${\bf\Sigma}_N^{}$ is given in (\ref{SIGMA}), 
and where ${\bf t}^T = (t_1,t_2,...,t_N)$ is the dual vector variable to 
${\bf x}^T = (x_1,x_2,...,x_N)$.
 The bilinear ${\bf t}$-form evaluates to
\begin{equation}\label{bilinearSIGMAform}
\begin{array}{llll} 
{\bf t}^T {\bf\Sigma}_N^{}{\bf t}
&=
(1-\sigma^{2})\frac{1}{N}\displaystyle
\sum\!\!\sum_{\hspace{-16pt}1\leq k<l\leq N} (t_k-t_l)^2 + \sigma^2\sum_{1\leq n \leq N} t_n^2 \\
&=
\displaystyle
\sum_{1\leq k\leq N}t_k^2-(1-\sigma^{2})\biggl({\textstyle{\frac{1}{\sqrt{N}}}}\sum_{1\leq k\leq N}\!t_k\!\biggr){\!}^{^{\scriptstyle{2}}}\!.
\end{array}
\end{equation}
\newpage

\section{Exact evaluation of $E_N^{}(z;\sigma)$ for arbitrary $N\in\Nset$}\label{exactRESULTSfiniteNb}
 When $\sigma=1$, i.e. in the case of i.i.d. random variables, (\ref{EXPECTmoreTHANoneX}) factors and yields a
simple closed-form expression valid for all $N\in\Nset$ and arbitrary $z\in\Cset$; viz.
\begin{equation}\label{EXPECTfONE}
E_N^{}(z;1)\! = 
\displaystyle
\int_{\Rset^N} 
\!\!\prod_{1\leq n\leq N} (x_n^2+z^2) \frac{e^{-\frac{1}{2} x_n^2}}{\sqrt{2\pi}}dx_n 
=
\displaystyle
(1+z^2)^N\quad \forall\; z\in\Cset;\quad N\in\Nset.
\end{equation}

 When $\sigma\neq 1$, we can evaluate (\ref{EXPECTmoreTHANoneX}) using (\ref{EXPECTofPOLYexpand}), 
which for multivariate normal random variables becomes explicitly computable because in this case we have
\begin{equation}\label{MOMENTSviaDIFFofM}
{\rm{Exp}}_N\!\left[{\textstyle\prod\limits_{m=1}^{N-j}} X_m^2\right]
=
\left[{\textstyle\prod\limits_{m=1}^{N-j}} \partial_{t_m}^2\right]M_N^{}({\bf{t}})\Big|^{}_{\bf t =\boldsymbol{0}},\quad j=0,...,N-1,
\end{equation}
with $M_N^{}({\bf{t}})$ given explicitly in (\ref{MOMgenFUNC}), (\ref{bilinearSIGMAform}).
 The first three $E_N(z;\sigma)$ evaluate to
\begin{equation} \begin{array}{lllll}
&\hspace{-12pt}\notag
E_1(z;\sigma) 
= z^{2} + z^0[\sigma^2+0], 
\\
&\hspace{-12pt}\notag
E_2 (z;\sigma) = z^4 +z^2 [\sigma^2+1] + z^0[\frac34\sigma^4 -\frac12\sigma^2 +\frac34] , \phantom{.............................}\hfill(49)
\\
&\hspace{-12pt}\notag
E_3 (z;\sigma) = z^6 +z^4[\sigma^2 +2] +z^2[\ \ \sigma^4 +0\sigma^2+ 2]\; +\, 
z^0[\frac59\sigma^6 -\frac23\sigma^4+\frac23\sigma^2+\frac49],\!\!\!\!\!\!\hspace{-20pt}
\end{array}\addtocounter{equation}{+1}
\end{equation}
but the expressions soon become unwieldy.
 Worse, formula (\ref{MOMENTSviaDIFFofM}), which yields a polynomial in $\sigma^2$ of degree $N-j$ with rational coefficients,
is less practical for computations than it may seem. 
 Since each derivative contributes a factor of 2 when counting the 
($M({\bf t})\times$ polynomials in ${\bf t}$)-factors which need to be multiplied before setting ${\bf t}={\boldsymbol{0}}$,
this leads to an exponential proliferation of terms.
 By hand this can be done only for very small $N$, and even MAPLE gave up when $N=8$. 
 Curiously, computing the moments by directly evaluating the integrals for l.h.s.(\ref{MOMENTSviaDIFFofM}), 
MAPLE was able to go up to about $N=12$. 
 While this was sufficient to obtain empirical evidence for the putative convergence of the finite-$N$ sequences to the
conjectured limiting curves, $N=12$ is a far cry from a ``large-$N$ regime,'' and
in the submitted version of this article I remarked that ``[a] more cleverly constructed evaluation scheme is needed to 
compute the relevant finite-$N$ expressions for $N$ beyond a dozen.
 Experts in multivariate normal random variables may know.''
 One of the referees responded to my remark by supplying such a more cleverly constructed evaluation 
of ${\rm{Exp}}_N\!\left[{\textstyle\prod_{m=1}^{N-j}} X_m^2\right]$.
 This  has yielded an improved 
\begin{theo}\label{thm1}
The integral $E_N(z;\sigma)$ evaluates to
\begin{equation}\label{EnDOUBLEsumFINITE}
E_N^{}(z;\sigma)
 = 
\sum\limits_{j=0}^N 
z^{2j} \genfrac{(}{)}{0pt}{0}{N}{j} 
\sum\limits_{k=0}^{N-j} 
\genfrac{(}{)}{0pt}{0}{N-j}{k}
\frac{(2k)!}{2^kk!}\left(\frac{\sigma^2-1}{N}\right)^{\!\!k} 
\end{equation}
\end{theo}

\smallskip
\noindent
\textit{Proof of Theorem \ref{thm1}}:

 The referee takes advantage of the fact that the sum of two independent normal random variables again is a normal random variable.
 The trick is to write $N$ dependent Gaussian random variables as linear combinations of $N+1$ independent standard normal random 
variables.
 Moreover, since we have from formula (\ref{MOMENTSviaDIFFofM}) that ${\rm{Exp}}_N\!\left[{\textstyle\prod_{m=1}^{N-j}} X_m^2\right]$
is a polynomial in $\sigma^2$ of degree $N-j$ with rational coefficients, it suffices to evaluate (\ref{MOMENTSviaDIFFofM}) for $\sigma\geq 1$.
 Thus, suppose $\sigma\geq 1$, and let $Y_0,Y_1,...,Y_N$ denote $N+1$ i.i.d. standard normal random variables.
 Now write each of the $N$ normal random variables $X_k$ as a weighted sum of $Y_0$ and $Y_k$,
namely $X_k = Y_k + \sqrt{\frac{\sigma^2-1}{N}}Y_0$, $k=1,...,N$.
 The $X_k$ are no longer independent; they are precisely our $N$-variate Gaussian random variables!
 Thus ${\rm{Exp}}_N\!\left[{\textstyle\prod_{m=1}^{N-j}} X_m^2\right]$ equals
\begin{equation}\label{eq:ENperREF}
{\rm{Exp}}_{\mathcal{N}(1;0)^{N+1}}
\!\left[{\textstyle\prod\limits_{m=1}^{N-j}} \Big(Y_m + \sqrt{\frac{\sigma^2-1}{N}}Y_0\Big)^{\!2}\right]
=
\sum\limits_{k=0}^{N-j} 
{{\genfrac{(}{)}{0pt}{0}{N-j}{k}}}
\dfrac{(2k)!}{2^kk!}\left(\dfrac{\sigma^2-1}{N}\right)^{\!\!k} ,
\end{equation}
where ${\rm{Exp}}_{\mathcal{N}(1;0)^{N+1}}[\cdots]$ denotes expected value w.r.t. the $N+1$ independent standard normal random variables
$Y_0,...,Y_N$; r.h.s.(\ref{eq:ENperREF}) follows by direct calculation.
\hfill Q.E.D.
\medskip

 Formula (\ref{EnDOUBLEsumFINITE}) can be plotted with the help of MAPLE easily for up to $N=1,000$ (over the range of $z$ and $\sigma$
values depicted in this paper). 
 This vast improvement over the direct MAPLE integration (feasible only for up to $N=12$) of the Gaussian integrals defining 
${\rm{Exp}}_N\!\left[{\textstyle\prod_{m=1}^{N-j}} X_m^2\right]$ has confirmed that the curves with $N=12$ are already remarkably 
close to the putative limiting curves.

\section{Large-$N$ limit of $E_N(z/\surd{N};\sigma)/E_N(0;\sigma)$ for $z\in\Cset$}\label{ASYMPzCOMPLEXrescaled}


 In this section we prove Theorem \ref{thmCOMPLEXzSCALED} --- except that we invoke the special case $z=0$ of a more general theorem 
for arbitrary $z\in\Rset$ which we prove in section \ref{ASYMPzREAL}.

\smallskip
\noindent
\textit{Proof of Theorem \ref{thmCOMPLEXzSCALED}}:

 The expressions $E_N(z;\sigma)/E_N(0;\sigma)$ are polynomials~in~$z^2$, viz.
\begin{equation}\label{ENforREzTAYLORscaled}
\begin{array}{lll}
\displaystyle \frac{E_N^{}(z;\sigma)}{E_N^{}(0;\sigma)}
&=& 
 \frac{\displaystyle\int_{\Rset^N}\!
\prod\!\!\prod_{\hspace{-16pt}1\leq k<l\leq N}\!\! e^{-\frac{1}{2N}(1-\sigma^{-2}) (x_k-x_l)^2}
\!\!\!\!\!\prod_{1\leq n\leq N}\!\!\!\!(x_n^2+z^2) {e^{-\frac{1}{2\sigma^2} x_n^2}}dx_n}
{\displaystyle\int_{\Rset^N}\!
\prod\!\!\prod_{\hspace{-16pt}1\leq k<l\leq N}\!\! e^{-\frac{1}{2N}(1-\sigma^{-2}) (x_k-x_l)^2}
\!\!\!\!\!\prod_{1\leq n\leq N}\!\!\!\!x_n^2 {e^{-\frac{1}{2\sigma^2} x_n^2}}dx_n}\\
&=&
1 + {\textstyle\sum\limits_{j=1}^{N}} z^{2j}{\begin{pmatrix} {\textstyle{N}}\cr {\textstyle{j}} \end{pmatrix}}\,
\displaystyle\int_{\Rset^j} \,{\textstyle\prod\limits_{k=1}^{j}} x_k^{-2} \mu_{0;\sigma}^{(j|N)}(x_1,...,x_j) d^jx,
\end{array}
\end{equation}
where $\mu_{0;\sigma}^{(j|N)}(x_1,...,x_j)$ 
is the $j$-th marginal measure of $\mu_{z;\sigma}^{(N)}(x_1,...,x_N)$ with $z=0$,
and $\mu_{z;\sigma}^{(N)}$ is given by the integrand of $E_N^{}(z;\sigma)$ divided by $E_N^{}(z;\sigma)$ itself.
 In the next section (see Corollary \ref{limPTSofMEASURES}) we will prove that for each $z\in\Rset$ and $j\in\Nset$, the sequence
$N\mapsto\mu_{z;\sigma}^{(j|N)}(x_1,...,x_j)$ converges  to either $\nu_{z;\sigma}^{(0)}(x_1)\cdots\nu_{z;\sigma}^{(0)}(x_j)$
or to $\frac12[\nu_{z;\sigma}^{(+)}(x_1)\cdots\nu_{z;\sigma}^{(+)}(x_j) +\nu_{z;\sigma}^{(-)}(x_1)\cdots\nu_{z;\sigma}^{(-)}(x_j)]$,
depending on whether $z^2\geq 2\sigma^2-3$ or $z^2 < 2\sigma^2-3$;
the probability densities $\nu_{z;\sigma}^{(0)}(x_1)$ and $\nu_{z;\sigma}^{(\pm)}(x_1)$ are defined in the text ensuing (\ref{rhoEULERLAGRANGEe}).
 The special case $z=0$ of Corollary \ref{limPTSofMEASURES} is here stated as
\begin{lemm}\label{lemmaRHOzNULL}
 The probability densities  $\nu_{0;\sigma}^{(0)}(x_1)$ and $\nu_{0;\sigma}^{(\pm)}(x_1)$ are given by
\begin{equation}\label{rhoZnull}
\nu_{0;\sigma}^{(\times)} (x_1) = \frac{x_1^2e^{-\frac12  x_1^2 + (1-\frac{1}{\sigma^2})m_\times^{} x_1^{}}}
{\int_\Rset \tilde{x}_1^2 e^{-\frac12  \tilde{x}_1^2 + (1-\frac{1}{\sigma^2}) m_\times^{}\tilde{x}_1^{}}d\tilde{x}_1},
\end{equation}
where $\nu_{0;\sigma}^{(\times)}$ stands for $\nu_{0;\sigma}^{(0)}$ when $\sigma^2\leq 3/2$, and for either
$\nu_{0;\sigma}^{(+)}$ or $\nu_{0;\sigma}^{(-)}$ when $\sigma^2> 3/2$.
 When $\sigma^2\leq \frac32$, then $m=m_0^{}:=0$,  and if $\sigma^2>3/2$, then $m=m_\pm^{}$ with
\begin{equation}\label{mPMzNULL}
m^{2}_\pm: = \sigma^4 \frac{2(\sigma^2 -1)-1}{(\sigma^2-1)^2}.
\end{equation}
\end{lemm}

 Note that $\nu_{0;\sigma}^{(0)}(x_1)$ is an even function of $x_1$, whereas the $\nu_{0;\sigma}^{(\pm)}(x_1)$ are not 
(both are mirror images of each other, though). 

 Now scaling $z\to z/\surd{N}$ and noting that 
\begin{equation}\label{combi}
\frac{1}{N^{j}}{\begin{pmatrix} {\textstyle{N}}\cr {\textstyle{j}} \end{pmatrix}} 
= \frac{N(N-1)\cdots(N-j+1)}{N^{j}} \frac{1}{j!}
= \frac{1}{j!} \prod_{k=1}^{j-1} \left(1-\frac{k}{N}\right) \stackrel{\scriptstyle{N\to\infty}}{\longrightarrow } \frac{1}{j!},
\end{equation}
we can let $N\to\infty$ term by term in the expansion (\ref{ENforREzTAYLORscaled}) (with $z\to z/\surd{N}$) to find, 
for the $n$-th partial sum, 
\begin{equation}\label{ENforREzTAYLORscaledNtoINFINITYpartialSUM} \hspace{-20pt}
\begin{array}{lll}
1\!+\!{\textstyle\sum\limits_{j=1}^{n}} {\displaystyle\frac{z^{2j}}{N^j}}{\begin{pmatrix} {\textstyle{N}}\cr {\textstyle{j}} \end{pmatrix}}\!\!
\displaystyle\int_{\Rset^j}\!{\textstyle\prod\limits_{k=1}^{j}} x_k^{-2} \mu_{0;\sigma}^{(j|N)}(x_1,...,x_j) d^jx
\stackrel{\scriptstyle{N\to\infty}}{\longrightarrow } 
1 \!+\! {\textstyle\sum\limits_{j=1}^{n}} \frac{1}{j!}
\!\displaystyle\left(\!\!z^{2}\!\!\int_{\Rset}  x_1^{-2} \nu_{0;\sigma}^{(\times)}(x_1^{}) dx_1^{}\right)^{\!\!j}\!\!.\!\!\!\!\!
\end{array}
\end{equation}
 The Gaussian integrals $\int_{\Rset}  x_1^{-2} \nu_{0;\sigma}^{(\times)}(x_1^{}) dx_1$ are easy to carry out.
 The simplest way is to notice that 
$\int_{\Rset}  x_1^{-2} \nu_{0;\sigma}^{(\times)}(x_1^{}) dx_1^{} = 1/(\mbox{Var}_\times(X_\times)+ \mbox{Exp}_\times(X_\times)^2)$,
where $X_\times$ is a normal random variable with mean $(1-\sigma^{-2})m_\times^{}$ and variance 1; here Exp$_\times$ and Var$_\times$ stand 
for the mean and variance computed w.r.t. the p.d.f. of $X_\times$.
 Thus we have 
\begin{equation}\label{GAUSSintegrals}
\int_{\Rset}  x_1^{-2} \nu_{0;\sigma}^{(\times)}(x_1^{}) dx_1^{} 
= \frac{1}{1+m_\times^2 (1-\sigma^{-2})^2} 
= \left\{ \begin{array}{lcr} \hspace{-10pt}
&\hspace{-10pt} 1 \quad &\mbox{if}\quad \sigma^2 \leq 3/2 \cr
&\hspace{-10pt} \frac{1}{2(\sigma^2 -1)} \quad &\mbox{if}\quad \sigma^2 > 3/2 \cr
 \end{array}
\right.
\end{equation}
 (Note that $\int_{\Rset}  x_1^{-2} \nu_{0;\sigma}^{(+)}(x_1^{}) dx_1^{} =\int_{\Rset}  x_1^{-2} \nu_{0;\sigma}^{(-)}(x_1^{}) dx_1^{}$.)

 Now letting $n\to\infty$ yields Theorem~\ref{thmCOMPLEXzSCALED}. \hfill QED

\newpage

 We end this subsection by illustrating Theorem \ref{thmCOMPLEXzSCALED} with the parameter choice $\sigma=4$, in which case
$2(\sigma^2-1)=30$, once for $z=x\in\Rset$ and once for $z=iy;\ y\in\Rset$.

\begin{figure}[H]
\centering
\includegraphics[scale=0.5]{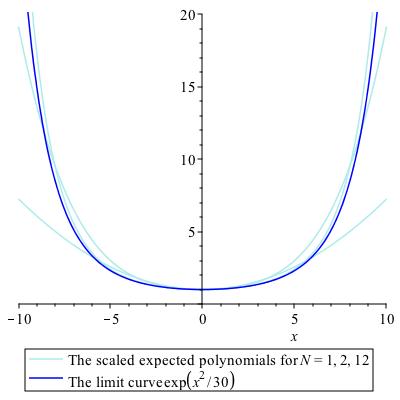}
\vspace{-10pt}
\caption{
\small Graphs of $x\mapsto E_N^{}(z/\surd{N};\sigma=4)/E_N^{}(0;4)$ with $N\in\{1,2,12\}$, together with the graph of
$x\mapsto \exp(z^2/30)$, for $z=x+i0$ and $x\in(-10,10)$.}
\label{ENforREALzSCALEDsigma4}
\end{figure} 

\begin{figure}[H]
\centering
\includegraphics[scale=0.5]{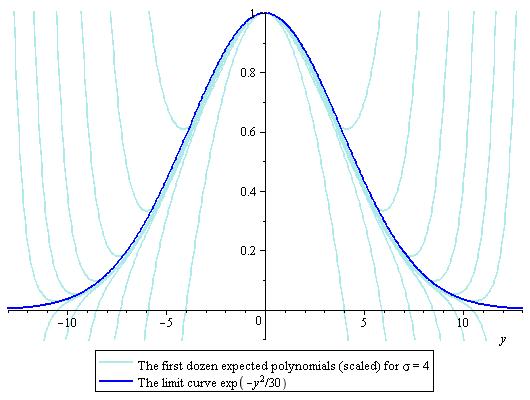}
\vspace{-10pt}
\caption{
\small Graphs of $y\mapsto E_N^{}(z/\surd{N};\sigma=4)/E_N^{}(0;4)$ with $N\in\{1,...,12\}$, together with the graph of
$y\mapsto \exp(z^2/30)$, for $z=0+iy$ and $y\in(-13,13)$.}
\label{ENforIMAGzSCALEDsigma4}
\end{figure}

\section{Large-$N$ limit of $E_N^{1/N}(z;\sigma)$ when $z\in\Rset$}\label{ASYMPzREAL}\vspace{-15pt}

 We first give a simple convergence proof for $E_N(z;\sigma)^{\frac1N}$ when $N\to\infty$ and $z=x+i0$ with
$x\in\Rset$.
 Although this proof doesn't reveal the limit, it simplifies the subsequent proof that it is given by 
(\ref{lnEXPECTfONEfTWOasNtoINFINITYexistsA}), respectively (\ref{lnEXPECTfONEfTWOasNtoINFINITYexistsB}).

\begin{theo}\label{thm2}
 If $\sigma>0$ and $z\in\Rset$, then there exists an $L(z;\sigma)\in\Rset_+$ such that \vspace{-10pt}
\begin{equation}\label{lnEXPECTfONEfTWOasNtoINFINITYexistsAGAIN}
\displaystyle
\left[{\textstyle\frac{1}{\sigma}}\int_{\Rset^N} 
\prod\!\!\prod_{\hspace{-16pt}1\leq k<l\leq N} e^{-\frac{1}{2N}(1-\sigma^{-2}) (x_k-x_l)^2}
\!\!\!\prod_{1\leq n\leq N}(x_n^2+z^2) \frac{e^{-\frac{1}{2\sigma^2} x_n^2}}{\sqrt{2\pi}}dx_n\right]{\!\!}^{^{\frac1N}}
 \stackrel{\scriptstyle{N\to\infty}}{\longrightarrow }
L(z;\sigma).
\end{equation}
\end{theo}

\smallskip
\noindent
\textit{Proof of Theorem \ref{thm2}}:

 Let $N>1$ and set $N=N_1+N_2$, with $N_1\in\Nset$ and $N_2\in\Nset$.
 Then Jensen's inequality applied to the last term in (\ref{harmonicIDENTITY}) gives us
\begin{equation}\label{Jensen}
\Bigl({\textstyle{\frac{1}{\sqrt{N}}\sum\limits_{1\leq k \leq N}}} x_k\Bigr)^2
\leq
\Bigl({\textstyle{\frac{1}{\sqrt{N_1}}\sum\limits_{1\leq k \leq N_1}}} x_k\Bigr)^2
+
\Bigl({\textstyle{\frac{1}{\sqrt{N_2}}\sum\limits_{N_1+1\leq k \leq N}}} x_k\Bigr)^2,
\end{equation}
with ``$=$'' iff $N_1=N_2$ and $\sum\limits_{1\leq k \leq N_1} x_k = \sum\limits_{N_1+1\leq k \leq N} x_k$.
 With the help of this inequality, when $N_1>1$, $N_2>1$, and $N=N_1+N_2$,
we now find from (\ref{harmonicIDENTITY}) that 
\begin{equation}\label{harmonicESTIMATE}
\frac{1}{N}
\sum\!\!\sum_{\hspace{-16pt}1\leq k<l\leq N} (x_k-x_l)^2
\geq 
\frac{1}{N_1}
\sum\!\!\sum_{\hspace{-16pt}1\leq k<l\leq N_1} (x_k-x_l)^2
+
\frac{1}{N_2}\;\;\;
\sum\!\!\!\!\sum_{\hspace{-20pt}N_1+1\leq k<l\leq N} (x_k-x_l)^2.
\end{equation}
 Therefore, since $z\in\Rset$ and $\sigma>0$, when $N_1>1$, $N_2>1$, and $N_1+N_2=N$, we have
\begin{equation}\label{SUPERsubPRODUCTIVITYposZETA}
\sigma E_N^{}(z;\sigma)
\left\{\begin{array}{lll} \hspace{-10pt}
&\hspace{-10pt} > \;\sigma E_{N_1}^{}(z;\sigma)\,\sigma E_{N_2}^{}(z;\sigma)\quad \mbox{if}\quad \sigma<1\cr
&\hspace{-10pt} = \;\sigma E_{N_1}^{}(z;\sigma)\,\sigma E_{N_2}^{}(z;\sigma)\quad \mbox{if}\quad \sigma=1\cr
&\hspace{-10pt} < \;\sigma E_{N_1}^{}(z;\sigma)\,\sigma E_{N_2}^{}(z;\sigma)\quad \mbox{if}\quad \sigma>1\cr
\end{array} \right.;
\end{equation}
thus, when $z\in\Rset$ and $\sigma>0$, the sequence $N\mapsto\ln(\sigma E_N^{}(z;\sigma))$ restricted to $N_1>1$, $N_2>1$, and $N=N_1+N_2$,
is superadditive for $\sigma<1$ and subadditive for $\sigma>1$.
 And since $\Bigl({\textstyle{\frac{1}{\sqrt{N}}\sum\limits_{1\leq k \leq N}}} x_k\Bigr)^2\geq 0$ 
(with ``$=$'' iff $\sum_kx_k=0$), we also have the estimates\footnote{Estimates in the opposite direction follow
from $(x_k-x_l)^2\geq 0$ with ``$=$'' iff $x_k=x_l$, thus
\begin{equation}\label{SUPERsubBOUNDS}
\sigma E_N^{}(z;\sigma)\;
\left\{\begin{array}{lll} 
&\hspace{-10pt} >\quad \sigma^{N}(\sigma^2+z^2)^N \quad \mbox{if}\quad \sigma<1\cr
&\hspace{-10pt} <\quad \sigma^{N}(\sigma^2+z^2)^N \quad \mbox{if}\quad \sigma>1\cr
\end{array} \right..
\end{equation}}
\begin{equation}\label{SUBsuperBOUNDS}
\sigma E_N^{}(z;\sigma)\;
\left\{\begin{array}{lll} \hspace{-10pt} 
&\hspace{-10pt}  <\; (1+z^2)^N \quad \mbox{if}\quad \sigma<1\cr
&\hspace{-10pt}  =\; (1+z^2)^N \quad \mbox{if}\quad \sigma=1\cr
&\hspace{-10pt}  >\; (1+z^2)^N \quad \mbox{if}\quad \sigma>1\cr
\end{array} \right..
\end{equation}
 Therefore, by Fekete's subadditivity lemma, we can conclude that 
\begin{equation}\label{Glimits}
(\sigma E_N^{}(z;\sigma))^{\frac1N} 
\stackrel{\scriptstyle{N\to\infty}}{\longrightarrow }
\left\{\begin{array}{lll} \hspace{-10pt}
&\hspace{-10pt} \  \sup_N (\sigma E_N^{}(z;\sigma))^{\frac1N} \quad \mbox{if}\quad \sigma<1\cr
&\hspace{-10pt} \qquad\qquad 1+z^2 \hspace{22pt} \quad       \mbox{if}\quad \sigma=1\cr
&\hspace{-10pt} \ \ \inf_N (\sigma E_N^{}(z;\sigma))^{\frac1N}  \quad \mbox{if}\quad \sigma>1\cr
\end{array} \right..
\end{equation}
 Of course, $\sigma^{1/N} \stackrel{\scriptstyle{N\to\infty}}{\longrightarrow } 1$, and Theorem \ref{thm2} is proved.
\hfill QED
\smallskip

 Note that (\ref{Glimits}) states that $E_N^{1/N}(z;\sigma)$ converges to $\sup_N (\sigma E_N^{}(z;\sigma))^{\frac1N}$ if $\sigma<1$
and to $\inf_N (\sigma E_N^{}(z;\sigma))^{\frac1N}$ if $\sigma>1$, which \emph{a-priori} is not the same as 
$\sup_N E_N^{}(z;\sigma)^{\frac1N}$ if $\sigma<1$, resp. $\inf_N E_N^{}(z;\sigma)^{\frac1N}$ if $\sigma>1$; note also that
for $\sigma=1$, (\ref{Glimits}) already exhibits the limit $L(z;1)$ explicitly.
 We next state the limit $L(z;\sigma)$ for general $\sigma> 0$.

\begin{theo}\label{thm3} Let $z\in\Rset$ and let $L(z;\sigma)$ be defined as in Thm. \ref{thm2}.

 Then, whenever $z^2\geq 2\sigma^2-3$, one has
\begin{equation}\label{lnEXPECTfONEfTWOasNtoINFINITYa}
L(z;\sigma) = {1+z^2}.
\end{equation}
 
 On the other hand, if $\sigma^2>3/2$ and $z^2 < 2\sigma^2 -3$, one has
\begin{equation}\label{lnEXPECTfONEfTWOasNtoINFINITYb}
L(z;\sigma) =  2(\sigma^2 -1) \exp\left( \frac{1+z^2}{2(\sigma^2-1)} -1\right).
\end{equation}
\end{theo}

\smallskip
\begin{rema}
 Somewhat surprisingly, $L(z;\sigma)\equiv L(z;1)$ when $z^2\geq 2\sigma^2-3$. 
 Thus, if $z^2\geq 2\sigma^2-3$, with $z\in\Rset$, then $L(z;\sigma)$ cannot be used to distinguish
the multi-variate normal random variables with $\sigma\neq 1$ from the i.i.d. standard normal random variables ($\sigma=1$)
in the limit $N\to\infty$.
 With $z\in\Rset$ any dependence of the multi-variate normal random variables with $\sigma\neq 1$ is visible
in the limit $N\to\infty$ only when $\sigma^2>3/2$ and $z^2< 2\sigma^2-3$.
\end{rema}

\smallskip
\noindent
\textit{Proof of Theorem \ref{thm3}}:

 We will show that the logarithm of the l.h.s.(\ref{lnEXPECTfONEfTWOasNtoINFINITYexistsAGAIN})
converges to the logarithm of the r.h.s.(\ref{lnEXPECTfONEfTWOasNtoINFINITYa}), respectively 
of r.h.s.(\ref{lnEXPECTfONEfTWOasNtoINFINITYb}).
 More precisely, we will closely follow \cite{MesserSpohn,KiesslingCPAM,KiesslingSpohnCMP} to show that 
$\frac1N\ln E_N^{}(z;\sigma) \stackrel{\scriptstyle{N\to\infty}}{\longrightarrow} \max\limits_\nu  \gFUNC_{z;\sigma}^{}(\nu)$, with
\begin{equation}\label{rhoFUNCTIONAL}
\begin{array}{ll} \hspace{-10pt}
\hspace{-8pt}\gFUNC_{z;\sigma}^{}(\nu) = 
\displaystyle -\int_{\Rset} \nu(x_1) \ln \frac{\nu(x_1)}{\nu_\sigma^{}(x_1)} dx_1 & + 
\displaystyle\int_{\Rset} \nu(x_1) \ln[\sigma(x_1^2+z^2)]  dx_1 \cr
&-\tfrac{\sigma^2-1}{4\sigma^2}\displaystyle \iint_{\!\!\!\Rset\times\Rset}\! \nu (x_1)\nu(x_2) (x_1 -x_2)^2 dx_1 dx_2,\!\!
\end{array}
\end{equation}
where $\nu_\sigma^{}$ is the probability density of a mean-zero normal random variable with variance $\sigma^2$, 
and the maximum is w.r.t. absolutely continuous probability densities $\nu$ on $\Rset$ for which $\gFUNC_{z;\sigma}^{}(\nu)$ exists. 
 The maximum will be explicitly computed near the end of our proof  of Theorem~\ref{thm3} 
(see equations (\ref{rhoEULERLAGRANGE})--(\ref{lnEXPECTfONEfTWOasNtoINFINITYc})),
and found to be $\ln L(z;\sigma)$ as given by (\ref{lnEXPECTfONEfTWOasNtoINFINITYa}) and~(\ref{lnEXPECTfONEfTWOasNtoINFINITYb}).

\begin{rema}
 It is a relatively straightforward exercise in functional analysis to show that $\gFUNC_{z;\sigma}^{}(\nu)$ does have a maximum.
 However, it is not necessary to show this up front, because the convergence proof for $\ln E_N^{}(z;\sigma)^{1/N}$ will yield 
this result as a by-product.
\end{rema}

\begin{prop}\label{lowGbd}
 For $z\in\Rset$ and $\sigma>0$,
\begin{equation}\label{muFUNCTIONALestimatedBYrhoFUNCTIONALliminf}
\lim_N N^{-1}\ln E_N^{}(z;\sigma)\geq \sup_\nu \gFUNC_{z;\sigma}^{}(\nu).
\end{equation}
\end{prop}

\smallskip\noindent
\textit{Proof of Proposition \ref{lowGbd}}:

 We begin with Gibbs' finite-$N$ variational principle \cite{GibbsBOOK}, which for us becomes
\begin{equation}\label{muFUNCTIONALmaximized}
\ln E_N^{}(z;\sigma) = \max_{\rho^{(N)}} {\cal G}_{z;\sigma}^{(N)}(\rho^{(N)}),
\end{equation}
with 
\begin{equation}\label{muFUNCTIONAL}
{\cal G}_{z;\sigma}^{(N)}(\rho^{(N)}) = 
-\int_{\Rset^N} \rho^{(N)} \ln \frac{\rho^{(N)}}{\nu_\sigma^{(N)}} d^Nx 
 + \int_{\Rset^N} \rho^{(N)}(x_1,...,x_N) \ln\!\!\!\prod_{1\leq n\leq N}(x_n^2+z^2)d^Nx,
\end{equation}
and where
$\nu_\sigma^{(N)}(x_1,...,x_N) = \frac{1}{\sigma}(\frac{1}{\sqrt{2\pi}})^N
\prod\!\!\!\prod\limits_{\hspace{-14pt}1\leq k<l\leq N} e^{-\frac{1}{2N}(1-\sigma^{-2}) (x_k-x_l)^2}
\!\!\!\!\prod\limits_{1\leq n\leq N}\!\!\!{e^{-\frac{1}{2\sigma^2} x_n^2}}$
is our multi-variate normal probability density; the functional is maximized over the set of absolutely continuous 
(w.r.t. Lebesgue measure) permutation-symmetric $N$-point probability densities
$\rho^{(N)}$ for which the 
relative entropy functional (with physicists' sign convention) exists, and which are integrable against 
$\ln \prod_{1\leq n\leq N}(x_n^2+z^2 )$.
 By the Gibbs inequality, the unique maximizer $\rho_{z;\sigma}^{(N)}$ of ${\cal G}_{z;\sigma}^{(N)}(\rho^{(N)})$ amongst these 
$N$-point probability measures is given by the integrand of $E_N^{}(z;\sigma)$ divided by the manifestly positive $E_N^{}(z;\sigma)$ itself; 
a simple computation then yields (\ref{muFUNCTIONALmaximized}).

 Lower bounds to $\max_\rho{\cal G}_{z;\sigma}^{(N)}(\rho^{(N)})$ are obtained by evaluating 
${\cal G}_{z;\sigma}^{(N)}(\rho^{(N)})$ with any particular symmetric product of admissible
one-point probability densities, viz. $\rho^{(N)}(x_1,...,x_N) = \nu(x_1)\cdots\nu(x_N)$, 
where ``admissible'' means that $\nu$ has finite relative entropy and finite second moment.  
 Thus, we have
\begin{equation}\label{muFUNCTIONALestimatedBYrhoFUNCTIONAL}
\hspace{-16pt}
\begin{array}{lll}
\max\limits_\rho {\cal G}_{z;\sigma}^{(N)}(\rho^{(N)}) 
&\!\!\!\geq\!\!\!\! & N\gFUNC_{z;\sigma}^{}(\nu) - 
\ln \sigma + (1-\tfrac{1}{\sigma^2}){\textstyle\frac14}\displaystyle\iint_{\!\!\!\Rset\times\Rset}\!\!\! \nu (x_1)\nu(x_2) (x_1 -x_2)^2 dx_1 dx_2.
\end{array}\!\!\!
\end{equation}
 Dividing by $N$, then letting $N\to\infty$, we find that 
\begin{equation}\label{muFUNCTIONALestimatedBYrhoFUNCTIONALanyRHO}
\lim_N N^{-1}\ln E_N^{}(z;\sigma)
\geq
\gFUNC_{z;\sigma}^{}(\nu)
\end{equation}
for any admissible $\nu$. 
 This proves Proposition \ref{lowGbd}. 
\hfill QED

\begin{rema}
 Note that (\ref{muFUNCTIONALestimatedBYrhoFUNCTIONALanyRHO}) together with Theorem \ref{thm2}
proves that $\gFUNC_{z;\sigma}^{}(\nu)$ does have a finite supremum over the stipulated set of probability densities $\nu$.
\end{rema}

We next prove a complementary estimate in the opposite direction, viz.
\begin{prop}\label{upGbd}
 For $z\in\Rset$ and $\sigma>0$,
\begin{equation}\label{muFUNCTIONALestimatedBYrhoFUNCTIONALlimsup}
\lim_N N^{-1}\ln E_N^{}(z;\sigma) \leq \sup_\nu \gFUNC_{z;\sigma}^{}(\nu).
\end{equation}
\end{prop}

\smallskip\noindent
\textit{Proof of Proposition \ref{upGbd}}: 

 As before, let $\rho_{z;\sigma}^{(N)}$ denote the maximizer of ${\cal G}_{z;\sigma}^{(N)}(\rho^{(N)})$. 
 We introduce the notation $\rho_{z;\sigma}^{(n|N)}$ for the $n$-th marginal density of $\rho_{z;\sigma}^{(N)}$ obtained by 
integrating $\rho_{z;\sigma}^{(N)}$ over $N-n$ of the $x$-variables; by the permutation invariance we may stipulate to integrate
over the $x$-variables indexed by $n+1,...,N$. 
\begin{lemm}\label{lemmaBOUNDandTIGHT}
 For each $n\in\Nset$, $z\in\Rset$, and $\sigma>1$ there are constants $C_n(z;\sigma)>0$ and  $K_n(z;\sigma)\in\Nset$ such that,
whenever $N>K_n(z;\sigma)$, we have
\begin{equation}\label{marginalBOUNDandTIGHT}
\rho_{z;\sigma}^{(n|N)} \leq C_n(z;\sigma) \prod_{k=1}^n(x_k^2+z^2)e^{-\frac{n-1}{4n}x^2_k}.
\end{equation}
\end{lemm}
 We will supply the proof of Lemma \ref{lemmaBOUNDandTIGHT} after finishing the main line of reasoning in the proof of Proposition \ref{upGbd},
which we now continue.
 
 Lemma \ref{lemmaBOUNDandTIGHT} implies that the sequence $N\mapsto \rho_{z;\sigma}^{(n|N)}$ 
is uniformly bounded in each $L^p(\Rset^n)$, that each and every of its moments is uniformly bounded, and that it is tight.
 Thus, for each $n\in\Nset$ the sequence $N\mapsto \rho_{z;\sigma}^{(n|N)}$ has weak limit points in the set of probability 
measures on $\Rset^n$, and the limit points are in every $L^p$ space, having finite moments of arbitrary order.
 Let $\tilde\rho_{z;\sigma}^{(n)}$ denote such a limit point, and let $n'<n$; 
then $\tilde\rho_{z;\sigma}^{(n')}:=\int_{\Rset^{n-n'}}\tilde\rho_{z;\sigma}^{(n)}d^{n-n'}x$ is a 
compatible limit point of $N\mapsto\rho_{z;\sigma}^{(n'|N)}$.

 With the help of the marginals we can rewrite ${\cal G}_{z;\sigma}^{(N)}(\rho_{z;\sigma}^{(N)})$ as follows,
\begin{equation}\label{muFUNCTIONALrewrite}
\begin{array}{lll}\hspace{-10pt}
\frac1N{\cal G}_{z;\sigma}^{(N)}(\rho_{z;\sigma}^{(N)}) &\!\!\!=\!\!\!& 
-\frac1N\displaystyle\int_{\Rset^N} \rho_{z;\sigma}^{(N)} \ln \frac{\rho_{z;\sigma}^{(N)}}{\nu_\sigma^{\otimes{N}}} d^Nx + 
\displaystyle\int_{\Rset} \rho_{z;\sigma}^{(1|N)}(x_1) \ln (z^2+x_1^2) dx_1 - \\
&\!\!\!\!\!\!& 
(1-\tfrac1N){\textstyle\frac{\sigma^{2}-1}{4\sigma^{2}}}\displaystyle\iint_{\Rset^2}\rho_{z;\sigma}^{(2|N)}(x_1,x_2)(x_1-x_2)^2dx_1dx_2 + 
(1-\tfrac1N)\ln\sigma .
\end{array}
\end{equation}
 We know the left and (hence) the right hand sides in (\ref{muFUNCTIONALrewrite}) converge as $N\to\infty$. 
 We now estimate r.h.s.(\ref{muFUNCTIONALrewrite}) in terms of the limit points of the sequence of marginals.

 First, by Lemma \ref{lemmaBOUNDandTIGHT}, for a convergent subsequence $N'\mapsto \rho_{z;\sigma}^{(n|N')}$ we have
\begin{equation}
\lim_{N'\to\infty}\iint_{\!\!\!\Rset^2} (x_1-x_2)^2 \rho_{z;\sigma}^{(2|N')}(x_1,x_2)
 d^2x = \iint_{\!\!\!\Rset^2} (x_1-x_2)^2 \tilde\rho_{z;\sigma}^{(2)}(x_1,x_2) d^2x ,
\end{equation}
and
\begin{equation}
\lim_{N'\to\infty}\int_{\Rset} \ln(z^2+x_1^2) \rho_{z;\sigma}^{(1|N')}(x_1)dx_1=\int_{\Rset} \ln(z^2+x_1^2) \tilde\rho_{z;\sigma}^{(1)}(x_1) dx_1 .
\end{equation}

 Second, by the subadditivity of the entropy functional for probability measures, and its weak upper semi-continuity, we have for any $n$,
\begin{equation}
\limsup_N -\frac1N\displaystyle\int_{\Rset^N} \rho_{z;\sigma}^{(N)} \ln \frac{\rho_{z;\sigma}^{(N)}}{\nu_\sigma^{\otimes{N}}} d^Nx \leq 
-\frac1n\displaystyle\int_{\Rset^n} \rho_{z;\sigma}^{(n|N)} \ln \frac{\rho_{z;\sigma}^{(n|N)}}{\nu_\sigma^{\otimes{n}}} d^nx .
\end{equation}
 Let $n\mapsto \tilde\rho_{z;\sigma}^{(n)}$, $n\in\Nset$, be a sequence of compatible limit points, and
let $\tilde\rho_{z;\sigma}$ denote any limit point of such a sequence of compatible marginal measures. 
 Then, as shown by Robinson and Ruelle \cite{RobinsonRuelle}, the ``mean entropy of this $N=\infty$ state'' is well-defined by 
\begin{equation}
S(\tilde\rho_{z;\sigma}) := \lim_{n\to\infty}
-\frac1n\displaystyle\int_{\Rset^n} \tilde\rho_{z;\sigma}^{(n)} \ln \frac{\tilde\rho_{z;\sigma}^{(n)}}{\nu_\sigma^{\otimes{n}}} d^nx .
\end{equation}
 Thus, we can conclude that
\begin{equation}\label{muFUNCTIONALestimatedBYrhoFUNCTIONALlimsupB} 
\begin{array}{lll}
\lim_N \textstyle\frac1N {\cal G}_{z;\sigma}^{(N)}(\rho_{z;\sigma}^{(N)}) \leq S(\tilde\rho_{z;\sigma}^{}) 
&\!\!\! + \displaystyle\int_{\Rset} \ln[\sigma (z^2+x_1^2)] \tilde\rho_{z;\sigma}^{(1)}(x_1) dx_1 \\
&- {\textstyle\frac{\sigma^{2}-1}{4\sigma^2}}\!\displaystyle\iint_{\!\!\!\!\Rset^2}\!\!  (x_1-x_2)^2 \tilde\rho_{z;\sigma}^{(2)}(x_1,x_2) dx_1dx_2 .
\end{array}
\end{equation}
 Now by the Hewitt--Savage extreme point decomposition of $\tilde\rho_{z;\sigma}^{}$, for $n\in\Nset$ we have
\begin{equation}\label{HewittSavage} 
\tilde\rho_{z;\sigma}^{(n)} (x_1^{},...,x_n^{})
= \int \nu(x_1^{})\cdots\nu(x_n^{}) \omega(d\nu|\tilde\rho_{z;\sigma}^{}) ;
\end{equation}
here, $\omega(d\nu|\tilde\rho_{z;\sigma}^{})$ is the Hewitt--Savage decomposition measure of $\tilde\rho_{z;\sigma}^{}$, 
a probability measure on the set of probability measures $\nu$ on $\Rset$; see \cite{HewittSavage}.
 By the linearity in $\tilde\rho_{z;\sigma}^{}$
of the second and third integrals in (\ref{muFUNCTIONALestimatedBYrhoFUNCTIONALlimsupB}), and by the affine linearity of the
mean entropy functional \cite{RobinsonRuelle}, we now have
\begin{equation}\label{HewittSavageENTROPY} 
\mbox{r.h.s.}(\ref{muFUNCTIONALestimatedBYrhoFUNCTIONALlimsupB})= \int \gFUNC_{z;\sigma}^{}(\nu)\omega(d\nu|\tilde\rho_{z;\sigma}^{}) 
\leq \sup_\nu \gFUNC_{z;\sigma}^{}(\nu),
\end{equation}
where the inequality is obvious.

 To finish the proof of  Proposition \ref{upGbd} it remains to prove Lemma \ref{lemmaBOUNDandTIGHT}.
\newpage

\smallskip\noindent
\textit{Proof of Lemma \ref{lemmaBOUNDandTIGHT}}: 

 When $n=1$ the Lemma is obviously true. 
 Hence, let $n>1$. 
 With the help of (\ref{multiVARnormalREWRITE}) and (\ref{harmonicIDENTITYconcl}), and using that $1= \frac1n + \frac{n-1}{2n} + \frac{n-1}{2n}$
to distribute $\sum_{k=1}^nx_k^2$ over three places, we rewrite $\rho_{z;\sigma}^{(n|N)}(x_1,...,x_j)$ as
follows,
\begin{equation}\label{ENforREzTAYLOR}
\begin{array}{lll}
\rho_{z;\sigma}^{(n|N)}(x_1^{},...,x_n^{})
&\!\!\!\!:=\!\!\!&\!\!\!\!\!
 \frac{\displaystyle\int_{\Rset^{N-n}}\!
\prod\!\!\prod_{\hspace{-16pt}1\leq j<k\leq N}\!\! e^{-\frac{1}{2N}(1-\sigma^{-2}) (x_j-x_k)^2}
\!\!\!\!\prod_{1\leq l\leq N}\!\!\!\!(x_l^2+z^2) {e^{-\frac{1}{2\sigma^2} x_l^2}}dx_{n+1}^{}\!\!\cdots dx_N^{}}
{\displaystyle\int_{\Rset^N}\!
\prod\!\!\prod_{\hspace{-16pt}1\leq j<k\leq N}\!\! e^{-\frac{1}{2N}(1-\sigma^{-2}) (x_j-x_k)^2}
\!\!\!\!
\prod_{1\leq l\leq N}\!\!\!\!(x_l^2+z^2) {e^{-\frac{1}{2\sigma^2} x_l^2}}dx_l} \!\!\!\!\!\!\! \\
&=&\!
\displaystyle \prod_{k=1}^n(x_k^2+z^2)e^{-\frac{n-1}{4n}x^2_k} R_{z;\sigma}^{(n|N)}(x_1^{},...,x_n^{})
\end{array}\!\!\!\!\!\!\!\!\!\!
\end{equation} 
where $R_{z;\sigma}^{(n|N)}(x_1^{},...,x_n^{})$ is defined by (\ref{ENforREzTAYLOR}).
 We now estimate the numerator of  $R_{z;\sigma}^{(n|N)}(x_1^{},...,x_n^{})$: we use that,
if $\sigma\leq 1$, and also when $N>n^2$ if $\sigma>1$, then
\begin{equation}
\frac1n\sum_{k=1}^n x_k^2 - \frac{\sigma^{2}-1}{\sigma^2}\frac1N\left(\sum_{k=1}^nx_k^{} \right)^2\geq 0;
\end{equation} 
we also use that 
\begin{equation}
\sum_{k=1}^n \left(\frac{n-1}{4n}x_k^2 -  x_k^{}\frac{\sigma^{2}-1}{\sigma^2}\frac1N \sum_{l=n+1}^Nx_l^{}\right) \geq 
- \frac{n^2}{n-1}\left(\frac{\sigma^{2}-1}{\sigma^2}\frac1N \sum_{k=n+1}^Nx_k^{}\right)^2.
\end{equation} 
 This yields $R_{z;\sigma}^{(n|N)}(x_1^{},...,x_n^{})\leq T_{z;\sigma}^{(n|N)}$, where  
\begin{equation}\label{T}
\begin{array}{lll}
T_{z;\sigma}^{(n|N)}
&\!\!\!:=\!\!\!&
 \frac{\displaystyle\int_{\Rset^{N-n}}\!\!\!\!\!\!
 e^{\frac{n^2}{n-1}\Bigl(\frac{\sigma^{2}-1}{\sigma^2}\frac1N \sum\limits_{k=n+1}^Nx_k^{}\Bigr)^2}
\!\!\!\prod\!\!\!\!\!\prod_{\hspace{-16pt}n+1\leq j<k\leq N}\!\!\!\!\! e^{-\frac{1}{2N}(1-\sigma^{-2}) (x_j-x_k)^2}
\!\!\!\!\!\!\!\prod_{n+1\leq l\leq N}\!\!\!\!\!\!(x_l^2+z^2) {e^{-\frac{1}{2\sigma^2} x_l^2}}dx_{l}^{}} 
{\displaystyle\int_{\Rset^N}\!
\prod\!\!\prod_{\hspace{-16pt}1\leq j<k\leq N}\!\! e^{-\frac{1}{2N}(1-\sigma^{-2}) (x_j-x_k)^2}
\!\!\!\!\prod_{1\leq l\leq N}\!\!\!\!(x_l^2+z^2) {e^{-\frac{1}{2\sigma^2} x_l^2}}dx_l} 
\end{array}
\end{equation}
 The denominator of $T_{z;\sigma}^{(n|N)}\!$ can be estimated from below by Jensen's inequality when averaging w.r.t. the
probability density $(\sqrt{2\pi}(1+z^2))^{-n} \prod_{k=1}^n (x_k^2+z^2) e^{-\frac12 x_k^2}$, thus
\begin{equation}\label{TJensen}
\begin{array}{rlr}
\displaystyle\int_{\Rset^n}
\prod\!\!\prod_{\hspace{-16pt}1\leq j<k\leq N}\!\! e^{-\frac{1}{2N}(1-\sigma^{-2}) (x_j-x_k)^2}
\!\!\prod_{1\leq l\leq n}\!\!(x_l^2+z^2) {e^{-\frac{1}{2\sigma^2} x_l^2}}dx_l 
 &\!\!\!\geq \!\!\!& \\ 
 {\displaystyle 
(\sqrt{2\pi}(1+z^2))^n 
e^{ \frac{\sigma^{2}-1}{2\sigma^2}\frac{n}{N} (3+z^2)}
\prod\!\!\!\!\!\prod_{\hspace{-16pt}n+1\leq j<k\leq N}\!\!\!\!\! e^{-\frac{1}{2N}(1-\sigma^{-2}) (x_j-x_k)^2}
}&&
\end{array}
\end{equation}
\vspace{-5pt}

\noindent
This gives \vspace{-10pt}
\begin{equation}\label{TJensenFINAL}
\hspace{-23pt}
\begin{array}{lll}
T_{z;\sigma}^{(n|N)}
&\!\!\!\!\leq\!\! \!\!\!\!&\!\!\!
 \frac{\displaystyle\int_{\Rset^{N-n}}\!\!\!\!\!\!
 e^{\frac{n^2}{n-1}\Bigl(\frac{\sigma^{2}-1}{\sigma^2}\frac1N \sum\limits_{k=n+1}^Nx_k^{}\Bigr)^2}
\!\!\!\prod\!\!\!\!\!\prod_{\hspace{-16pt}n+1\leq j<k\leq N}\!\!\!\!\!\! e^{-\frac{1}{2N}(1-\sigma^{-2}) (x_j-x_k)^2}
\!\!\!\!\!\!\!\prod_{n+1\leq l\leq N}\!\!\!\!\!\!(x_l^2+z^2) {e^{-\frac{1}{2\sigma^2} x_l^2}}dx_{l}^{}} 
{\!\!(\sqrt{2\pi}(1+z^2))^n 
e^{\frac{\sigma^{2}-1}{2\sigma^2}\frac{n}{N} (3+z^2)}\!\!\!\!
\displaystyle\int_{\Rset^{N-n}}\!
\prod\!\!\!\!\prod_{\hspace{-16pt}n+1\leq j<k\leq N}\!\!\!\!\!\! e^{-\frac{1}{2N}(1-\sigma^{-2}) (x_j-x_k)^2}
\!\!\!\!\!\!\!\prod_{n+1\leq l\leq N}\!\!\!\!\!\!(x_l^2+z^2) {e^{-\frac{1}{2\sigma^2} x_l^2}}\!dx_l\!} 
 \\ &\!\!\!= \!\!\!& 
 \frac{\displaystyle\int_{\Rset^{N-n}}\!\!\!\!\!\!
 e^{\Big[1+\frac{2n^2}{n-1}\frac{\sigma^{2}-1}{\sigma^2}\frac1N \Big]\frac{\sigma^{2}-1}{2\sigma^2}\frac1N \Bigl(\sum\limits_{k=n+1}^Nx_k^{}\Bigr)^2}
\!\!\!\!\!\!\!\prod_{n+1\leq l\leq N}\!\!\!\!\!\!(x_l^2+z^2) {e^{-\frac{1}{2} x_l^2}}dx_{l}^{}} 
{\!\!(\sqrt{2\pi}(1+z^2))^n 
e^{\frac{\sigma^{2}-1}{2\sigma^2}\frac{n}{N} (3+z^2)}\!\!\!\!
\displaystyle\int_{\Rset^{N-n}}\!
 e^{\frac{\sigma^{2}-1}{2\sigma^2}\frac1N \Bigl(\sum\limits_{k=n+1}^Nx_k^{}\Bigr)^2}
\!\!\!\!\!\!\!\prod_{n+1\leq l\leq N}\!\!\!\!\!\!(x_l^2+z^2) {e^{-\frac{1}{2} x_l^2}}dx_l} 
\end{array}\hspace{-20pt}\vspace{-10pt}
\end{equation}
where the equality was obtained with the help of (\ref{multiVARnormalREWRITE}) and (\ref{harmonicIDENTITYconcl}); 
the numerator integral at r.h.s. is finite when $\sigma\leq 1$ if $N> \frac{2n^2}{n-1}\frac{1-\sigma^2}{\sigma^2}$, and also when
$\sigma>1$ if $N>(\sigma^2-1)(\frac{2n}{n-1}\frac{\sigma^2-1}{\sigma^2}-1)n$, while
the denominator integral at r.h.s. is always finite.
 The ratio of these two integrals is the reciprocal of an obviously defined expected value 
Ave$\Big[\exp\Big({-\frac{n^2}{n-1}\Bigl(\frac{\sigma^{2}-1}{\sigma^2}\frac1N \sum_{k=n+1}^Nx_k^{}\Bigr)^2}\Big)\Big]$.
 Applying Jensen's inequality, 
\begin{equation}\label{TJensenFINALfinal}
\begin{array}{lll}
T_{z;\sigma}^{(n|N)}
&\!\!\!\leq \!\!\!&
\displaystyle \frac{e^{\frac{n^2}{n-1}\bigl(\frac{\sigma^{2}-1}{\sigma^2}\bigr)^2\bigl(1-\frac{n}{N}\bigr)^2
{\rm{Ave}}\Big[\Bigl(\frac{1}{N-n}\sum\limits_{k=n+1}^Nx_k^{}\Bigr)^2\Big]}}
{(\sqrt{2\pi}(1+z^2))^n e^{\frac{\sigma^{2}-1}{2\sigma^2}\frac{n}{N} (3+z^2)}}.
\end{array}\vspace{-10pt}
\end{equation}
 Now, ${\rm{Ave}}\Big[\Bigl(\frac{1}{N-n}\!\sum_{k=n+1}^N\!x_k^{}\Bigr){}^{^{\!\!2}}\Big] = \frac{d}{d\beta}\bf{F}(\beta)$ at
$\beta = \Big[1+\frac{2n^2}{n-1}\frac{\sigma^{2}-1}{\sigma^2}\frac1N \Big]\frac{\sigma^{2}-1}{2\sigma^2}\Big(1-\frac{n}{N}\Big)\frac1N$, 
with
\begin{equation}
{\bf{F}}(\beta) := \frac{1}{N-n}\ln \displaystyle\int_{\Rset^{N-n}}\!\!\!\!\!\!
 \exp\Big({\beta \Bigl(\frac{1}{\surd{(N-n)}}\!{\textstyle\sum\limits_{k=n+1}^N}\!\! x_k^{}\Bigr)^2}\Big)
\!\!\prod_{n+1\leq l\leq N}\!\!\!\!\!\!(x_l^2+z^2) {e^{-\frac{1}{2} x_l^2}}dx_{l}^{} 
\end{equation}
is defined  for all $\beta<\frac12$; in fact,
\begin{equation}
{\bf{F}}(\beta) \leq \ln\left[ \sqrt{\frac{2\pi}{1-2\beta}}\left(z^2 + \frac{1}{1-2\beta}\right)\right].
\end{equation}
 Moreover, $\beta\mapsto {\bf{F}}$ has arbitrarily many derivatives 
on its $\beta$-domain of definition. 
 In particular, both the first and the second $\beta$ derivative are manifestly positive. 
 Thus,  $\beta\mapsto {\bf{F}}$ is an increasing convex function, and we just found that $\bf{F}(\beta)$ 
has a convex increasing upper bound \emph{uniformly in} $N$ on its $\beta$-domain of definition.  
 Note that for large $N$ the $\beta$ derivative needs to be taken essentially for $\beta\approx 0$; therefore as explained in
\cite{KiesslingCPAM}, proof of Lemma~3, 
the $\beta$ derivative of  $\bf{F}(\beta)$ for $\beta\approx 0$ is bounded \emph{uniformly in}~$N$. 
 
 This proves Lemma \ref{lemmaBOUNDandTIGHT}. \hfill QED

 Proposition \ref{upGbd} is proved. \hfill QED

 By Proposition \ref{lowGbd} and Proposition \ref{upGbd}, we conclude that
\begin{equation}\label{varPRINCIPLElim}
\lim_N N^{-1}\ln E_N^{}(z;\sigma) = \sup_\nu \gFUNC_{z;\sigma}^{}(\nu).
\end{equation}
 Moreover, suppose supp$(\omega(d\nu|\tilde\rho_{z;\sigma}^{}))$ does not consist entirely of
maximizers of $\gFUNC_{z;\sigma}^{}(\nu)$; then strict inequality holds in (\ref{HewittSavageENTROPY}), 
violating (\ref{varPRINCIPLElim}). 
 So $\sup_\nu \gFUNC_{z;\sigma}^{}(\nu) = \max_\nu \gFUNC_{z;\sigma}^{}(\nu)$.

 To find the maximum of $\gFUNC_{z;\sigma}^{}(\nu)$ is a standard problem in variational calculus.
 The maximum is taken at a critical point of $\gFUNC_{z;\sigma}^{}(\nu)$, i.e. its Gateaux derivative at a maximizer 
vanishes in all directions, which gives the Euler--Lagrange equation
\begin{equation}\label{rhoEULERLAGRANGE}
\nu (x_1) = \frac{(x_1^2+ z^2)e^{-\frac{1}{2\sigma^2} x_1^2-\frac12(1-\frac{1}{\sigma^2})\int_\Rset (x_1-\tilde{x})^2\nu(\tilde{x})d\tilde{x}}}
{\int_\Rset(\hat{x}^2+z^2)e^{-\frac{1}{2\sigma^2}\hat{x}^2-\frac12(1-\frac{1}{\sigma^2})\int_\Rset (\hat{x}-\tilde{x})^2\nu(\tilde{x})d\tilde{x}}d\hat{x}}.
\end{equation}
 The fixed point equation (\ref{rhoEULERLAGRANGE}) yields the functional form of $\nu(x_1)$ explicitly,
\begin{equation}\label{rhoEULERLAGRANGEb}
\nu (x_1) = \frac{(x_1^2+z^2)e^{-\frac12  x_1^2 + (1-\frac{1}{\sigma^2})m x_1^{}}}
{\int_\Rset (\tilde{x}^2+ z^2)e^{-\frac12  \tilde{x}^2 + (1-\frac{1}{\sigma^2}) m\tilde{x}}d\tilde{x}};
\end{equation}
here, $m:= \int_{\Rset} \tilde{x} \nu(\tilde{x}) d\tilde{x}$ is the mean of $\nu$, which obeys its own fixed point equation, obtained 
by multiplying (\ref{rhoEULERLAGRANGEb}) by $x_1$ and integrating, which yields
\begin{equation}\label{rhoEULERLAGRANGEc}
  m  = m (1-\tfrac{1}{\sigma^2})\, \frac{z^2  + 3 +  (1-\frac{1}{\sigma^2})^2 m^2}{z^2  +1 +  (1-\frac{1}{\sigma^2})^2 m^2}.
\end{equation}
 The fixed point equation (\ref{rhoEULERLAGRANGEc}) is always solved by $m=0\ (=:m_0^{})$, but real solutions
$m\neq 0$ may exist as well --- they need to satisfy
\begin{equation}\label{rhoEULERLAGRANGEd}
\frac{\sigma^2}{\sigma^{2}-1} =  \frac{z^2  + 3 +  (1-\frac{1}{\sigma^2})^2 m^2}{z^2  +1 +  (1-\frac{1}{\sigma^2})^2 m^2}.
\end{equation}
 Since $z^2\geq 0$, r.h.s.(\ref{rhoEULERLAGRANGEd})$\leq 0$ iff $\sigma\leq 1$, and 
then no real solution of (\ref{rhoEULERLAGRANGEd}) exists.
 Yet, \emph{iff} $\sigma^2>3/2$, then two real solutions, $m_+ = - m_- >0$, do exist \emph{iff} $z^2 <2\sigma^2-3$,
\begin{equation}\label{rhoEULERLAGRANGEe}
m^{2}_\pm  = \sigma^4 \frac{2(\sigma^2 -1)-1-z^2}{(\sigma^2-1)^2}.
\end{equation}

 Accordingly, the pertinent solutions of the Euler--Lagrange equation for $\nu$ are denoted by 
$\nu_{z;\sigma}^{(0)}$ 
and 
$\nu_{z;\sigma}^{(\pm)}$,
 respectively.
 Note that $\nu_{z;\sigma}^{(0)}(x_1)$ is an even function of $x_1$, while $\nu_{z;\sigma}^{(\pm)}(x_1)$ is not
($\nu_{z;\sigma}^{(+)}(x_1)$ and $\nu_{z;\sigma}^{(-)}(x_1)$ are mirror images of each other, though). 
 In the region of $z;\sigma$ parameter space in which $m =0$ is \emph{the only solution} to (\ref{rhoEULERLAGRANGEc})
(recall: $m=0$ is \emph{always a solution} to (\ref{rhoEULERLAGRANGEc})), we have 
$\max_\nu \gFUNC_{z;\sigma}^{}(\nu) = \gFUNC_{z;\sigma}^{}(\nu_{z;\sigma}^{(0)})$, 
with\vspace{-5pt}
\begin{equation}\label{lnEXPECTfONEfTWOasNtoINFINITYaa}
\gFUNC_{z;\sigma}^{}(\nu_{z;\sigma}^{(0)})  = \ln (1 + {z^2}),\vspace{-5pt}
\end{equation}
while in the region in which beside $m = 0$ 
also $m = m_\pm^{} \neq 0$ solves (\ref{rhoEULERLAGRANGEc}) [i.e. when $\sigma^2>3/2$ and $0\leq z^2 <2\sigma^2 -3$], we need
to compare $ \gFUNC_{z;\sigma}^{}(\nu_{z;\sigma}^{(\pm)})$ to $\gFUNC_{z;\sigma}^{}(\nu_{z;\sigma}^{(0)})$, with \vspace{-5pt}
\begin{equation}\label{lnEXPECTfONEfTWOasNtoINFINITYbb}
\gFUNC_{z;\sigma}^{}(\nu_{z;\sigma}^{(\pm)})  = \ln [2(\sigma^2 -1)] + \frac{1+z^2}{2(\sigma^2-1)} -1.\vspace{-5pt}
\end{equation}
 Setting $1-\frac{1+z^2}{2(\sigma^2-1)}=:\eta$ and using the Maclaurin series of $\ln(1-\eta)$ we obtain 
\begin{equation}\label{lnEXPECTfONEfTWOasNtoINFINITYc}
\gFUNC_{z;\sigma}^{}(\nu_{z;\sigma}^{(\pm)}) -\gFUNC_{z;\sigma}^{}(\nu_{z;\sigma}^{(0)}) 
= {\textstyle\sum\limits_{n=2}^\infty} \frac1n \Big[\frac{2\sigma^2-3-z^2}{2(\sigma^2-1)}\Big]^n,
\end{equation}
which, for $\sigma^2>3/2$, is manifestly positive when $0\leq z^2<2\sigma^2-3$, and we conclude that in this case
$\max_\nu \gFUNC_{z;\sigma}^{}(\nu) = \gFUNC_{z;\sigma}^{}(\nu_{z;\sigma}^{(\pm)})$.

 Theorem \ref{thm3} is proved.
\hfill QED

 The proof of Theorem \ref{thm3} also supplies the deferred argument in the proof of Theorem \ref{thmCOMPLEXzSCALED}.
 Namely, as a consequence of the proof of Proposition \ref{upGbd} we have
\begin{coro}\label{limPTSofMEASURES} 
 For each $z\in\Rset$ and $j\in\Nset$, the sequence
$N\mapsto\rho_{z;\sigma}^{(j|N)}(x_1,...,x_j)$ converges  to either $\nu_{z;\sigma}^{(0)}(x_1)\cdots\nu_{z;\sigma}^{(0)}(x_j)$
or to $\frac12[\nu_{z;\sigma}^{(+)}(x_1)\cdots\nu_{z;\sigma}^{(+)}(x_j) +\nu_{z;\sigma}^{(-)}(x_1)\cdots\nu_{z;\sigma}^{(-)}(x_j)]$,
depending on whether $z^2\geq 2\sigma^2-3$ or $z^2 < 2\sigma^2-3$.
\end{coro}

\noindent
\textit{Proof of Corollary \ref{limPTSofMEASURES}}: 

By the proven tightness of the sequence $N\mapsto\rho_{z;\sigma}^{(j|N)}(x_1,...,x_j)$,
every subsequence of this sequence is also tight.
 Therefore, every subsequence of $N\mapsto\rho_{z;\sigma}^{(j|N)}(x_1,...,x_j)$ has a convergent subsequence, which converges 
to $\nu_{z;\sigma}^{(0)}(x_1)\cdots\nu_{z;\sigma}^{(0)}(x_j)$ if $z^2\geq 2\sigma^2-3$,
and to some $p\nu_{z;\sigma}^{(+)}(x_1)\cdots\nu_{z;\sigma}^{(+)}(x_j) + (1-p)\nu_{z;\sigma}^{(-)}(x_1)\cdots\nu_{z;\sigma}^{(-)}(x_j)$, with $p\in(0,1)$
independent of $j$, if $z^2 < 2\sigma^2-3$; the invariance of $\rho_{z;\sigma}^{(1|N)}(x_1)$ under reflection $x_1\to-x_1$ implies that $p=\frac12$.
  Therefore the sequence $N\mapsto\rho_{z;\sigma}^{(j|N)}(x_1,...,x_j)$ converges to either 
$\nu_{z;\sigma}^{(0)}(x_1)\cdots\nu_{z;\sigma}^{(0)}(x_j)$ or to 
$\frac12[\nu_{z;\sigma}^{(+)}(x_1)\cdots\nu_{z;\sigma}^{(+)}(x_j) +\nu_{z;\sigma}^{(-)}(x_1)\cdots\nu_{z;\sigma}^{(-)}(x_j)]$,
according as $z^2\geq 2\sigma^2-3$ or $z^2 < 2\sigma^2-3$. 
\hfill QED

 We illustrate Thm.~\ref{thm3} with
two figures show the $z$-dependence of $E_N^{}(z;\sigma)^{\frac1N}$ 
for one choice of $\sigma^2<3/2$ and one of $\sigma^2>3/2$.
 Note the different scales.

\vspace{-10pt}
\begin{figure}[H]
\centering
\includegraphics[scale=0.4]{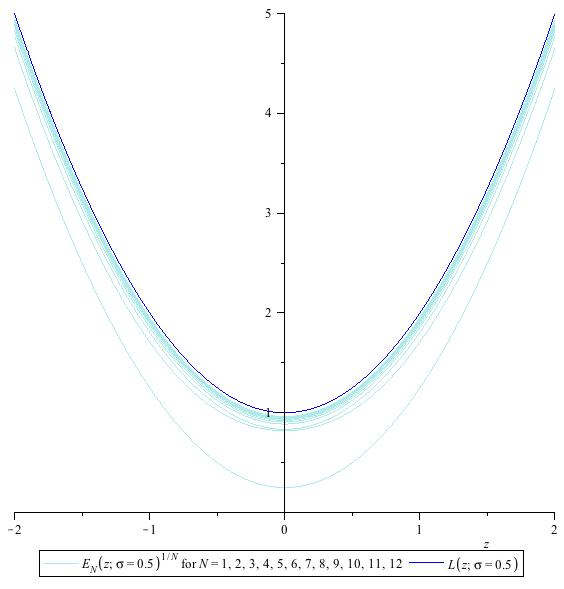}
\vspace{-5pt}
\caption{
\small Graphs of $z\mapsto E_N^{1/N}(z;\sigma=1/2)$ with $N\in\{1,...,12\}$ (in turquoise), together with the graph of
$z\mapsto L(z;\sigma=1/2)$ (in dark blue), for $z\in(-2,2)$.}\label{ENforREALzSigHALF}
\medskip
\centering
\includegraphics[scale=0.5]{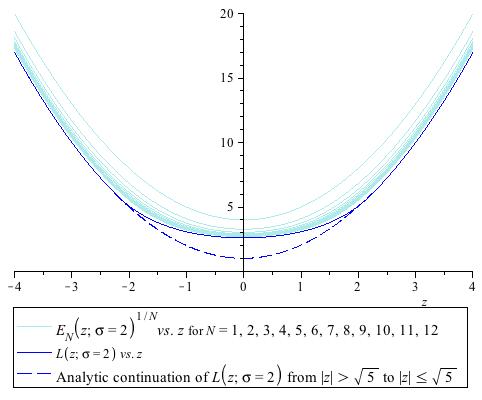}
\vspace{-10pt}
\caption{
\small Graphs of $z\mapsto E_N^{1/N}(z;\sigma=2)$ with $N\in\{1,...,12\}$ (continuous, in turquoise), together with the graph of
$z\mapsto L(z;\sigma=2)$ (continuous, in blue), for $z\in(-4,4)$.
 Also shown (in blue; dashed) is the analytic continuation of $L(z;2)$ from $|z|>\surd{5}$ to $|z|\leq\surd{5}$.}
\label{ENforREALzSigTWO}
\end{figure} 
\vspace{-10pt}

\newpage

 Quite remarkably, not more than the first dozen $N$ are needed in each of the two figures to nicely illustrate
the convergence of $E_N^{1/N}(z;\sigma)$ to $L(z;\sigma)$ when $z\in\Rset$.

 In addition to illustrating the proven convergence, the two figures also hint at some finer details 
which do not follow from our proofs, and while not unambiguously visible in the displayed figures, they are
discernable in their blowups using MAPLE.
 Namely, convergence appears to be monotone up in Fig.~\ref{ENforREALzSigHALF} and monotone down in Fig.~\ref{ENforREALzSigTWO}.
 This monotone ordering of $N\mapsto E_N^{1/N}(z;1/2)$ and of $N\mapsto E_N^{1/N}(z;2)$ 
goes beyond what is proved in (\ref{Glimits}), namely that the limiting $L$ curve is the 
supremum of the family of finite-$N$ curves for $\sigma^2=1/2$ in Fig.~\ref{ENforREALzSigHALF},
the infimum of the finite-$N$ curves for $\sigma^2=4$ in Fig.~\ref{ENforREALzSigTWO}. 

 Two-dimensional figures cannot show that the finite-$N$ curves converge to the curve of maximal 
value of $\exp(\gFUNC_{z;\sigma}(\nu))$ amongst all bounded normalized continuum densities $\nu$ with finite second moments. 
 Yet, in Fig.~\ref{ENforREALzSigTWO} a consequence of this maximum-entropy principle is illustrated by plotting in addition to $L(z;2)$ also 
the curve $z\mapsto \exp(\gFUNC_{z;2}(\nu_{z;2}^{(0)}))$ obtained by restricting the maximization of $\gFUNC_{z;2}(\nu)$ 
to reflection-symmetric densities $\nu$. 
 For $z^2>5$ this curve coincides with $L(z;2)$ (dark blue, continuous), while for $z^2<5$ this curve (dark blue, dashed) runs below 
$L(z;2)$ --- this shows that the critical points of $\gFUNC_{z;\sigma}(\nu)$ with broken reflection symmetry have 
higher relative entropy than the reflection-symmetric ones when $z^2<5$.

\begin{rema}
 The Gibbs variational principle 
allows us to give an equilibrium statistical mechanics re-interpretation of our multi-variate normal expected polynomials as the
``configurational canonical partition function'' of a one-dimensional physical (toy) model of $N$ point particles which have 
harmonic pair interactions, are confined in an external ``double well potential'' whose overall width is controlled by $\sigma^2$
and its central height by $\ln z^{-2}$, and which are in contact with a heat bath at a temperature $\propto N$.
 The harmonic pair interactions have coupling constant $\propto(1-\frac{1}{\sigma^2})$ which makes them attractive for $\sigma>1$ and
repulsive for $\sigma\in(0,1)$; the confining potential offers them two preferred locations to center on --- plus an energetically 
less preferential but still stationary location in the middle; lastly, due to the thermal motions the particles tend to spread out.
 For repulsive pair interactions the law of large numbers $N\to\infty$ yields only a unique, hence symmetric, ``thermodynamic'' phase 
independently of the height of the central peak of the confining double-well potential.
 For sufficiently attractive pair interactions (i.e. $\sigma^2>3/2$), condensation wins over spreading when the central peak of the
double-well potential is sufficiently large, viz. $z^2$ is sufficiently small, in which case the system chooses amongst two 
symmetrically located centers for condensation on the $x$-axis.
 This symmetry-breaking bifurcation is  a second-order phase transition. 
 This re-interpretation works only for $z\in\Rset$.
\end{rema}
\newpage

\section{Large-$N$ limit points of $E_N(z;\sigma)^{1/N}$ with $iz\in \Rset$}\label{ASYMPzIMAG}\vspace{-10pt}

 In the introduction we explained that a discussion of the large-$N$ asymptotics of (\ref{EXPECTmoreTHANoneX}) with $iz\in\Rset$ 
requires considering the even-$N$ and odd-$N$ subsequences separately. 
 We recall: First of all, $\{E_N^{}(z;\sigma)\}_{N\in\Nset}^{}$ --- which is well-defined $\forall\;z\in\Cset$ and is $\in\Rset$
if $z\in\Rset$ and if $iz\in\Rset$ --- may change
sign alternatingly with $N$ for certain subsets of $iz\in\Rset$, as it does when the $\{X_k\}_{k=1...N}$
are i.i.d. random variables, and so may not converge at all (rescaling being irrelevant here).
 Secondly, only for odd $N=2K-1$, $K\in\Nset$, is $E_{2K-1}^{}(z;\sigma)^{1/(2K-1)}$ a-priori well-defined if $iz\in\Rset$;
if a negative sign does occur for certain $iz\in\Rset$ when $N=2K$ 
then $\{E_{2K}^{}(z;\sigma)^{1/2K}\}_{K\in\Nset}^{}$ would a-priori not be defined for those $iz\in\Rset$
(although it could be defined by analytic continuation, involving families of Riemann surfaces)
---  we will show in subsection \ref{ASYMPzIMAGevenNpositivity} that this complication does not occur.
 In subsection \ref{ASYMPzIMAGevenNoddNnumerics} we will then  jointly present our numerical study of the even-$N$ and odd-$N$ 
subsequences of $\{E_N^{1/N}(z;\sigma)\}_{N\in\Nset}^{}$.\vspace{-10pt}

\subsection{$\{E_N(z;\sigma)^{1/N}\}_{N\in\Nset}^{}$ is well-defined when $iz\in \Rset$}\label{ASYMPzIMAGevenNpositivity}

  In this subsection we prove that $E_{2K}^{}(z;\sigma)\geq 0$ for $K\in\Nset$ and $iz\in\Rset$.
 This establishes that the even-$N$ subsequence of $\{E_N^{1/N}(z;\sigma)\}_{N\in\Nset}^{}$ is well-defined when $iz\in\Rset$; 
recall that the odd-$N$ subsequence is a-priori well-defined.

\begin{theo}\label{thmPOSevenN}
 Let $K\in\Nset$.
Then for $\sigma>0$ and $z^2\in\Rset$ we have $E_{2K}(z;\sigma)\geq 0$, with
$E_{2K}(z;\sigma)= 0$ iff $\sigma=1$ and $z^2=-1$.
\end{theo}

\noindent
\textit{Proof of Theorem \ref{thmPOSevenN}}:

 We begin by defining an abbreviation for the integrand of ${E_N^{}(z;\sigma)}$, viz.
\begin{equation}\label{maximizingMU}
\vareps^{(N)}_{z;\sigma}({\bf{x}}) : = 
\textstyle\frac{1}{\sigma}\frac{1}{\sqrt{2\pi}^N}
\; \prod\!\!\!\prod\limits_{\hspace{-14pt}1\leq k<l\leq N} e^{-\frac{1}{2N}(1-\frac{1}{\sigma^2}) (x_k-x_l)^2}
\!\!\!\!\prod\limits_{1\leq n\leq N}\!\!\!(x_n^2+z^2){e^{-\frac{1}{2\sigma^2} x_n^2}};
\end{equation}
if ${E_N^{}(z;\sigma)}\neq 0$, then $\dot\varsigma^{(N)}_{z;\sigma}:=\vareps^{(N)}_{z;\sigma}/{E_N^{}(z;\sigma)}$
extends our earlier definition of $\rho^{(N)}_{z;\sigma}$ from $z^2\geq 0$ to $z^2\in\Rset$.

 By (\ref{harmonicIDENTITYconcl}) we have 
\begin{equation}\label{harmonicIDENTITYconclSPLIT}
\textstyle(\sigma^{2}-1)\frac{1}{2K}\;
\sum\!\!\sum\limits_{\hspace{-18pt}1\leq k<l\leq 2K} (x_k-x_l)^2 +\!\!\sum\limits_{1\leq n \leq 2K}\!\!\!\! x_n^2 
=
\sigma^{2}\!\!\!\sum\limits_{1\leq k \leq 2K}\!\!\!\! x_k^2 - 
(\sigma^{2}-1)\Bigl({\textstyle{\frac{1}{\sqrt{2K}}}}\!\!\sum\limits_{1\leq k \leq 2K}\!\!\!\! x_k\Bigr)^2\!.
\end{equation}
 We now split the $N=2K$ variables into two disjoint sets of size $K$, keeping the notation $x_n^{}$ for $n=1,...,K$ and 
renaming $x_n^{}=:\tilde{x}_k^{}$ if $n=K+k$ with $k=1,...,K$. 
 Accordingly, at r.h.s.(\ref{harmonicIDENTITYconclSPLIT}) we rewrite
$\sum_{k=1}^{2K}f(x_k) = \sum_{k=1}^{K}\big(f(x_k)+f(\tilde{x}_k)\big)$, 
and also add and subtract
$(\sigma^{2}-1)\Big[\big({\textstyle{\frac{1}{\sqrt{K}}}}\sum_{k=1}^{K}\! x_k\big)^2\! +
\big({\textstyle{\frac{1}{\sqrt{K}}}}\sum_{k=1}^{K}\! \tilde{x}_k\big)^2\Big]$ 
to r.h.s.(\ref{harmonicIDENTITYconclSPLIT}). 
 With the help of (\ref{harmonicIDENTITYconcl}) we then find 
\begin{equation}\label{harmonicIDENTITYconclSPLITrejoin}\hspace{-14pt}
\begin{array}{lll}
\mbox{l.h.s.}(\ref{harmonicIDENTITYconclSPLIT}) = &
\textstyle(\sigma^{2}-1)\frac{1}{K}\;
\sum\!\!\sum\limits_{\hspace{-18pt}1\leq k<l\leq K} \Big[(x_k-x_l)^2 + (\tilde{x}_k-\tilde{x}_l)^2 \Big]
+\!\!\sum\limits_{1\leq n \leq K}\!\!\Big[ x_n^2 +\tilde{x}_n^2\Big] \\
&\!\!\! +
(\sigma^{2}-1)\Big[\big({\textstyle{\frac{1}{\sqrt{K}}}}\!\!\sum\limits_{1\leq k \leq K}\!\!\!\! x_k\big)^2\! +
\big({\textstyle{\frac{1}{\sqrt{K}}}}\!\!\sum\limits_{1\leq k \leq K}\!\!\!\! \tilde{x}_k\big)^2
- 
\Bigl({\textstyle{\frac{1}{\sqrt{2K}}}}\!\!\sum\limits_{1\leq k \leq K}\!\!\!\! (x_k+\tilde{x}_k)\Bigr)^2\Big].\!\!\!\!
\end{array}
\end{equation}
 The expression in the $2^{\mathrm{nd}}$ line at r.h.s.(\ref{harmonicIDENTITYconclSPLITrejoin}) simplifies to 
$(\sigma^2-1)\Bigl({\textstyle{\frac{1}{\sqrt{2K}}}}\!\!\sum\limits_{1\leq k \leq K}\!\!\!\! (x_k -\tilde{x}_k)\Bigr)^2$.
 Setting $(x_1^{},...,x_K^{})=:\vec{x}$ and $(\tilde{x}_1^{},...,\tilde{x}_K^{})=:\vec{\tilde{x}}$, and defining
${\vec{u}}:=\frac{1}{\sqrt{K}}(1,...,1)\in\Rset^K$ (the unit vector along the diagonal of the first $2^K$-ant), in total we obtain 
\begin{equation}\label{E2KasEKkernelEK}
E_{2K}^{}(z;\sigma)\! = 
\sigma
\displaystyle
\int_{\Rset^K}\!\! \int_{\Rset^K} 
e^{-\frac{1}{4}(1-\frac{1}{\sigma^2}) \left({\vec{u}}\cdot(\vec{x}-\vec{\tilde{x}})\right)^2}
\vareps^{(K)}_{z;\sigma}(\vec{x})
\vareps^{(K)}_{z;\sigma}(\vec{\tilde{x}})
d^{^K}\!\!x\,d^{^K}\!\!\tilde{x}.
\end{equation}

 Now, as long as $\sigma>1$ we have
\begin{equation}\label{kernelFOURIERtransf}
e^{-\frac{1}{4}(1-\frac{1}{\sigma^2}) \left({\vec{u}}\cdot(\vec{x}-\vec{\tilde{x}})\right)^2}
=
\displaystyle
\int_{\Rset}
e^{\pm i\xi {\vec{u}}\cdot(\vec{x}-\vec{\tilde{x}})}
\textstyle{\frac{\sigma e^{-\frac{\sigma^2}{\sigma^2-1}\xi^2}}{\sqrt{\pi (\sigma^{2}-1)}}}
d\xi;
\end{equation}
and so, upon inserting either version of (\ref{kernelFOURIERtransf}) into (\ref{E2KasEKkernelEK}) and 
carrying out the two $\Rset^K$ integrations first, and denoting Fourier transform by ($\widehat{\phantom{B}}$),
we obtain
\begin{equation}\label{EXPECTmoreTHANoneXYinFOURIER}
E_{2K}^{}(z;\sigma)\! = 
\displaystyle
\sigma^2\int_{\Rset}\!
\Big|\widehat{\vareps^{(K)}_{z;\sigma}}(\xi \vec{u})\Big|^2
\textstyle{\frac{e^{-\frac{\sigma^2}{\sigma^2-1}\xi^2}}{\sqrt{\pi (\sigma^{2}-1)}}}d\xi,
\end{equation}
which manifestly shows that $E_{2K}^{}(z;\sigma)\! > 0$ when $\sigma>1$. 

 Next, as long as $0<\sigma<1$ we have
\begin{equation}\label{kernelLAPLACEtransf}
e^{-\frac{1}{4}(1-\frac{1}{\sigma^2}) \left({\vec{u}}\cdot(\vec{x}-\vec{\tilde{x}})\right)^2}
=
\displaystyle
\int_{\Rset}
e^{\pm\xi {\vec{u}}\cdot(\vec{x}-\vec{\tilde{x}})}
\textstyle{\frac{\sigma e^{-\frac{\sigma^2}{1-\sigma^2}\xi^2}}{\sqrt{\pi (1-\sigma^{2})}}}d\xi;
\end{equation}
and so, upon inserting any one of the two possible versions of (\ref{kernelLAPLACEtransf}) into (\ref{E2KasEKkernelEK}) 
and carrying out the two $\Rset^K$ integrations first, denoting the 
(double-sided) Laplace transform by ($\widetilde{\phantom{B}}$), 
and noting that $\widetilde{\vareps^{(K)}_{z;\sigma}}(\xi \vec{u})=\widetilde{\vareps^{(K)}_{z;\sigma}}(-\xi \vec{u})$, we obtain
\begin{equation}\label{EXPECTmoreTHANoneXYinLAPLACE}
E_{2K}^{}(z;\sigma)\! = 
\displaystyle
\sigma^2\int_{\Rset}\!
\Big|\widetilde{\vareps^{(K)}_{z;\sigma}}(\xi \vec{u})\Big|^2
\textstyle{\frac{e^{-\frac{\sigma^2}{1-\sigma^2}\xi^2}}{\sqrt{\pi (1-\sigma^{2})}}}
d\xi,
\end{equation}
which manifestly shows that $E_{2K}^{}(z;\sigma)\! > 0$ when $0<\sigma<1$. 

 Finally, taking $\sigma\to 1$ in identity (\ref{EXPECTmoreTHANoneXYinFOURIER}) or (\ref{EXPECTmoreTHANoneXYinLAPLACE}) we obtain
$E_{2K}^{}(z;1)\! = E_{K}^{2}(z;1)$.
  This is nothing new for us, for we know that $E_{N}^{}(z;1)\! = (1+z^2)^N$, which
shows that $E_{N}^{}(\pm i;1)\! = 0$, while $E_{2K}^{}(z;1)\! = (1+z^2)^{2K}>0$ for $z^2\neq -1$.
\hfill QED

\begin{rema}
 In the appendix we will present an alternative, more elementary proof of the non-negativity of $E_{2K}^{}(z;\sigma)$ for
$z^2\in\Rset$ which, however,
is restricted to $\sigma^2\geq 1$.
\end{rema}

 Our Theorem \ref{thmPOSevenN} and our proof of Theorem \ref{thm3} also
prove the existence of \emph{limit points} of the sequence $N\mapsto |E_N^{}(z;\sigma)|^{1/N}$ when $iz\in\Rset$.
 Indeed, by Theorem \ref{thmPOSevenN} the sequence $N\mapsto E_N^{}(z;\sigma)^{1/N}$ is well-defined for $iz\in\Rset$,
and so we may pull the absolute value inside the integral and repeat
the sub/super-additivity estimates to get the existence of upper bounds uniformly in $N$ for any $\sigma>0$ --- the claim follows.

\subsection{The even- and odd-$N$ subsequences of $E_N^{1/N}(z;\sigma)$; $iz\in \Rset$}\label{NrootIMAGzSEQUENCEsub}\vspace{-5pt}

\subsubsection{Heuristic considerations about the subsequences}\label{NrootIMAGzSEQUENCEsubHEURISTICS}\vspace{-5pt}
 When $z^2<0$ the only $\sigma$ value for which something is rigorously known about 
the limit points of $E_N^{}(z;\sigma)^{1/N}$ when $N\to\infty$ is $\sigma=1$.
 In this case the multi-variate normal random variables are i.i.d. standard normal, and for the 
even-$N$ subsequence we have $E_{2K}^{}(z;\sigma)^{1/2K}\to |1+z^2|$ while 
we have $E_{2K-1}^{}(z;\sigma)^{1/2K-1}\to 1+z^2$ as $K\to\infty$  for the odd-$N$ sequence.
 The case $\sigma=1$ will play an important role as reference point for regime $\sigma\neq 1$ when $z^2<0$.

 Namely, it is impossible not to note that $1+z^2$ for $z^2<0$ is not only the analytical continuation to $iz\in\Rset$ of the real-$z$ limit
$E_{N}^{}(z;1)^{1/N}\to 1+z^2$ as $N\to\infty$ --- see (\ref{Glimits}) ---; when $z\in\Rset$ we also have
$E_{N}^{}(z;\sigma)^{1/N}\to 1+z^2$ as $N\to\infty$ for all $\sigma^2\leq 3/2$ --- recall Theorem \ref{thm3}.
 This suggests that there may be an open neighborhood of $\sigma=1$ such that the odd-$N$ subsequence of $E_N^{}(z;\sigma)^{1/N}$ with 
$iz\in\Rset$ converges to $1+z^2$, while the even-$N$ subsequence converges to the absolute value thereof. 
 
 More generally when $\sigma\neq 1$, the analytic continuation to $iz\in\Rset$ of the $z\in\Rset$ limit of $E_{N}^{}(z;\sigma)^{1/N}$, 
and its absolute value, would seem to be the right place to start looking for possible limit points of the sequence 
$E_{N}^{}(z;\sigma)^{1/N}$, $z=iy$, $y\in\Rset$.

 More precisely, we need to consider the analytic continuations, and absolute values, of the two competing real analytical functions that
feature in Theorem \ref{thm3}.
 For convenience we now set $z=iy$ and rename the r.h.s.s of (\ref{lnEXPECTfONEfTWOasNtoINFINITYa}) and (\ref{lnEXPECTfONEfTWOasNtoINFINITYb}) as follows,
\begin{eqnarray}\label{Lone}
L_1(y) &:=& {1-y^2},\\
\label{Ltwo}
L_2(y;\sigma) &:=&  2(\sigma^2 -1) \exp\left( \frac{1-y^2}{2(\sigma^2-1)} -1\right); \qquad \sigma\neq 1.
\end{eqnarray}
 Incidentally, the subscripts at $L$ may serve as a reminder that their right-hand sides are the analytical continuations from $z=x+i0$ to $z=0+iy$
of the limit functions originally computed, for $z\in\Rset$,  with the single-phase, respectively
double-phase solutions to the fixed point equation  (\ref{rhoEULERLAGRANGEd}) that we derived so far only for the regime $z^2\geq 0$.
 We recall that when $z\in\Rset$, the double-phase regime exists only when $\sigma^2>3/2$ and $z^2< 2\sigma^2-3$, which restricts the validity of
the $z\in\Rset$-counterpart of (\ref{Ltwo}) to this parameter regime even though the formulas for $L$ itself does not hint at such a restriction. 
 We should therefore expect that some similar restrictions apply to (\ref{Lone}) and (\ref{Ltwo}). 
 For now, though, in absence of any theoretical knowledge of such restrictions when $iz\in\Rset$, we will
operate at a purely heuristic level and allow (\ref{Lone}) and (\ref{Ltwo}) without a-priori
restriction on the $\sigma$ or $y$ values.

\begin{figure}[H]
\centering
\includegraphics[scale=0.44]{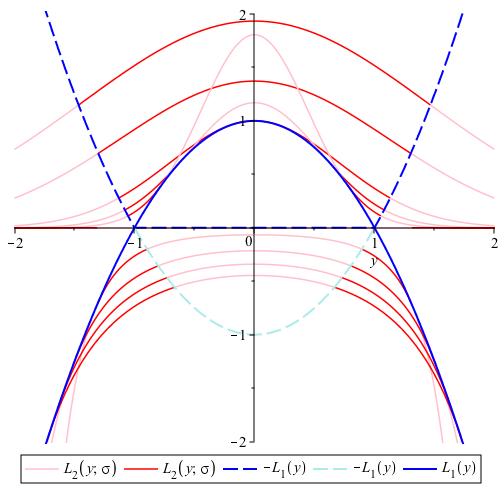}
\caption{
\small For $z=iy$ with $y\in (-2,2)$ the figure shows graphs of the functions $L_1(y)$ (in dark blue, continuous) 
and $-L_1(y)$ (in light and dark blue, dashed), as well as graphs of the functions $L_2(y;\sigma)$ (in dark and light red, continuous) 
for variances $\sigma\in\{0,0.4,0.6,0.8,1.095,1.14,\sqrt{3/2},1.5,1.75\}$ (ordered bottom-up at $y=1$).
 The $L_2$ curves for $\sigma<1$ are negative, those for $\sigma>1$ are positive; the $L_1$ curve ($\sigma=1$) 
obviously takes positive as well as negative values, and so does the dashed $-L_1$ curve.
 The light red portions of $L_2(y;\sigma)$ do not seem to appear as $N\to\infty$ limit points of $E_{N}^{}(z;\sigma)^{1/N}$, $z=iy$, $y\in\Rset$;
only the dark red portions of $L_2(y;\sigma)$ do.
 Also the dark portions of the (continuous or dashed) blue curves appear as limit point curves, 
while the light blue dashed curve never seems to appear.
 The absolute value of the negative dark colored continuous curves also captures limit point curves --- they are 
not shown, in order not to overload the picture.}
\label{NrootENforIMAGzL1L2curves}
\end{figure}

 In the next subsection we will graphically compare the evaluation of $E_{N}^{}(z;\sigma)^{1/N}$, $z=iy$, $y\in\Rset$, for $N=1,...,12$ 
for a selection of $\sigma$-values with the family of curves (\ref{Ltwo}), and with (\ref{Lone}) and their absolute values.
 A representative selection from the family of curves (\ref{Ltwo}) is shown in Fig.~\ref{NrootENforIMAGzL1L2curves}, 
together with $\pm$(\ref{Lone}).
 \vspace{-5pt}
 We observe that for $\sigma^2\geq 3/2$ the curves $y\mapsto L_2(y;\sigma)$ form a $\sigma^2$-ordered family above the curve $y\mapsto L_1(y)$.
 For $\sigma^2< 3/2$ each of the curves $y\mapsto L_2(y;\sigma)$ touches the curve $y\mapsto L_1(y)$ at two symmetric locations, namely
when $y^2 =3-2\sigma^2$; thus, the smaller $\sigma^2$, the further out these two points of contact are located.

 Beside these points of contact, the two symmetric points of intersection $\pm y_*$ of $L_2(y;\sigma)$ with $-L_1(y)$ will play an 
important role.
 They are computed as follows.
  Define $y^2_*(\sigma^2) := 1+2\kappa_*^{} (\sigma^2-1) $, where
$\kappa_*$ is the unique solution of the fixed point equation $\kappa = e^{-\kappa-1}$; numerically, $\kappa_*^{}=0.27846454276...$. 
 We remark that $\sigma^2\mapsto y_*^2$ is monotonic increasing, with $y_*^2 = 1-2\kappa_*^{}>0$ for $\sigma^2=0$,
with $1-2\kappa_*^{}< y_*^2<1$ for $0< \sigma^2< 1$, with $y_*^2=1$ exactly when $\sigma^2= 1$, and with $y_*^2>1$ for $\sigma^2> 1$.

 Note also that the one-sided limits $\lim_{\sigma\to 1^\pm} L_2(y;\sigma) \neq L_1(y)$.
 In fact, the curve $y\mapsto\max\{L_1(y),0\}=\inf\{L_2(y;\sigma)|\sigma>1\}$ 
is the lower extremal curve of the family of curves $\{y\mapsto L_2(y;\sigma)|\sigma >1\}$,
while $y\mapsto\min\{L_1(y),0\}=\sup\{L_2(y;\sigma)|0\leq\sigma<1\}$ 
is the upper extremal curve of $\{y\mapsto L_2(y;\sigma)|0\leq \sigma <1\}$.

\subsubsection{Computer-Generated Graphical Evidence for Conjecture \ref{conjIMz}}\label{ASYMPzIMAGevenNoddNnumerics} 

 We have evaluated $E_N^{}(iy;\sigma)$ algebraically for $N$ up to a dozen using MAPLE, and graphed its $N$th root versus $y$ for 
a selection of $\sigma$ values, respectively vs. $\sigma$ for various values of $y$; see below.
 A comparison of these curves with the curves in Fig.~\ref{NrootENforIMAGzL1L2curves}, and their absolute value curves, is reported in 
the ensuing subsections.

 Of course, since the case $\sigma=1$ is elementary, a MAPLE evaluation of $E_N^{}(iy;1)$ would be pointless and simply reproduce the
continuous black curve in Fig.~\ref{NrootENforIMAGzL1L2curves} for the odd-$N$ subsequence, and (depending on $y$) the larger of the 
continuous and the broken black curves in Fig.~\ref{NrootENforIMAGzL1L2curves} for the even-$N$ subsequence.
 We therefore consider only one example each from the three remaining parameter regions: $\sigma^2\geq 3/2$, $3/2>\sigma^2>1$, and 
$1>\sigma^2\geq 0$.
 All three figures unequivocally support our Conjecture \ref{conjIMz}.

\smallskip\noindent{\bf{ 6.2.2a. Graphical evidence for $\sigma^2 \geq 3/2$}}\smallskip

 The case $\sigma^2=3/2$ is representative for $\sigma^2\geq 3/2$.
 The next Figure (Fig.~\ref{NrootENzIMAGsigSQR3half}) shows the odd- and even-$N$ subsequences of $E_N^{}(iy;\sqrt{3/2})^{1/N}$ together 
with portions of the curves $L_2(y;\sqrt{3/2})$ and, respectively, $L_1(y)$ or $-L_1(y)$.

 Fig.~\ref{NrootENzIMAGsigSQR3half} supports the our Conjecture \ref{conjIMz} for $\sigma^2\geq 3/2$.
 It suggests that for $|y| < \sqrt{1+\kappa_*^{}}$ the full sequence $N\mapsto E_N^{1/N}(iy; \sqrt{3/2})$ converges to
$L_2(y;\sqrt{3/2})$, while for $|y| > \sqrt{1+\kappa_*^{}}$ the odd-$N$ and even-$N$ subsequences of $N\mapsto E_N^{1/N}(iy; \sqrt{3/2})$ 
converge separately to $L_1(y)$ and $|L_1(y)|=-L_1(y)$, respectively.

\begin{figure}[H]
\centering
\includegraphics[scale=0.5]{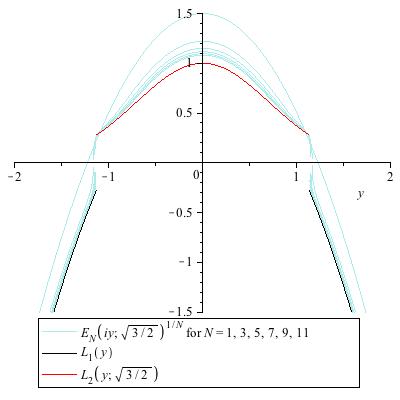}
\includegraphics[scale=0.5]{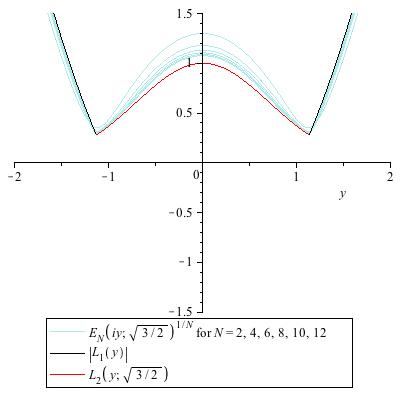}
\vspace{-15pt}
\caption{
\small Graphs (in turquoise) of $y\mapsto E_N^{1/N}(iy; \sqrt{3/2})$ for $N\in\{1,3,5,7,9,11\}$ (left panel) and $N\in\{2,4,6,8,10,12\}$ (right panel)
together with the graphs (black) of $y\mapsto \pm L_1(y)$ (respectively) for $|y|>\sqrt{1+\kappa_*^{}}$, and of 
$y\mapsto L_2(y;\sqrt{3/2})$ (in red) for $|y|<\sqrt{1+\kappa_*^{}}$.}
\vspace{-10pt}
\label{NrootENzIMAGsigSQR3half}
\end{figure} 

\smallskip\noindent{\bf{ 6.2.2b. Graphical evidence for $3/2 > \sigma^2 > 1$}}\smallskip

 The case $\sigma=1.15$ is representative for the regime $3/2 > \sigma^2 > 1$ when $iz\in\Rset$.
\begin{figure}[H]
\centering
\includegraphics[scale=0.5]{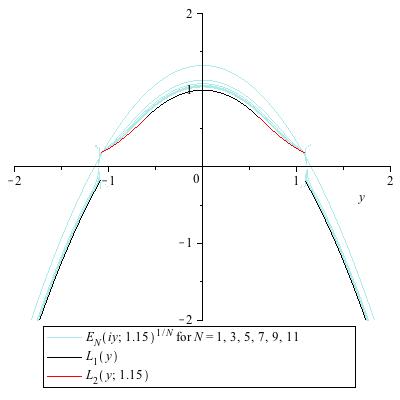}
\includegraphics[scale=0.5]{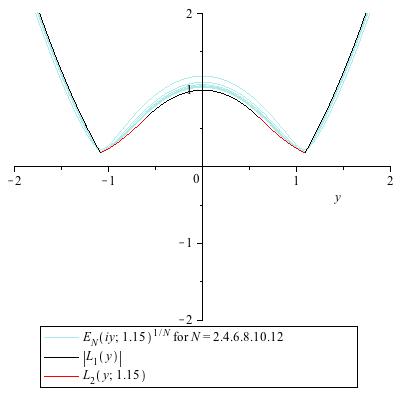}
\vspace{-15pt}
\caption{
\small Graphs (in turquoise) of $y\mapsto E_N^{1/N}(iy; 1.15)$ for $N\in\{1,3,5,7,9,11\}$ (left panel) and $N\in\{2,4,6,8,10,12\}$ (right panel)
together with the graphs (black) of $y\mapsto L_1(y)$ (odd-$N$), respectively $y\mapsto |L_1(y)|$ (even-$N$), 
for $|y| < \sqrt{3-2\cdot1.15^2}$ and for $|y|> \sqrt{1+2\kappa_*^{}(1.15^2-1)}$, plus the map 
$y\mapsto L_2(y;1.15)$ (in red) for $\sqrt{3-2\cdot1.15^2}<|y|<\sqrt{1+2\kappa_*^{}(1.15^2-1)}$.
 The tiny spikes at $|y| =\sqrt{1+2\kappa_*^{}(1.15^2-1)}$ are numerical artefacts.}
\vspace{-10pt}
\label{NrootENzIMAGsigONEptONEfive}
\end{figure} 

 Figure~\ref{NrootENzIMAGsigONEptONEfive} shows the odd-$N$ and the even-$N$ subsequences of $E_N^{}(iy;1.15)^{1/N}$ together with 
portions of the curves $L_2(y;1.15)$ and, respectively, $L_1(y)$ or $-L_1(y)$.

 Fig.~\ref{NrootENzIMAGsigONEptONEfive} supports our Conjecture \ref{conjIMz} for $1< \sigma^2 < 3/2$.
 It suggests that for $|y| < \sqrt{1+2\kappa_*^{}(1.15^2-1)}$ 
the full sequence $N\mapsto E_N^{1/N}(iy; 1.15)$ converges:
to $L_1(y)$ for $|y|\!<\! \sqrt{3-2\cdot1.15^2}$ and to $L_2(y;1.15)$
for  $\sqrt{3-2\cdot1.15^2} <|y|\!<\! \sqrt{1+2\kappa_*^{}(1.15^2-1)}$, 
while for $|y| > \sqrt{1+2\kappa_*^{}(1.15^2-1)}$ the odd- and even-$N$ subsequences of $N\mapsto E_N^{1/N}(iy; 1.15)$ 
converge separately to $L_1(y)$ and $|L_1(y)|=-L_1(y)$, respectively.

\smallskip\noindent{\bf{ 6.2.2c. Graphical evidence for $1 > \sigma^2 \geq 0$}}\smallskip

 The case $\sigma^2=1/2$ is representative for the parameter regime $1 > \sigma^2 \geq 0$ when $iz\in\Rset$.
 Figure~9 shows the odd- and even-$N$ subsequences of $E_N^{}(iy;1/\surd{2})^{1/N}$ and portions of the curves
$L_2(y;1/\surd{2})$ and $L_1(y)$, respectively $-L_2(y;1/\surd{2})$ and $-L_1(y)$.

\begin{figure}[H]
\centering
\includegraphics[scale=0.5]{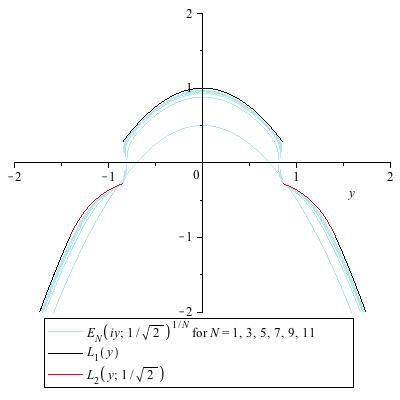}
\includegraphics[scale=0.5]{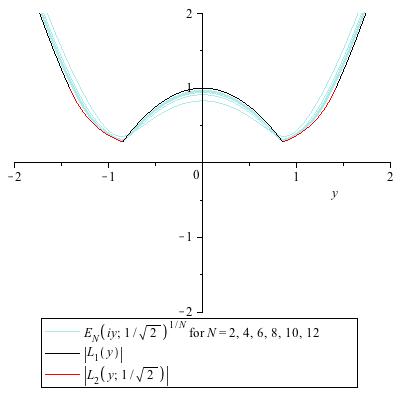}
\vspace{-15pt}
\caption{
\small Graphs (in turquoise) of $y\mapsto E_N^{1/N}(iy; \sqrt{1/2})$ for $N\in\{1,3,5,7,9,11\}$ (left panel) and $N\in\{2,4,6,8,10,12\}$ (right panel)
together with the graphs (black) of $y\mapsto L_1(y)$ for $0<y<\sqrt{1-\kappa_*^{}}$ and $y>\sqrt{2}$ (left panel), respectively
$y\mapsto L_1(y)$ for $0<y<\sqrt{1-\kappa_*^{}}$ and $y\mapsto -L_1(y)$ for $y>\sqrt{2}$ (right panel), as well as (in red)
$y\mapsto \pm L_2(y;\sqrt{1/2})$ ($+$: odd-$N$; $-$: even-$N$) for $\sqrt{1-\kappa_*^{}}<y<\sqrt{2}$.}
\vspace{-10pt}
\label{NrootENzIMAGsigSQR1half}
\end{figure} 

\subsection{The ramifications of the fixed points at $z=\pm i\surd{2}$, $\sigma=1$}\label{notSURE}

 When $z=\pm i\surd{2}$ then $E_N^{}(\pm i\surd{2};1) = (-1)^N$.
 Thus, the even-$N$ subsequences of $\{\sigma\mapsto E_N^{}(\pm i\surd{2};\sigma)\}_{N\in\Nset}^{}$ 
and of $\{\sigma\mapsto E_N^{1/N}(\pm i\surd{2};\sigma)\}_{N\in\Nset}^{}$ both have the fixed point $1$, 
and the odd-$N$ subsequences of $\{\sigma\mapsto E_N^{}(\pm i\surd{2};\sigma)\}_{N\in\Nset}^{}$ 
and of $\{\sigma\mapsto E_N^{1/N}(\pm i\surd{2};\sigma)\}_{N\in\Nset}^{}$ both have the fixed point $-1$.
 
 The upshot is that there might exist an open neighborhood of $\sigma=1$ in which not only the even-$N$ and odd-$N$
subsequences of $\{\sigma\mapsto E_N^{1/N}(\pm i\surd{2};\sigma)\}_{N\in\Nset}^{}$ converge (for which we have presented
empirical evidence in the previous three figures) but also odd-$N$ and even-$N$ subsequences of
$\{\sigma\mapsto E_N^{}(\pm i\surd{2};\sigma)\}_{N\in\Nset}^{}$.
 Not both of these can converge to nontrivial limits.
 Since our previous three figures suggest that for some $\sigma\neq 1$ the limit points of
the subsequences of $N\mapsto E_N^{1/N}(iy;\sigma)$ are indistinguishable from those of $N\mapsto E_N^{1/N}(iy;1)$, it is
clear that the limit points of
$\{\sigma\mapsto E_N^{1/N}(\pm i\surd{2};\sigma)\}_{N\in\Nset}^{}$ will be $\pm 1$ in some $\sigma$-neighborhood of $\sigma=1$.
 However, the subsequences of $N\mapsto E_N^{}(i\surd{2};\sigma)$ may converge to nontrivial limits there.

\begin{figure}[H]
\centering
\includegraphics[scale=0.5]{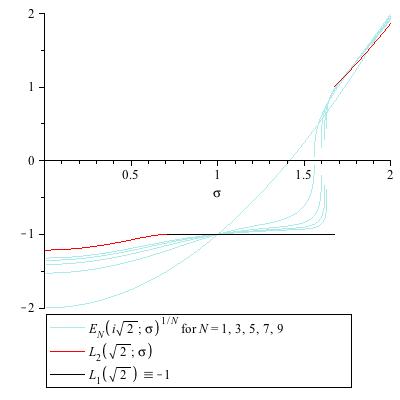}
\includegraphics[scale=0.5]{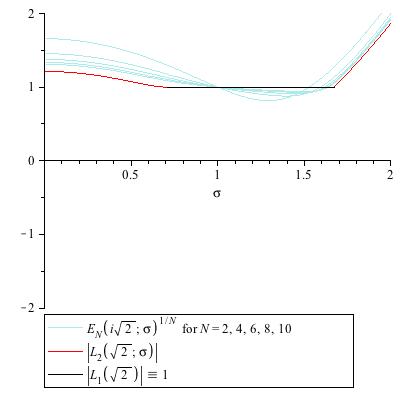}
\caption{
\small Graphs (in turquoise) of $\sigma\mapsto (E_N^{}(z=i\surd{2};\sigma))^{1/N}$ for $N\in\{1,3,5,7,9\}$ (left panel)
and for $N\in\{2,4,6,8,10\}$ (right panel), together with the graphs (red) of $\sigma\mapsto L_2(\surd{2};\sigma)$ (left panel) and
$\sigma\mapsto |L_2(\surd{2};\sigma)|$ (right panel) --- both for $\sigma\not\in (\sqrt{1/2},\sqrt{1+1/2\kappa_*^{}})$ ---, 
plus the graphs (black) of $\sigma\mapsto L_1(\surd{2})\equiv-1$ (left panel) and
$\sigma\mapsto |L_1(\surd{2})|\equiv 1$ (right panel), both for $\sigma\in (\sqrt{1/2},\sqrt{1+1/2\kappa_*^{}})$.}
\label{NrootENzIMAGsigPLOT} 
\end{figure}

\begin{figure}[H]
\centering
\includegraphics[scale=0.5]{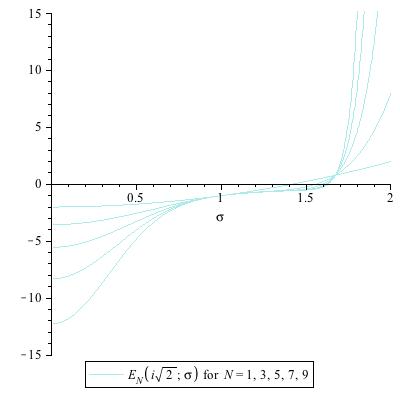} 
\includegraphics[scale=0.5]{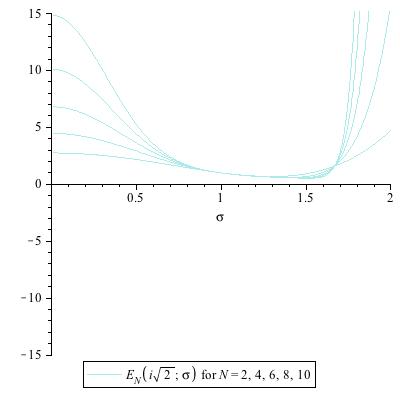}
\caption{
\small Graphs (in turquoise) of $\sigma\mapsto E_N^{}(i\surd{2};\sigma)$ for $N\in\{1,3,5,7,9\}$ (left panel) and 
for $N\in\{2,4,6,8,10\}$ (right panel) with $0<\sigma<2$.}
\label{ENzIMAGsigPLOT}
\end{figure}

The next two figures show, separately for odd $N$ and even $N\leq 10$, 
first: $\sigma\mapsto E_N^{1/N}(i\surd{2};\sigma)$, and second: $\sigma\mapsto E_N^{}(i\surd{2};\sigma)$.
 To facilitate the comparison of Fig.~\ref{NrootENzIMAGsigPLOT} with Fig.~\ref{ENzIMAGsigPLOT}
we note that in Fig.~\ref{NrootENzIMAGsigPLOT} the $N$-ordering of the curves at $\sigma=0$ is bottom-up in the
left, and top-down in the right panel, while in Fig.~\ref{ENzIMAGsigPLOT} it is reversed. 

 Fig.~\ref{NrootENzIMAGsigPLOT} shows the open neighborhood of $\sigma=1$ where
$\limpt E_N^{1/N}(\pm i\surd{2};\sigma) = \{\pm 1\}$, which is the interval $(\sqrt{1/2},\sqrt{1+1/2\kappa_*^{}})$, plus
some left and right neighborhoods of this interval. 
 Flat parts of the putative limiting curves indicate that the scaling is not sensitive enough to resolve the finer details.
 Indeed, the putative convergence of the odd-$N$ and even-$N$ subsequences of the different scaling sequence
$\{\sigma\mapsto E_N^{}(\pm i\surd{2};\sigma)\}_{N\in\Nset}^{}$ for $\sigma\in (\sqrt{1/2},\sqrt{1+1/2\kappa_*^{}})$ 
is clearly discernible in Fig.~\ref{ENzIMAGsigPLOT}.
 Note that in this scaling, for $\sigma\not\in (\sqrt{1/2},\sqrt{1+1/2\kappa_*^{}})$ the 
odd-$N$ and even-$N$ subsequences of $\{\sigma\mapsto E_N^{}(\pm i\surd{2};\sigma)\}_{N\in\Nset}^{}$ must diverge to $\infty$ in magnitude ---
if indeed the odd-$N$ and even-$N$ subsequences of $\{\sigma\mapsto E_N^{1/N}(\pm i\surd{2};\sigma)\}_{N\in\Nset}^{}$ converge 
to $L_2(\sqrt{2},\sigma)$, respectively to its magnitude, when $N\to\infty$, as featured in Fig.~\ref{NrootENzIMAGsigPLOT}.

\begin{rema}\label{rema:PSEUDOfixPT}
 Interestingly, roughly at $\sigma\approx 1.68$ all the $E_N^{}(i\surd{2};\sigma)$
curves for odd $N$, and those for even $N$, appear to intersect at a single point, respectively; alas, this 
appearance is due to the limited numerical resolution of the pictures. 
 A blow-up shows that these curves do not all intersect at the same single point (for odd, respectively even $N$).
 However, to explain such a near miss, it is reasonable to suspect that for some {\sc{nearby}} imaginary $z$-value (say, $iy_\bullet$) there is a 
special symmetry which forces all even-$N$, respectively odd-$N$, curves to intersect at (their own) single point --- the near
misses for nearby $iy$ values then follow by continuity of the parameter dependence.
 By the same token, for such a putative $iy_\bullet$ in the neighborhood of $\pm i\surd{2}$ the numbers $\pm1$ will no longer be fixed
points of the even-$N$ or odd-$N$ subsequences of $E_N^{}(iy;\sigma)$, but by the continuity of the parameter dependence the
nearby values $1-y_\bullet^2$ and their negatives could have the (misleading) appearance of fixed points.
\end{rema}

\newpage

\section{\hspace{-10pt} Relative Entropy with Signed A-Priori Measures}\label{RelEntSignMeas}
 Since all the putative limiting curves for the even-$N$ and odd-$N$ subsequences of $N\mapsto E_N^{1/N}(iy;\sigma)$ 
shown in section \ref{ASYMPzIMAGevenNoddNnumerics} are piecewise real analytic functions ``patched together'' 
from selected portions of the analytic continuations of the limits of $N\mapsto E_N^{1/N}(z;\sigma)$ for real $z$, and since these real-$z$
limit functions in turn were obtained by solving the Euler--Lagrange equations of the maximum relative entropy principles 
for probability densities, it is very suggestive to look at the analytical extension of these Euler--Lagrange equations and 
try to formulate a relative entropy principle from which they are obtained. 
 We will introduce such ``\emph{entropy principles relative to a signed a-priori measure}'' below.
 We distinguish several variations on this theme.
 We will see that \emph{complex measures} are needed to get all the empirical results of section \ref{ASYMPzIMAG} --- 
note though that the a-priori measure is always a signed measure.

 Recall that $\vareps^{(N)}_{z;\sigma} = \frac{1}{\sigma}(\frac{1}{\sqrt{2\pi}})^N
\prod\!\!\!\prod\limits_{\hspace{-14pt}1\leq k<l\leq N} e^{-\frac{1}{2N}(1-\frac{1}{\sigma^2}) (x_k-x_l)^2}
\!\!\!\!\prod\limits_{1\leq n\leq N}\!\!\!(x_n^2+z^2){e^{-\frac{1}{2\sigma^2} x_n^2}}$
is the integrand of $E_N^{}(z;\sigma)$, $N>1$, which for negative $z^2$ is the density of a \emph{signed (a-priori) measure}.
 Note that $\vareps^{(N)}_{z;\sigma}$ is not normalized. 
 For $z=iy$, $y\in\Rset$, we state
\begin{defi}\label{DEFsignRELentro}
 For $N>1$ we define what we call a \emph{signed relative entropy} as
\begin{equation}\label{muFUNCTIONALsigned}
{\cal H}^{(N)}_{y;\sigma}(\dot\varsigma^{(N)}) := 
-\int_{\Rset^N} \dot\varsigma^{(N)} \ln \frac{\dot\varsigma^{(N)}}{\vareps_{iy;\sigma}^{(N)}} d^Nx ,
\end{equation}
where the functional is defined on the \emph{admissible set} of absolutely continuous 
(w.r.t. Lebesgue measure) densities $\dot\varsigma^{(N)}$ of permutation-symmetric $N$-point \emph{signed measures} $\varsigma^{(N)}$ which 
are normalized up to sign (viz.: integrate to either $+1$ or $-1$), with Radon--Nikodym derivative 
${\dot\varsigma^{(N)}}/{\vareps_{iy;\sigma}^{(N)}} > 0$, and with
$|{\cal H}^{(N)}_{y;\sigma}(\dot\varsigma^{(N)})|<\infty$ and ${\cal H}^{(N)}_{y;\sigma}(\dot\varsigma^{(N)})\in\Rset$.
\end{defi}

 As for the conventional relative entropy functional of probability measures, we call $\dot\varsigma^{(N)}_*$ a \emph{critical point} of
${\cal H}^{(N)}_{y;\sigma}(\dot\varsigma^{(N)})$ if all its Gateaux derivatives\footnote{Here: directional derivatives in the
direction of any $\psi=\psi_++\psi_-$, with $\psi_\pm\in C^\infty$ compactly supported inside the support of the positive/negative part
of the a-priori measure, respectively, and with $\psi$ integrating to zero to preserve the normalization of $\dot\varsigma^{(N)}$.}
at $\dot\varsigma^{(N)}_*$ vanish.
 If $\dot\varsigma^{(N)}_*$ is a critical point of ${\cal H}^{(N)}_{y;\sigma}(\dot\varsigma^{(N)})$, we call 
${\cal H}^{(N)}_{y;\sigma}(\dot\varsigma^{(N)}_*)$ a \emph{critical value} of ${\cal H}^{(N)}_{y;\sigma}(\dot\varsigma^{(N)})$, 
and the set of all critical values is 
written ${\mathrm{crit}}_{\varsigma^{(N)}} {\cal H}^{(N)}_{y;\sigma}(\dot\varsigma^{(N)})$.

 It is easy to see that 
there is a unique critical point of ${\cal H}^{(N)}_{y;\sigma}(\dot\varsigma^{(N)})$, 
given by $\dot\varsigma^{(N)}_{y;\sigma}= \vareps^{(N)}_{iy;\sigma} /|E_N^{}(iy;\sigma)|$, 
which is the \emph{normalized} (up to sign) density of our signed a-priori measure, 
$\vareps^{(N)}_{z;\sigma}$, manifestly integrating to one of $\pm1$.
 By direct computation, ${\cal H}_{y;\sigma}^{(N)}(\dot\varsigma^{(N)}_{y;\sigma}) = \sign(E_N^{}(iy;\sigma))\ln |E_N^{}(iy;\sigma)|$.
 Thus we have an analogue for $\ln |E_N^{}(iy;\sigma)|$ of Gibbs' finite-$N$ variational principle, i.e.
\begin{equation}\label{muFUNCTIONALprinciple}
\sign(E_N^{}(iy;\sigma))
 \ln |E_N^{}(iy;\sigma)|
\in {\mathrm{crit}}_{\varsigma^{(N)}} {\cal H}^{(N)}_{y;\sigma}(\dot\varsigma^{(N)});
\end{equation}
note that ${\mathrm{crit}}_{\varsigma^{(N)}} {\cal H}^{(N)}_{y;\sigma}(\dot\varsigma^{(N)})$ here contains a single value.

\begin{rema}
 Note that Gibbs' variational principle (\ref{muFUNCTIONALmaximized}) for the finite-$N$ canonical ensemble probability measures 
(i.e. when $z=x+i0$ with $x\in\Rset$) obviously implies
\begin{equation}\label{muFUNCTIONALcriticalPT}
\ln E_N^{}(x;\sigma) \in \mathrm{crit}_{\rho^{(N)}} {\cal G}_{x;\sigma}^{(N)}(\rho^{(N)}),
\end{equation}
but (\ref{muFUNCTIONALcriticalPT}) is actually equivalent to (\ref{muFUNCTIONALmaximized}) because of the concavity of the map
$\rho^{(N)}\mapsto {\cal G}_{x;\sigma}^{(N)}(\rho^{(N)})$.
  The signed relative entropy functional does not inherit concavity from Gibbs' relative entropy functional.
  Its critical points are saddle points in the set of admissible signed measures.
 This is easily seen by considering variations restricted to either the positive or negative supports of the
a-priori signed measure.
\end{rema}

\begin{rema}
 By Theorem \ref{thmPOSevenN} we have $E_N^{}(iy;\sigma)\!>\!0$ for  even $N$ as long as $\sigma\neq1$.
 As found empirically in section \ref{ASYMPzIMAG}, in some regions in $(\sigma,y)$ space $E_N^{}(iy;\sigma)>0$ also for odd $N$.
 In all these cases $\sign(E_N^{}(iy;\sigma))=+1$ and we can drop the absolute value bars at l.h.s.(\ref{muFUNCTIONALprinciple})
to get the perfect analog of (\ref{muFUNCTIONALcriticalPT}), viz.
\begin{equation}\label{sigmaFUNCTIONALcriticalPT}
\ln E_N^{}(iy;\sigma)\in {\mathrm{crit}}_{\varsigma^{(N)}} {\cal H}^{(N)}_{y;\sigma}(\dot\varsigma^{(N)}).
\end{equation}
\end{rema}

 We can obtain the more agreeable (\ref{sigmaFUNCTIONALcriticalPT}) also for the regions in $(\sigma,y)$ space in which 
$E_N^{}(iy;\sigma)<0$ for odd $N$, provided we allow the relative entropy to be complex, thus
\begin{defi}\label{DEFcomplexRELentroSIGNED}
 For $N>1$ we define what we call a \emph{complex entropy relative to an a-priori signed measure} as given by (\ref{muFUNCTIONALsigned}),
but with ${\dot\varsigma^{(N)}}/{\vareps_{iy;\sigma}^{(N)}} > 0$ replaced with ${\dot\varsigma^{(N)}}/{\vareps_{iy;\sigma}^{(N)}} \neq 0$,
with $\dot\varsigma^{(N)}$ normalized to $+1$, and with ${\cal H}^{(N)}_{y;\sigma}(\dot\varsigma^{(N)})\in\Cset$.
\end{defi}

\begin{rema}\label{ipi}
 The critical point of ${\cal H}^{(N)}_{y;\sigma}(\dot\varsigma^{(N)})$ as defined in Definition \ref{DEFcomplexRELentroSIGNED}
is then given by the density 
$\dot\varsigma^{(N)}_{y;\sigma}= \vareps^{(N)}_{iy;\sigma} /E_N^{}(iy;\sigma)$ of a real signed measure, yet
when $E_N^{}(iy;\sigma)<0$ its logarithm is simply one of the complex numbers 
$\ln |E_N^{}(iy;\sigma)| + i (2l-1)\pi$, $l\in\Zset$.
 The particular value of $l$ can be left undetermined as long as
we are interested only in $E_N^{}(iy;\sigma)$ and not in its natural logarithm.

 To properly define $\ln E_N^{}(iy;\sigma)$ one has to admit suitable densities $\dot\varsigma^{(N)}$ of normalized complex measures,
and invoke an analytical continuation analysis. 
 We don't need this for our present purposes.
 Also an extension to a-priori complex measures is possible, but will not be pursued in this paper. 
 However, normalized complex 1-point measures will feature in the asymptotic evaluation of $E_N^{1/N}(iy;\sigma)$.
\end{rema}

 Guided by the proof of Theorem \ref{thm3}
we next look for critical points of ${\cal H}^{(N)}_{y;\sigma}(\dot\varsigma^{(N)})$ on a restricted set of admissible (see above)
signed $N$-point densities $\dot\varsigma^{(N)}$ formed by symmetric products of signed one-point densities, 
$\dot\varsigma^{(N)}(x_1,...,x_N) = \dot\varsigma(x_1)\cdots\dot\varsigma(x_N)$ ---
this is possible because the sign-changing part 
$\prod_{1\leq n\leq N}(x_n^2-y^2)$ of $\vareps^{(N)}_{iy;\sigma}$ is itself a symmetric product of signed one-point functions.
 Analogously to (\ref{muFUNCTIONALestimatedBYrhoFUNCTIONAL}) we have
\begin{equation}\label{muFUNCTIONALrestrictedTOrhoPRODUCT}
\hspace{-10pt}
\frac1N {\cal H}^{(N)}_{y;\sigma}(\dot\varsigma^{\otimes N}) 
= \hFUNC_{y;\sigma}^{}(\dot\varsigma) - {\textstyle\frac1N}\!
\left(\ln \sigma -\tfrac{\sigma^2-1}{4\sigma^2}\int_{\Rset} \int_{\Rset} \dot\varsigma (x_1)\dot\varsigma(x_2) (x_1 -x_2)^2 dx_1 dx_2\right)
\end{equation}
with 
\begin{equation}\label{rhoFUNCTIONALsigned}
\hspace{-8pt} \hFUNC_{y;\sigma}(\dot\varsigma) = 
-\int_{\Rset} \dot\varsigma(x_1) \ln \frac{\dot\varsigma(x_1)}{\nu_{iy;\sigma}^{}(x_1)} dx_1  
- \tfrac{\sigma^2-1}{4\sigma^2}\int_{\Rset} \int_{\Rset} \dot\varsigma (x_1)\dot\varsigma(x_2) (x_1 -x_2)^2 dx_1 dx_2,\!\!
\end{equation}
where $\nu_{iy;\sigma}^{}(x_1) = (x_1^2-y^2){e^{-\frac{1}{2\sigma^2} x_1^2}}/\sqrt{2\pi}$ is the density of the signed a-priori measure. 
 For any fixed $\dot\varsigma$ in the admitted class we then have
\begin{equation}\label{muFUNCTIONALtendsTOrhoFUNCTIONALforFIXEDrho}
\lim_{N\to\infty} \frac1N {\cal H}_{y;\sigma}^{(N)}(\dot\varsigma^{\otimes N}) = \hFUNC_{y;\sigma}^{}(\dot\varsigma).
\end{equation}

 Still inspired by the proof of Theorem \ref{thm3} we may now adopt the working hypothesis that the limit points of 
$\{\frac1N\ln E_N^{}(iy;\sigma)\}_{N\in\Nset}^{}$ along positive subsequences of $N\mapsto E_N^{}(iy;\sigma)$ are captured by the
critical values of $\hFUNC_{y;\sigma}^{}(\dot\varsigma)$ over the set of absolutely continuous densities $\dot\varsigma$ of \emph{signed measures} 
$\varsigma$ with finite second moment and for which the relative signed entropy 
$-\int_{\Rset} \dot\varsigma(x_1) \ln \frac{\dot\varsigma(x_1)}{\nu_{iy;\sigma}^{}(x_1)} dx_1 \in\Rset$
(cf. Definition \ref{DEFsignRELentro} for ${\cal H}_{y;\sigma}^{(N)}(\dot\varsigma^{(N)})$). 
 However, as we will see in a moment, this working hypothesis will lead to the putative limiting curves shown in section \ref{ASYMPzIMAG}
only when $\sigma>1$, yet empirically it turns out to be false when $0<\sigma<1$;
recall that the even-$N$ subsequences are positive for all $\sigma>0$.

 Note that at $\sigma=1$ the coefficient of the ``interaction term'' in (\ref{rhoFUNCTIONALsigned}) changes sign.

 Interestingly enough, the Euler--Lagrange equations for the real critical points of $\hFUNC_{y;\sigma}^{}(\dot\varsigma)$ also
have complex solutions of the type $e^{i ax}\times$ an admissible signed density (up to normalization), with $a\in\Rset$,
and if we adopt the generalized working hypothesis that the limit points of 
$\{\frac1N\ln E_N^{}(iy;\sigma)\}_{N\in\Nset}^{}$ along positive subsequences of $N\mapsto E_N^{}(iy;\sigma)$ are captured by the 
real parts (denoted $\Re\eett$) of $\hFUNC_{y;\sigma}^{}(\dot\varsigma)$ evaluated with the
densities $\dot\varsigma$ of such \emph{complex solutions} to the Euler--Lagrange equations,
then we will be able to find all the positive putative limiting curves displayed in section \ref{ASYMPzIMAG}.
 We thus have to enlarge the domain of definition of the functional (\ref{rhoFUNCTIONALsigned}) in order to include complex
critical points. 

\begin{defi}\label{complFREEen}
 We define the \emph{complex continuum free-energy functional} $\hFUNC_{y;\sigma}^{}(\dot\varsigma)$ to be given 
by (\ref{rhoFUNCTIONALsigned}) with $\frac{\dot\varsigma(x_1)}{\nu_{iy;\sigma}^{}(x_1)} = \iota(x_1)\rho(x_1)$ for 
$\iota(x_1) \in U(1)$ and $\rho(x_1)\in\Rset_+$, with $\dot\varsigma(x_1)$ normalized to 1, having finite second moment, 
and having what we call a \emph{complex entropy relative to a signed measure} given by 
\begin{equation}\label{eq:entropyCOMPEXsigned}
{\cal H}^{(1)}_{y;\sigma}(\dot\varsigma) 
:=  -\int_{\Rset} \dot\varsigma(x_1) \ln \frac{\dot\varsigma(x_1)}{\nu_{iy;\sigma}^{}(x_1)} dx_1 \in\Cset.
\end{equation}
\end{defi}
 
\begin{rema}
 Given $\iota(x_1) = e^{i\vartheta(x_1)}$ for some $\vartheta(x_1)\in\Rset$, the complex entropy (\ref{eq:entropyCOMPEXsigned}) is 
defined unambiguously, i.e. 
\begin{equation}\label{eq:entropyCOMPEXsignedEVAL}
{\cal H}^{(1)}_{y;\sigma}(\dot\varsigma) 
=
-\int_{\Rset} \dot\varsigma(x_1) \big[\ln \rho(x_1) + i\vartheta(x_1)\big] dx_1 .
\end{equation}
 However, $\!\vartheta$ is generally defined by $\dot\varsigma$ only up to an additive odd-integer multiple of~$i\pi$.
\end{rema}

 With the help of Definitions \ref{DEFcomplexRELentroSIGNED} and \ref{complFREEen} we
now state, and then vindicate, our first \emph{``complex entropy principle relative to a signed measure''} --- about the set 
of possible limit points, denoted ``limpt,'' of the sequences $\{\frac1N\ln E_N^{}(iy;\sigma)\}_{N\in\Nset}^{}$, $y\in\Rset$:
\begin{conj}\label{conjSIGNEDentropy} 
 For each positive subsequence of $\{E_N^{1/N}(iy;\sigma)\}_{N\in\Nset}^{}$, $y\in\Rset$, we have
\begin{equation}\label{eq:entropyCRITptPOS}
\limpt
{\textstyle\frac{1}{N'}}{\mathrm{crit}}_{\varsigma^{(N')}}
{\cal H}_{y;\sigma}^{(N')}(\dot\varsigma^{(N')})
\subset \Re\eett( {\mathrm{crit}}_\varsigma\, \hFUNC_{y;\sigma}^{}(\dot\varsigma)),
\end{equation} 
while for each negative subsequence of $\{E_N^{1/N}(iy;\sigma)\}_{N\in\Nset}^{}$, $y\in\Rset$, we have
\begin{equation}\label{eq:entropyCRITptNEG}
\limpt
{\textstyle\frac{1}{N'}}{\mathrm{crit}}_{\varsigma^{(N')}}
{\cal H}_{y;\sigma}^{(N')}(\dot\varsigma^{(N')})
\subset  {\mathrm{crit}}_\varsigma\, \hFUNC_{y;\sigma}^{}(\dot\varsigma).
\end{equation} 
\end{conj}

\begin{rema}
 While ${\cal H}_{y;\sigma}^{(N)}(\dot\varsigma^{(N)})$ has a unique critical point, and unique limits presumably exist
for the positive and the negative subsequences of $\{E_N^{1/N}(iy;\sigma)\}_{N\in\Nset}^{}$, respectively, the 
natural logarithm of the negative subsequences is would be given only up to an undetermined additive odd multiple of $i\pi$. 
 Also $\hFUNC_{y;\sigma}^{}(\dot\varsigma)$ ``suffers'' from the same ``additive $i\pi$ non-uniqueness.'' 
 In addition, the critical points of $\hFUNC_{y;\sigma}^{}(\dot\varsigma)$ are generally not unique;
hence, the set-theoretical inclusion in (\ref{eq:entropyCRITptPOS}) and (\ref{eq:entropyCRITptNEG}).
\end{rema}

 We offer a vindication for why we call Conjecture \ref{conjSIGNEDentropy} a ``conjecture'' and not merely a ``surmise.''
 Computing the Euler--Lagrange equation for $\hFUNC_{y;\sigma}^{}(\dot\varsigma)$ we obtain
\begin{equation}\label{rhoEULERLAGRANGEagain}
\dot\varsigma (x_1) = 
\frac{(x_1^2-y^2)e^{-\frac{1}{2\sigma^2} x_1^2 - \frac12(1-\frac{1}{\sigma^2})\int_\Rset(x_1-\tilde{x})^2\dot\varsigma(\tilde{x})d\tilde{x}}}
{\int_\Rset(\hat{x}^2-y^2)e^{-\frac{1}{2\sigma^2}\hat{x}^2-\frac12(1-\frac{1}{\sigma^2})\int_\Rset(\hat{x}-\tilde{x})^2\dot\varsigma(\tilde{x})
d\tilde{x}}d\hat{x}},
\end{equation}
which is precisely (\ref{rhoEULERLAGRANGE}) with $z=iy$ for $y\in\Rset$.
 Since the solutions of the Euler--Lagrange equation (\ref{rhoEULERLAGRANGEagain}) are generally no longer probability densities,
all the calculations from section \ref{ASYMPzREAL} apply formally, but the meaning of the symbols is altered. 
 For instance, (\ref{rhoEULERLAGRANGEagain}) can be simplified to yield the functional form of $\dot\varsigma(x_1)$ explicitly,
\begin{equation}\label{rhoEULERLAGRANGEbAGAIN}
\dot\varsigma (x_1) = \frac{(x_1^2-y^2)e^{-\frac12  x_1^2 + (1-\frac{1}{\sigma^2})m x_1^{}}}
{\int_\Rset (\hat{x}^2-y^2)e^{-\frac12  \hat{x}^2 + (1-\frac{1}{\sigma^2}) m\hat{x}}d\hat{x}};
\end{equation}
however, $m:=\int \tilde{x}\dot\varsigma(\tilde{x}){\rm{d}}\tilde{x}$ is not a ``mean of $\dot\varsigma$'' since $\dot\varsigma$ is generally complex
or at least changes sign.
 Therefore, $m$ now is rather more akin to a dipole moment.

 As before, $m$ obeys its own fixed point equation, obtained by multiplying (\ref{rhoEULERLAGRANGEbAGAIN}) by $x_1$ and integrating, 
which yields
\begin{equation}\label{rhoEULERLAGRANGEcAGAIN}
  m  = m (1-\tfrac{1}{\sigma^2})\, \frac{3-y^2 +  (1-\frac{1}{\sigma^2})^2 m^2}{1-y^2 +  (1-\frac{1}{\sigma^2})^2 m^2}.
\end{equation}
 This is {exactly} the fixed point equation (\ref{rhoEULERLAGRANGEc}) with $z=iy$, but since $\dot\varsigma$ no longer needs to be positive,
we don't need to look only for real solutions of (\ref{rhoEULERLAGRANGEcAGAIN}).
 Once again, like (\ref{rhoEULERLAGRANGEc}) so also (\ref{rhoEULERLAGRANGEcAGAIN}) is always solved by $m=0\ (=:m_0^{})$.
 However, if $\sigma^2\neq 1$ and $y^2 \neq 3-2\sigma^2$, then solutions $m\neq 0$ do exist as well, satisfying
\begin{equation}\label{rhoEULERLAGRANGEdAGAIN}
\frac{\sigma^2}{\sigma^{2}-1}  =  \frac{\sigma^4(3-y^2)  +  (\sigma^{2}-1)^2 m^2}{\sigma^4(1-y^2) +  (\sigma^{2}-1)^2 m^2};
\end{equation}
these are given by
\begin{equation}\label{rhoEULERLAGRANGEeAGAIN}
m^{2}_\pm  = \sigma^4 \frac{2(\sigma^2 -1)-1+y^2}{(\sigma^2-1)^2}.
\end{equation}
 If $y^2 >3-2\sigma^2$, then $m_+ = - m_- >0$ is real.
 But if $y^2 < 3-2\sigma^2$, which is possible iff $\sigma^2<3/2$, 
then purely imaginary $m_+ = - m_-$ exist, being complex conjugates of each other.
 We will admit all of these solutions.

 Accordingly, the pertinent solutions of the Euler--Lagrange equation for $\dot\varsigma$ are now denoted by 
$\dot\varsigma_{iy;\sigma}^{(0)}$ 
and 
$\dot\varsigma_{iy;\sigma}^{(\pm)}$,
 respectively.
 Note that $\dot\varsigma_{iy;\sigma}^{(0)}(x_1)$ is an even function of $x_1$, 
while the $\dot\varsigma_{iy;\sigma}^{(\pm)}(x_1)$ are not --- both are mirror images of each other if $m_\pm$ are real, and complex 
conjugates of each other if $m_\pm$ are purely imaginary. 

  Next we evaluate the complex free-energy functional for the  solutions $\dot\varsigma_{iy;\sigma}^{(\times)}$. 
 Using nothing more than the fixed point equations yields the identity
\begin{equation}\label{hFUNCevalCRITpt}
\hFUNC_{y;\sigma}^{}(\dot\varsigma_{iy;\sigma}^{(\times)})  = 
\ln {\tfrac{1}{\sqrt{2\pi}}\int_\Rset (x_1^2-y^2)e^{-\frac12  x_1^2 + (1-\frac{1}{\sigma^2}) m_\times^{}x_1^{}}dx_1} 
- 
\tfrac12 (1-\tfrac{1}{\sigma^2})m_\times^2.
\end{equation}
 Straightforward computation yields for the above integrals 
\begin{eqnarray}
\tfrac{1}{\sqrt{2\pi}}{\int_\Rset (x_1^2-y^2)e^{-\frac12  x_1^2 + (1-\frac{1}{\sigma^2}) m_0^{}x_1^{}}dx_1} 
&=&  1-y^2 ,\label{normalizingINTEGRALrhoNULL}\\
\tfrac{1}{\sqrt{2\pi}}{\int_\Rset (x_1^2-y^2)e^{-\frac12  x_1^2 + (1-\frac{1}{\sigma^2}) m_\pm^{}x_1^{}}dx_1} 
&=&  2 (\sigma^2-1) e^{\frac12(2\sigma^2-3+y^2)}.\label{normalizingINTEGRALrhoPM}
\end{eqnarray}
 Thus, for $m = 0$ we find
\begin{equation}\label{hFUNCnull}
\hFUNC_{y;\sigma}^{}(\dot\varsigma_{iy;\sigma}^{(0)})  =  \ln (1 - {y^2}),
\end{equation}
while for $m = m_\pm^{} \neq 0$ (which is possible iff $\sigma\neq 1$) we find
\begin{equation}\label{hFUNCpm}
\hFUNC_{y;\sigma}^{}(\dot\varsigma_{iy;\sigma}^{(\pm)})  = \ln [2(\sigma^2 -1)] + \frac{1-y^2}{2(\sigma^2-1)} -1.
\end{equation}

 Note that (\ref{normalizingINTEGRALrhoNULL}) changes sign at $y^2=1$, and (\ref{normalizingINTEGRALrhoPM}) changes sign at $\sigma^2=1$. 
 Since the evaluation of $\hFUNC_{y;\sigma}^{}(\dot\varsigma_{iy;\sigma}^{(\times)})$ involves the natural logarithm of the sign-changing normalizing 
integrals (\ref{normalizingINTEGRALrhoNULL}) and (\ref{normalizingINTEGRALrhoPM}), it is clear that its complex analytic continuation 
is needed when $m=0$ and $y^2>1$, or when $m= m_\pm$ and $\sigma^2<1$.
 Thus we have
\begin{equation}\label{hFUNCnullEXTENDED}
\ln (1 - {y^2}) = \ln |1 - {y^2}|  + i (2l-1)\pi \, {\boldsymbol{H}}(y^2-1),\ l\in\Zset,
\end{equation}\vspace{-15pt}
\begin{equation}\label{hFUNCpmEXTENDED}
 \ln (\sigma^2 -1) = \ln |\sigma^2 -1| + i (2l-1)\pi \, {\boldsymbol{H}}(1-\sigma^2),\ l\in\Zset,
\end{equation}
where $\boldsymbol{H}$ is Heaviside's function; again, $l$ remains undetermined (cf. Remark \ref{ipi}).

 With (\ref{hFUNCnull}), (\ref{hFUNCpm}),  and (\ref{hFUNCnullEXTENDED}), (\ref{hFUNCpmEXTENDED}), Conjecture \ref{conjSIGNEDentropy} 
is vindicated. 
 Indeed, note that by exponentiating left- and right-hand sides we capture the complete catalog of putative limiting curves
of convergent subsequences of $\{E_N^{1/N}(iy;\sigma)\}_{N\in\Nset}^{}$, $y\in\Rset$, which we have listed in Conjecture \ref{conjIMz},
and in section \ref{ASYMPzIMAG} in our empirical study of these sequences.
 In particular, Conjecture \ref{conjSIGNEDentropy} automatically produces the empirical ``sign flips'' listed in Conjecture \ref{conjIMz},
which do not follow by simply replacing $z=x$ with $z=iy$ in the real-$z$ formulas for $L(z;\sigma)$ obtained in section \ref{ASYMPzREAL}.

 Conjecture \ref{conjSIGNEDentropy} does not explain all of Conjecture \ref{conjIMz}.
 Namely, the portions of the curves $y\mapsto \exp(\Re\eett(\hFUNC_{y;\sigma}^{}(\dot\varsigma_{iy;\sigma}^{(\pm)})))$ 
and $y\mapsto \exp(\hFUNC_{y;\sigma}^{}(\dot\varsigma_{iy;\sigma}^{(\pm)}))$ colored in light red in Fig.~\ref{NrootENforIMAGzL1L2curves}
do not seem to be limiting curves of any subsequence of $\{E_N^{1/N}(iy;\sigma)\}_{N\in\Nset}^{}$, $y\in\Rset$. 
 These belong to the parameter regimes

(i) $1< \sigma^2$ and $y^2 >  1+2\kappa_*^{} (\sigma^2-1)$,

(ii) $1< \sigma^2<3/2$ and $y^2 < 3-2\sigma^2$,

(iii) $\sigma^2<1$ and $y^2\geq 3-2\sigma^2$ or $y^2 <  1+2\kappa_*^{} (\sigma^2-1)$,

\noindent
where the limit curve seems to correspond to $m=m_0^{} (=0)$ critical points. 
 Outside the parameter regimes listed in (i), (ii), (iii), the limit curve seems to be captured by $m=m_\pm^{}$ critical points. 
 Our heuristic entropy principle in Conjecture \ref{conjSIGNEDentropy} does not deliver a selection principle 
for whether an $m=0$ or $m=m_\pm^{}$ critical point should be picked.
 A tentative selection principle is implied by combining Conjecture \ref{conjSIGNEDentropy} with
the next conjecture, cf. Fig.~\ref{NrootENforIMAGzL1L2curves}
and the ensuing double figures in subsection \ref{ASYMPzIMAGevenNoddNnumerics}.
\begin{conj}\label{conjSELECTION}
The limit $N\to\infty$ of the even-$N$ subsequence of $\{E_N^{1/N}(iy;\sigma)\}_{N\in\Nset}^{}$ 
is a piecewise real analytical function $y\in\Rset$ with analyticity defects at the meeting points
of the two curves 
$y\mapsto\exp(\Re\eett(\hFUNC_{z;\sigma}^{}(\dot\varsigma_{iy;\sigma}^{(0)})))$ and 
$y\mapsto\exp(\Re\eett(\hFUNC_{z;\sigma}^{}(\dot\varsigma_{iy;\sigma}^{(\pm)})))$ 
(i.e. whenever they intersect or touch).
  Moreover, for sufficiently large negative $y$, the limit is given by
the larger one of the two curves in the prolate regime ($\sigma>1$), 
and by the smaller one in the oblate regime ($\sigma<1$).

 The limit $N\to\infty$ of the odd-$N$ subsequence of $E_N^{1/N}(iy;\sigma)$ is monotonic decreasing in $|y|$. 
 It coincides with the limit of the even-$N$ subsequence where that one is monotone decreasing in $|y|$, too, 
and it is the negative thereof where that one is increasing in $|y|$.
\end{conj}

 In Conjecture \ref{conjSELECTION} we have assumed that the even-$N$ and odd-$N$ subsequences of $\{E_N^{1/N}(iy;\sigma)\}_{N\in\Nset}^{}$ 
each have a limit $N\to\infty$.
 Note that Conjecture \ref{conjSIGNEDentropy} implies that the limit curves are piecewise analytic functions; thus
Conjecture \ref{conjSIGNEDentropy} and Conjecture \ref{conjSELECTION} in concert
imply that the limit of the even-$N$ subsequence is exchanged between the two curves 
$y\mapsto\exp(\Re\eett(\hFUNC_{z;\sigma}^{}(\dot\varsigma_{iy;\sigma}^{(0)})))$ and 
$y\mapsto\exp(\Re\eett(\hFUNC_{z;\sigma}^{}(\dot\varsigma_{iy;\sigma}^{(\pm)})))$ 
each time they meet, and that at these same meeting points also the 
limit of the odd-$N$ subsequence is exchanged between the two curves 
$y\mapsto\exp(\hFUNC_{z;\sigma}^{}(\dot\varsigma_{iy;\sigma}^{(0)}))$ and 
$y\mapsto\exp(\hFUNC_{z;\sigma}^{}(\dot\varsigma_{iy;\sigma}^{(\pm)}))$.
 Lastly, Conjecture \ref{conjSELECTION} implies that for sufficiently large $|y|$ the limit curves coincide with those obtained 
with i.i.d. standard normal random variables.

\begin{rema} The selection principle implied jointly by Conjectures \ref{conjSIGNEDentropy} and \ref{conjSELECTION} is called tentative because
it is tied to our multivariate normal random variable model. 
 Some of its features may be very specific and not generalize to other models.
\end{rema}

 Our heuristic reasoning in this section of the limit points of $\{E_N^{1/N}(iy;\sigma)\}_{N\in\Nset}^{}$, $y\in\Rset$,
has been inspired by our proof of Theorem \ref{thm3}.
 The same proof suggests to also address the limit points of the sequences of marginal measures. 
 Indeed, it may seem strange at first why complex solutions to the Euler--Lagrange equations should play any role at all in the 
evaluation of the limit points of the \emph{real} sequence of marginal measures defined by the partial integrals of the integrands of
$\{E_N^{}(iy;\sigma)\}_{N\in\Nset}^{}$, $y\in\Rset$.
 However, recalling our real-$z$ Corollary \ref{limPTSofMEASURES}, it is reasonable to analogously formulate
\begin{conj}\label{conjMARGINALS}
 For each $z=iy$ with $y\in\Rset$, and $j\in\Nset$, the limit points of the sequence
$N\mapsto\dot\varsigma_{iy;\sigma}^{(j|N)}(x_1,...,x_j)$ are given by either 
$\dot\varsigma_{iy;\sigma}^{(0)}(x_1)\cdots\dot\varsigma_{iy;\sigma}^{(0)}(x_j)$
or
$\frac12[\dot\varsigma_{iy;\sigma}^{(+)}(x_1)\cdots\dot\varsigma_{z;\sigma}^{(+)}(x_j) +\dot\varsigma_{iy;\sigma}^{(-)}(x_1)\cdots\dot\varsigma_{iy;\sigma}^{(-)}(x_j)]$.
\end{conj}

 Recall that the arithmetic means of complex conjugates of each other yield their real part; of course, these real limit points of the
signed measures are themselves signed measures. 
 In particular, when $n=1$, their arithmetic mean reads
\begin{equation}
\frac12\left[\dot\varsigma_{iy;\sigma}^{(+)}(x_1)+\dot\varsigma_{iy;\sigma}^{(-)}(x_1)\right] 
=
\frac{
            (x_1^2-y^2)e^{-\frac12  x_1^2}\cosh\left((1-\frac{1}{\sigma^2})m_+^{} x_1\right)}{\displaystyle
\int_{\Rset}(\hat{x}_1^2-y^2)e^{-\frac12  \hat{x}_1^2}\cosh\left((1-\tfrac{1}{\sigma^2})m_+^{} \hat{x}_1\right)d\hat{x}_1}.
\end{equation}
 Recall that $m_+^{}$ can be either real or purely imaginary, depending on $\sigma$.

\section{Summary and Outlook}\label{Outlook}\vspace{-10pt}

 Our partly rigorous and partly computer-assisted study of the complex expected polynomials 
${\rm Exp}_N^{}\!\left[\prod_{1\leq k\leq N} (X_k^2+z^2)\right]$ with multivariate normal zeros $\pm iX_k$ 
should leave no doubt that their large degree asymptotics when $iz\in\Rset$ is captured by
our heuristic ``signed relative entropy principle,'' Conjecture \ref{conjSIGNEDentropy}, and
more refined by Conjectures \ref{conjSIGNEDentropy} and \ref{conjSELECTION} in concert, 
involving the novel notion of an entropy of a signed or complex measure relative to a signed a-priori measure.

 To rigorously establish these signed relative entropy principles is the challenging goal.
 Our proof of the large degree asymptotics when $z\in\Rset$ makes it plain that new ideas will be needed
to accomplish this feat: namely, with $z\in\Rset$ we could make explicit use of the pointwise positivity 
of the integrand of $E_N^{}(z;\sigma)$, of the concavity of the relative entropy functional with a-priori probability measure,
and also of the subadditivity of this traditional entropy functional --- none of these technical ingredients are available when
the a-priori measure is not a positive measure!

 One of the referees asked whether ``there is some hope to prove that the solution conjectured in sections 6 and 7 is correct, 
bypassing the variational principle and obtaining directly the mean-field equation for the density ...?'' 
 This is certainly conceivable. 
 In \cite{CLMPa} the analogous feat was achieved for the prescribed Gauss curvature equation with non-sign-changing $K(\sV)$
(in our current context), and it would be surprising if it should not be possible for sign-changing expected value functionals.
 In particular, with the help of the other referee's trick to evaluate $E_N^{}(z;\sigma)$ by ``chipping in'' an extra Gaussian random
variable, it ought to be possible to obtain control of the finite-$N$ expressions for all the $n$-th marginal signed measures for our 
Gaussian ``signed ensembles,'' and to prove their convergence to the conjectured limits. 

 Our random polynomials with multivariate normal zeros only served the purpose of supplying an explicitly solvable test case for our general ideas.
 In particular, recall that the original problem which got this research started was the question by Alice Chang 
whether the statistical mechanics techniques used in \cite{CLMPa,KiesslingCPAM,CLMPb,KieLeb,ChaKieDMJ,KiesslingPHYSICA,KiesslingWangJSP}
to derive the prescribed Gauss curvature equation on $\Rset^2$ or $\Sset^2$ 
(see, e.g., \cite{KazdanWarner,ChangYangA,Han}) from Onsager's equilibrium ensemble of $N\in\Nset$ point vortices 
in the limit $N\to\infty$ could be generalized from Gauss curvature functions $K(\sV)\propto \pm e^{\psi(\sV)}$ 
(where $\psi$ is a real, given stream function of the point vortex model) to sign-changing Gauss curvature functions. 
 There is no doubt in my mind anymore that the answer will be ``Yes!'' 
 Also higher-dimensional $Q$-curvature problems \cite{KiesslingPHYSICA,Ali} should be tractable with this method.

 An interesting intermediate step, sort of but not literally ``half-way'' between our multivariate normal random zero problem and the 
point vortex problem with ``signed ensemble measure'' (pertinent to a sign-changing prescribed Gauss curvature problem), is to consider 
expected random polynomials 
 ${\rm Exp}_N^{}\!\left[\prod_{1\leq k\leq N} (X_k^2+z^2)\right]$ with random zeros $\pm iX_k$ 
distributed according to the eigenvalue laws for some random matrix ensembles \cite{MehtaBOOK,ForresterBOOK},
in particular the Gaussian orthogonal, unitary, or symplectic ensembles (usually abbreviated GOE, GUE, GSE).
 More precisely, if $\boldsymbol{R}$ is any $N\times N$ random matrix picked from, say, GOE or GUE, then the block-diagonal 
matrix $\boldsymbol{M}=\mbox{diag}(\boldsymbol{R},-\boldsymbol{R})$ has real
zeros which come in pairs $\pm X_k$, and its characteristic polynomial is 
$\det(\boldsymbol{M}-\lambda\boldsymbol{I}) = \prod_{1\leq k\leq N} (X_k^2-\lambda^2)$. 
 Thus we see that the expected polynomials  ${\rm Exp}_N^{}\!\left[\prod_{1\leq k\leq N} (X_k^2+z^2)\right]$ in this case
are essentially the expected characteristic polynomials of the block-diagonal $\boldsymbol{M}$ matrices. 
 In the GOE and GUE the $X_k$ are distributed by a probability measure (up to a normalizing factor) given by
$\prod\!\!\!\prod\limits_{\hspace{-14pt}1\leq k<l\leq N}\!\! |x_k-x_l|^\beta
\!\!\!\!\prod\limits_{1\leq n\leq N}\!\!\!{e^{-\frac{1}{2} x_n^2}}$, with $\beta=1$ for GOE and $\beta=2$ for GUE.
 A similar formula with $\beta=4$ holds for the GSE, and extensions to so-called $\beta$-ensembles with
arbitrary $\beta>0$ have been constructed in \cite{betaRM}.
 It should be possible to evaluate the finite-$N$ expressions exactly and to study whether they
converge in a limit $N\to\infty$ to critical points of the signed relative entropy functional that is associated with this problem in an analogous
manner as the signed relative entropy functional in our multivariate normal random zero problem.
 For this one needs to rescale $\beta\mapsto \beta/N$, and the rescaled $\beta$ would seem to play an analogous role to our $1/\sigma$.
 Whether this problem has anything more interesting to contribute than our multivariate normal random zero problem\footnote{Incidentally, 
 our study shows that the multivariate normal random variables with $\sigma\neq 1$ will not be amongst the laws asked for in {\bf{Q1}}.}
to answering the introductory questions {\bf{Q1}} -- {\bf{Q3}}, or {\bf{Q2}${}^\prime$}, is to be seen.
 In any event, since the eigenvalues are jointly distributed like point vortices in a canonical ensemble confined by some Gaussian a-priori 
measure, yet restricted to the line $\Rset\subset\Rset^2$, it is clear that the expected characteristic polynomial
problem brings us a step closer to our signed point vortex ensemble approach to 
the prescribed sign-changing Gauss curvature problem. 

 More generally, a relative entropy principle such as formulated in Conjecture \ref{conjSIGNEDentropy}, 
relative to signed or even complex a-priori measures, will presumably be useful whenever
one needs to evaluate the large-$N$ asymptotics of expected values of products
$\prod_{k=1}^{N} g(X_k)$ when $g$ is a real sign-changing or complex function, 
the expected value taken w.r.t. a permutation-symmetric law of $N$ non-i.i.d. random variables; see (\ref{ExpOFgOVERfPROD}).

 The prototypical example which leads to complex expressions is the characteristic function of the sum of those random variables, 
i.e. $g(X) = e^{itX}$.
 This ventures further than the examples discussed in this paper because the a-priori measure will then be complex.
 In that case analytical continuation of the moment generating function will
obviously play an important role; one referee noted \cite{ShamisZeitouni} as a recent example --- these authors point
to the Lee--Yang circle theorem as a motivating example.

 Analytic continuation also played a role in our study with signed a-priori measure, on the one hand because the logarithm of
a negative real number differed by some non-zero odd-integer multiple of $i\pi$ from the logarithm of its absolute value, 
on the other because we had to admit also complex solutions to the Euler--Lagrange equations for the critical points.
 At the same time, the more refined details such as those stated in Conjecture \ref{conjSELECTION} compared
with the relatively simple content of Theorem \ref{thm3} make it plain that it would be na$\ddot{\mbox{\i}}$ve to expect 
results to be given ``merely'' by analytical continuation of the maximum entropy results for probability measures. 
 The final form of a refined signed or complex entropy principle which delivers the selection criteria for the piecewise 
analytical branches can only be expected to emerge after studying many different models.
 At present our selection rules are only tentative because they are tied to our multivariate normal random variable model. 

 Other expected value problems of real sign-changing $g$ are: ``Gain vs. Loss'' in population dynamics, 
electric charge imbalances, etc.
 Even the weird idea of ``negative probabilities in quantum mechanics'' \cite{Dirac} 
might be translated into something meaningful using sign-changing expected values w.r.t. true probabilities.

 I end by relaying an observation made by both referees.
 In the words of one of the referees, our conjecture(s)
``reminds one of the replica trick that has been successfully used in the solution of some disordered statistical mechanics mean-field 
models, see for instance \cite{MPV}.
 In this case it happens often that one looks for stationary points and not maxima or minima of a suitable free energy functional.
 Is this only a coincidence or is there a more stringent connection?''
 Since I do not have an answer, I take the opportunity to pass this interesting question on to expert readers.

\newpage
\section*{Appendix}

\subsection*{A. Alternate proof of Corollary~\ref{coro:var}.}

\smallskip
\noindent
\textit{Proof of Corollary~\ref{coro:var}}:

By the permutation symmetry  in $\{x_1,...,x_N\}$ of the pdf given by the integrand of (\ref{EXPECTfTWO}),
for the pertinent random variables $\{X_1,...,X_N\}$ we have 
\begin{equation}\label{VarXasAVE}
N{\rm{Var}}_N\!\left[X_1\right] = {\textstyle{\sum\limits_{k=1}^N}}{\rm{Var}}_N\!\left[X_k\right],
\end{equation}
and for these (as for any) mean-zero random variables $\{X_1,...,X_N\}$, we have
\begin{equation}\label{VarXasExpXsqr}
{\textstyle{\sum\limits_{k=1}^N}}{\rm{Var}}_N\!\left[X_k\right] = {\rm{Exp}}_N\left[{\textstyle{\sum\limits_{k=1}^N}} X_k^2\right].
\end{equation}
 Under a Euclidean transformation (here: rotation) from $\{x_1,...,x_N\}$ to $\{y_1,...,y_N\}$, 
\begin{equation}\label{ExpXsqrExpYsqr}
{\rm{Exp}}_N\left[{\textstyle{\sum\limits_{k=1}^N}} X_k^2\right] = {\rm{Exp}}_N\left[{\textstyle{\sum\limits_{k=1}^N}} Y_k^2\right].
\end{equation}
 By the proof of Prop. 2.1, 
\begin{equation}\label{ExpYsqr}
{\rm{Exp}}_N\left[{\textstyle{\sum\limits_{k=1}^N}} Y_k^2\right] = {\rm{Exp}}_N\left[Y_1^2\right] + (N-1){\rm{Exp}}_N\left[Y_2^2\right]
= \sigma^2 + N-1.
\end{equation}
 The proof is complete. \hfill QED

\begin{rema}
There is no unique rotation in $\Rset^N$ which maps $\{x_1,x_2,...,x_N\}$ into  
$y_1:={\textstyle{\frac{1}{\sqrt{N}}}}\sum_{1\leq k \leq N} x_k$;
even after stipulating that the $x_1$ axis should be rotated into the $y_1$ axis 
we are left to choose an arbitrary $SO(N-1)$ rotation ``about the $y_1$-axis'' to fix the remaining $N-1$ axes.
 Such a freedom may be useful to simplify the expected value integrals in the $y$ coordinates, but we have not
explored this here. 
\end{rema}

\newpage

\subsection*{B. Alternate proof of the positivity of $E_{2K}^{}(z;\sigma);\; K\in\Nset$, $\sigma\geq1$}

\begin{prop}\label{propPOSevenN}
 Let $K\in\Nset$.
Then for $\sigma\geq 1$ and $z^2\in\Rset$ we have $E_{2K}(z;\sigma)\geq 0$, with ``$=0$'' iff $\sigma=1$ and $z^2=-1$.
\end{prop}

\smallskip
\noindent
\textit{Proof of Proposition \ref{propPOSevenN}}:

 We split the $N=2K$ variables into two disjoint sets of size $K$, keeping the notation $x_n^{}$ for $n=1,...,K$ and 
renaming $x_n^{}=:\tilde{x}_k^{}$ if $n=K+k$ with $k=1,...,K$.
 Note that
\begin{equation}\label{interactionXYsplitAGAIN}
\sum\!\sum_{\hspace{-18pt}1\leq k,l\leq 2K} (x_k-x_l)^2
=
\sum\!\sum_{\hspace{-16pt}1\leq k,l\leq K} (x_k-x_l)^2
+
\sum\!\sum_{\hspace{-16pt}1\leq k,l\leq K} (\tilde{x}_k-\tilde{x}_l)^2
+
2 \sum\!\sum_{\hspace{-16pt}1\leq k,l\leq K} (x_k-\tilde{x}_l)^2,
\end{equation}
and further that 
\begin{equation}\label{interactionXYsplitBagain}
 \sum\!\sum_{\hspace{-16pt}1\leq k,l\leq K} (x_k-\tilde{x}_l)^2
=
K\!\!\sum_{1\leq k\leq K}\!\!x_k^2 + K\!\!\sum_{1\leq l\leq K}\!\!\tilde{x}_l^2 - 2\sum\!\sum_{\hspace{-16pt}1\leq k,l\leq K}x_k\tilde{x}_l.
\end{equation}
 Again recalling that $(x_1^{},...,x_K^{})=:\vec{x}$, and defining
\begin{equation}
\displaystyle
\prod\!\!\prod_{\hspace{-16pt}1\leq k<l\leq K}\! e^{-\frac{1}{4K}(1-\frac{1}{\sigma^2}) (x_k-x_l)^2}\!\!\!
\prod_{1\leq n\leq K}\!\!(z^2+ x_n^2)e^{- \frac{1}{4}(1+\frac{1}{\sigma^2}) x_n^2}
=: \Upsilon(\vec{x})
\end{equation}
and
\begin{equation}
\displaystyle
\prod\!\!\prod_{\hspace{-16pt}1\leq k,l\leq K}\! e^{\frac{1}{2K}(1-\frac{1}{\sigma^2}) x_k\tilde{x}_l}
=: \Omega(\vec{x},\vec{\tilde{x}}),
\end{equation}
we have, for $K\in\Nset$,
\begin{equation}\label{EXPECTmoreTHANoneXY}
\begin{array}{rl}
\hspace{-20pt}
E_{2K}^{}(z;\sigma)\! = &  \!\!
{\textstyle\frac{1}{\sigma{({2\pi}})^K}}
\displaystyle
\int_{\Rset^K}\!\! \int_{\Rset^K} 
\Upsilon(\vec{x})
\Omega(\vec{x},\vec{\tilde{x}})
\Upsilon(\vec{\tilde{x}})
d^{^K}\!\!x\,d^{^K}\!\!\tilde{x}\\
= & \!\! {\textstyle\frac{1}{\sigma{({2\pi}})^K}} 
\sum\limits_{j=0}^\infty \frac{1}{j!} \left(\frac{1}{2K}(1-\frac{1}{\sigma^2})\right)^j
\displaystyle
\int_{\Rset^K}\!\! \int_{\Rset^K} 
\Upsilon(\vec{x})
\Upsilon(\vec{\tilde{x}})
\Bigl(\,\textstyle\sum\!\sum\limits_{\hspace{-14pt}1\leq k,l\leq K}\!\!x_k\tilde{x}_l\Bigr)^j\!\!
d^{^K}\!\!x\,d^{^K}\!\!\tilde{x}\!\!\!\!\!\\
= & \!\! {\textstyle\frac{1}{\sigma{({2\pi}})^K}} 
\sum\limits_{j=0}^\infty \frac{1}{j!} \left(\frac{1}{2K}(1-\frac{1}{\sigma^2})\right)^j
\displaystyle
\biggl(\int_{\Rset^K}
\Upsilon(\vec{x})
\Bigl(\textstyle\sum\limits_{\; 1\leq k\leq K}\!\!x_k\Bigr)^j
d^{^K}\!\!x\biggr)^2,\hspace{-10pt}
\end{array}
\end{equation}
which, since $\sigma\geq 1$, is manifestly $\geq0$.
 More precisely, r.h.s.(\ref{EXPECTmoreTHANoneXY}) is estimated from below by the $j=0$ contribution, evaluating to
$\frac{1}{\sigma}\Big[\sqrt{2\frac{\sigma^2}{\sigma^2+1} }(z^2 +2\frac{\sigma^2}{\sigma^2+1} )\Big]^{2K}\geq 0$, with ``$=0$'' iff
$\sigma=1$ and $z^2=-1$.
 When $\sigma =1$, the $j=0$ term is also the only contribution to r.h.s.(\ref{EXPECTmoreTHANoneXY}). \hfill QED

\newpage


\end{document}